\documentclass{mscs}

\edef\keptrmdefault{\rmdefault}
\edef\keptsfdefault{\sfdefault}
\edef\keptttdefault{\ttdefault}
\usepackage[math]{iwona}
\edef\rmdefault{\keptrmdefault}
\edef\sfdefault{\keptsfdefault}
\edef\ttdefault{\keptttdefault}

\usepackage{float}
\usepackage{mathpartir}
\usepackage{amsmath}
\usepackage{amssymb}
\usepackage{stmaryrd}
\usepackage{graphicx}
\usepackage{soul}
\usepackage[usenames,dvipsnames]{color}
\usepackage{xspace}
\usepackage{mathtools}
\usepackage{url}
\usepackage{hyperref}
\usepackage{breakurl}
\usepackage[T1]{fontenc}
\usepackage[scaled=0.75]{beramono}
\usepackage{upgreek}
\usepackage{pifont}
\usepackage{framed}
\usepackage{tikz}
\usepackage{dashundergaps}
\usepackage{scalerel}
\usepackage{bbold}
\usepackage{subcaption}
\usepackage{mathdots}
\usepackage{bbm}
\usepackage[export]{adjustbox}[2011/08/13]
\usepackage{accents}
\usetikzlibrary{matrix}
\usetikzlibrary{positioning}

\usetikzlibrary{decorations}
\pgfdeclaredecoration{simple line}{initial}{
  \state{initial}[width=\pgfdecoratedpathlength-1sp]{\pgfmoveto{\pgfpointorigin}}
  \state{final}{\pgflineto{\pgfpointorigin}}
}
\tikzset{
   shift left/.style={decorate,decoration={simple line,raise=#1}},
   shift right/.style={decorate,decoration={simple line,raise=-1*#1}},
}

\DeclareFontFamily{U}{MnSymbolC}{}
\DeclareFontShape{U}{MnSymbolC}{m}{n}{
    <-6>  MnSymbolC5
   <6-7>  MnSymbolC6
   <7-8>  MnSymbolC7
   <8-9>  MnSymbolC8
   <9-10> MnSymbolC9
  <10-12> MnSymbolC10
  <12->   MnSymbolC12}{}
\DeclareFontShape{U}{MnSymbolC}{b}{n}{
    <-6>  MnSymbolC-Bold5
   <6-7>  MnSymbolC-Bold6
   <7-8>  MnSymbolC-Bold7
   <8-9>  MnSymbolC-Bold8
   <9-10> MnSymbolC-Bold9
  <10-12> MnSymbolC-Bold10
  <12->   MnSymbolC-Bold12}{}
\DeclareSymbolFont{MnSyC}{U}{MnSymbolC}{m}{n}

\DeclareMathSymbol{\medsquare}{\mathord}{MnSyC}{106}

\DeclareFontFamily{U}{MnSymbolD}{}
\DeclareFontShape{U}{MnSymbolD}{m}{n}{
    <-6>  MnSymbolD5
   <6-7>  MnSymbolD6
   <7-8>  MnSymbolD7
   <8-9>  MnSymbolD8
   <9-10> MnSymbolD9
  <10-12> MnSymbolD10
  <12->   MnSymbolD12}{}
\DeclareFontShape{U}{MnSymbolD}{b}{n}{
    <-6>  MnSymbolD-Bold5
   <6-7>  MnSymbolD-Bold6
   <7-8>  MnSymbolD-Bold7
   <8-9>  MnSymbolD-Bold8
   <9-10> MnSymbolD-Bold9
  <10-12> MnSymbolD-Bold10
  <12->   MnSymbolD-Bold12}{}
\DeclareSymbolFont{MnSyD}{U}{MnSymbolD}{m}{n}

\DeclareMathSymbol{\Cong}{\mathrel}{MnSyD}{12}

\DeclareFontFamily{U}{matha}{\hyphenchar\font45}
\DeclareFontShape{U}{matha}{m}{n}{
      <5> <6> <7> <8> <9> <10> gen * matha
      <10.95> matha10 <12> <14.4> <17.28> <20.74> <24.88> matha12
      }{}
\DeclareSymbolFont{matha}{U}{matha}{m}{n}

\DeclareMathSymbol{\twoPrime}{3}{matha}{"32}
\DeclareMathSymbol{\threePrime}{3}{matha}{"33}
\DeclareMathSymbol{\fourPrime}{3}{matha}{"34}

\DeclareFontFamily{U}{MnSymbolC}{}
\DeclareFontShape{U}{MnSymbolC}{m}{n}{
    <-6>  MnSymbolC5
   <6-7>  MnSymbolC6
   <7-8>  MnSymbolC7
   <8-9>  MnSymbolC8
   <9-10> MnSymbolC9
  <10-12> MnSymbolC10
  <12->   MnSymbolC12}{}
\DeclareFontShape{U}{MnSymbolC}{b}{n}{
    <-6>  MnSymbolC-Bold5
   <6-7>  MnSymbolC-Bold6
   <7-8>  MnSymbolC-Bold7
   <8-9>  MnSymbolC-Bold8
   <9-10> MnSymbolC-Bold9
  <10-12> MnSymbolC-Bold10
  <12->   MnSymbolC-Bold12}{}
\DeclareSymbolFont{MnSyC}{U}{MnSymbolC}{m}{n}

\DeclareMathSymbol{\MnSymbolvdots}{\mathord}{MnSyC}{6}

\DeclareFontFamily{U}{MnSymbolC}{}
\DeclareFontShape{U}{MnSymbolC}{m}{n}{
    <-6>  MnSymbolC5
   <6-7>  MnSymbolC6
   <7-8>  MnSymbolC7
   <8-9>  MnSymbolC8
   <9-10> MnSymbolC9
  <10-12> MnSymbolC10
  <12->   MnSymbolC12}{}
\DeclareFontShape{U}{MnSymbolC}{b}{n}{
    <-6>  MnSymbolC-Bold5
   <6-7>  MnSymbolC-Bold6
   <7-8>  MnSymbolC-Bold7
   <8-9>  MnSymbolC-Bold8
   <9-10> MnSymbolC-Bold9
  <10-12> MnSymbolC-Bold10
  <12->   MnSymbolC-Bold12}{}
\DeclareSymbolFont{MnSyC}{U}{MnSymbolC}{m}{n}

\DeclareMathSymbol{\smalltriangleright}{\mathord}{MnSyC}{72}
\DeclareMathSymbol{\smalltriangleup}{\mathord}{MnSyC}{73}
\DeclareMathSymbol{\smalltriangleleft}{\mathord}{MnSyC}{74}
\DeclareMathSymbol{\smalltriangledown}{\mathord}{MnSyC}{75}
\DeclareMathSymbol{\filledtriangleright}{\mathord}{MnSyC}{76}
\DeclareMathSymbol{\filledtriangleup}{\mathord}{MnSyC}{77}
\DeclareMathSymbol{\filledtriangleleft}{\mathord}{MnSyC}{78}
\DeclareMathSymbol{\filledtriangledown}{\mathord}{MnSyC}{79}

\DeclareFontFamily{U}{MnSymbolC}{}
\DeclareFontShape{U}{MnSymbolC}{m}{n}{
    <-6>  MnSymbolC5
   <6-7>  MnSymbolC6
   <7-8>  MnSymbolC7
   <8-9>  MnSymbolC8
   <9-10> MnSymbolC9
  <10-12> MnSymbolC10
  <12->   MnSymbolC12}{}
\DeclareFontShape{U}{MnSymbolC}{b}{n}{
    <-6>  MnSymbolC-Bold5
   <6-7>  MnSymbolC-Bold6
   <7-8>  MnSymbolC-Bold7
   <8-9>  MnSymbolC-Bold8
   <9-10> MnSymbolC-Bold9
  <10-12> MnSymbolC-Bold10
  <12->   MnSymbolC-Bold12}{}
\DeclareSymbolFont{MnSyC}{U}{MnSymbolC}{m}{n}

\let\bowtie\relax
\DeclareMathSymbol{\bowtie}{\mathrel}{MnSyC}{38}

\usepackage{bm} 
\usepackage[authoryear]{natbib}
\usepackage[xcolor]{changebar}
\cbcolor{red}

\newcommand*{\lambdaCalculus}{$\lambda$-calculus\xspace}
\newcommand*{\lambdaCalculi}{$\lambda$-calculi\xspace}
\newcommand*{\piCalculus}{$\pi$-calculus\xspace}
\newcommand*{\piCalculi}{$\pi$-calculi\xspace}

\newcommand*{\etal}{et al\xspace}

\newcommand*{\ttt}[1]{\texttt{#1}}

\newcommand*{\eqdef}{\stackrel{\smash{\text{\tiny def}}}{=}}

\newcommand*{\Cdot}{\raisebox{-0.25ex}{\scalebox{1.4}{$\cdot$}}}
\newcommand*{\set}[1]{\{#1\}}

\DeclareSymbolFont{bbsymbol}{U}{bbold}{m}{n}
\DeclareMathSymbol{\bbsemi}{\mathbin}{bbsymbol}{"3B}

\newcommand*{\lowlight}[1]{\textcolor{darkgray}{#1}}
\newcommand*{\sub}[2]{#1_{\lowlight{#2}}}
\renewcommand*{\sup}[2]{#1^{\lowlight{#2}}}

\renewcommand*{\vdots}{\MnSymbolvdots}
\newcommand*{\param}{\cdot}

\newcommand*{\appref}[1]{Appendix~\ref{app:#1}}

\newcommand*{\defref}[1]{Definition~\ref{def:#1}}

\newcommand*{\exref}[1]{Example~\ref{ex:#1}}
\newcommand*{\exrefTwo}[2]{Examples~\ref{ex:#1} and \ref{ex:#2}}
\newcommand*{\figref}[1]{Figure~\ref{fig:#1}}

\newcommand*{\lemref}[1]{Lemma~\ref{lem:#1}}

\newcommand*{\secref}[1]{Section~\ref{sec:#1}}

\newcommand*{\Secref}[1]{\S\,\ref{sec:#1}}

\newcommand*{\thmref}[1]{Theorem~\ref{thm:#1}}

\newenvironment{nop}{}{}
\newenvironment{sdisplaymath}{
\begin{nop}\small\begin{displaymath}}{
\end{displaymath}\end{nop}\ignorespacesafterend}
\newenvironment{smathpar}{
\begin{nop}\small\begin{mathpar}}{
\end{mathpar}\end{nop}\ignorespacesafterend}

\newenvironment{mathfig}{\begin{sdisplaymath}}{\end{sdisplaymath}}
\newenvironment{syntaxfig}{\begin{mathfig}\begin{array}{@{}l@{\quad}r@{~~}c@{\quad}ll}}{\end{array}\end{mathfig}}

\makeatletter
\newbox\sf@box
  {\def\sf@one{#1}%
   \def\sf@two{#2}%
   \setbox\sf@box\hbox
     \bgroup}%
  { \egroup
   \ifx\@empty\sf@two\@empty\relax
     \def\sf@two{\@empty}
   \fi
   \ifx\@empty\sf@one\@empty\relax
     \subfloat[\sf@two]{\box\sf@box}%
   \else
     \subfloat[\sf@one][\sf@two]{\box\sf@box}%
   \fi}
\makeatother

\definecolor{highlightcolor}{rgb}{1.0,0.8,0.8}
\definecolor{shadecolor}{rgb}{0.9,0.9,0.9}
\definecolor{lightgray}{rgb}{0.8,0.8,0.8}
\newcommand*{\shadebox}[1]{\fcolorbox{lightgray}{shadecolor}{\raisebox{0pt}[0.60\baselineskip][0.05\baselineskip]{#1}}}

\newenvironment{nscenter}
 {\parskip=0pt\par\nopagebreak\centering}
 {\par\noindent\ignorespacesafterend}

\newsavebox{\vardisplaymathbox}

\newtheorem{theorem}{Theorem}

\newtheorem{lemma}{Lemma}

\newtheorem{definition}{Definition}

\newtheorem{example}{Example}

\everymath{\displaystyle}

\allowdisplaybreaks

\usepackage{microtype}

\hypersetup{ 
    colorlinks=false,
    pdfborder={0 0 0},
}

\newcommand{\crossrule}{\noindent\textcolor{lightgray}{\cleaders\hbox{.}\hfill}}
\newcommand{\smathparscale}{0.9}
\newcommand{\Paragraph}[1]{\paragraph{\textbf{#1}}}

\newcommand{\zerodisplayskips}{%
  \setlength{\abovedisplayshortskip}{0pt}
  \setlength{\belowdisplayshortskip}{0pt}}

\appto{\small}{\zerodisplayskips}

\newcommand*{\ruleName}[1]{\smash{\textnormal{\textcolor{blue}{#1}}}}
\newcommand*{\highlight}[1]{\textcolor{red}{#1}}

\newcommand*{\piBoundOutput}[1]{\compl{#1}}

\newcommand*{\piInput}[1]{\underline{#1}}

\newcommand*{\piOutput}[2]{\compl{#1}\langle#2\rangle}
\newcommand*{\piTau}{\tau}

\newcommand*{\piAction}[2]{#1.#2}
\newcommand*{\piChoice}[2]{{#1} + {#2}}
\newcommand*{\piChoiceL}[2]{{#1} + {#2}}
\newcommand*{\piChoiceR}[2]{{#1} + {#2}}
\newcommand*{\piPar}[2]{{#1} \mid {#2}}
\newcommand*{\piParL}[3]{{#2} \mathbin{\lowlight{^{#1}}{\mid}} {#3}}
\newcommand*{\piParR}[3]{{#2} \mathbin{\sup{\mid}{#1}} {#3}}
\newcommand*{\piParLNu}[2]{{#1} \mathbin{\sub{\mid}{\nu}} {#2}}
\newcommand*{\piParRNu}[2]{{#1} \mathbin{\lowlight{_{\nu}}{\mid}} {#2}}
\newcommand*{\piParLTau}[3]{{#1} \mathbin{\sub{\mid}{#3}} {#2}}
\newcommand*{\piParRTau}[3]{{#1} \mathbin{\lowlight{_{#3}}{\mid}} {#2}}
\newcommand*{\piReplicate}[1]{{!#1}}
\newcommand*{\piRestrictN}[2]{(\nu#1)\;#2}
\newcommand*{\piRestrict}[1]{\nu#1}
\newcommand*{\piRestrictA}[2]{\sup{\nu}{#1}#2}
\newcommand*{\piRestrictOutput}[1]{\compl{\nu}#1}
\newcommand*{\piZero}[0]{\mathbf{0}}

\newcommand*{\sourceF}{\textsf{src}}
\newcommand*{\targetF}{\textsf{tgt}}
\newcommand*{\source}[1]{\sourceF({#1})}
\newcommand*{\target}[1]{\targetF({#1})}

\newlength{\arrowlen}
\newcommand*{\myrightarrow}[1]{\xrightarrow{\mathmakebox[\arrowlen]{#1}}}

\newcommand*{\transition}[1]{\myrightarrow{\smash{#1}}}
\newcommand*{\transitionWithoutSmash}[1]{\myrightarrow{#1}}
\newcommand*{\compl}[1]{\overline{#1}}

\newcommand*{\cxtRaw}[2]{\sup{#1}{#2}}
\newcommand*{\concur}{\smile}
\newcommand*{\residual}[2]{{#1}\slash{#2}}
\newcommand*{\Proc}[1]{\textsf{Proc}\;{#1}}

\newcommand*{\plus}[2]{#1 + #2}
\newcommand*{\suc}[1]{\plus{#1}{1}}
\newcommand*{\ren}[2]{\renRaw{#1} {#2}}
\newcommand*{\renRaw}[1]{{#1}^*}

\newcommand*{\id}{\textsf{id}}
\newcommand*{\swap}[1]{\sub{\swapR}{#1}}
\newcommand*{\swapR}{\textsf{swap}}
\newcommand*{\pop}[2]{\sub{\popR}{{#1}}\;{#2}}
\newcommand*{\popR}{\textsf{pop}}
\newcommand*{\push}[1]{\sub{\pushR}{#1}}
\newcommand*{\pushR}{\textsf{push}}

\newcommand*{\Action}[1]{\textsf{Action}\;{#1}}

\newcommand*{\Actions}[1]{\textsf{Action}^*\;{#1}}
\newcommand*{\magnitude}[1]{|#1|}

\newcommand*{\EqRefl}[1]{#1_{=}}
\newcommand*{\congRestrictSwap}[1]{\sub{\nu\nu\text{-}\swapR}{#1}} 
\newcommand*{\Cofin}{\mathrel{\protect\rotatebox[origin=c]{90}{$\Join$}}}
\newcommand*{\cofin}[1]{\mathrel{\sub{\Cofin}{#1}}}
\newcommand*{\freeBraid}{\mathrel{\protect\rotatebox[origin=c]{90}{$\ltimes$}}}
\newcommand*{\FreeBraid}[1]{\sub{\freeBraid}{#1}}
\newcommand*{\boundBraid}{\mathrel{\protect\rotatebox[origin=c]{90}{$\rtimes$}}}
\newcommand*{\BoundBraid}{\boundBraid}
\newcommand*{\BoundBraidRefl}[1]{\sub{#1}{\BoundBraid}}

\newcommand*{\permEq}{\simeq}
\newcommand*{\permEqNil}[1]{\sub{\nil}{#1}}
\newcommand*{\permEqCons}[2]{{#1} \cons {#2}}
\newcommand*{\permEqTrans}[2]{{#1} \circ {#2}}
\newcommand*{\permEqSwap}[3]{{({#1} \concur {#2})} \cons {#3}}
\newcommand*{\cons}{\mathrel{\Cdot}}
\newcommand*{\nil}{\varepsilon}

\newcommand*{\braiding}[1]{\sub{\gamma}{#1}}

\renewcommand*{\vec}[1]{\bm{#1}}
\newcommand*{\twoDim}[1]{\ring{#1}}

\newcommand*{\after}{\circ}

\renewcommand*{\to}{\longrightarrow}
\newcommand{\typename}[1]{\textsf{#1}}

\renewcommand*{\secref}[1]{\Secref{#1}} 

\newcommand*{\exampleend}{\hfill$\blacksquare$}

\let\cite=\citep
\bibpunct{[}{]}{;}{a}{}{,}
\nochangebars

\title{\begin{changebar}Proof-relevant \piCalculus: a constructive \\
account of concurrency and causality\end{changebar}}

\pubyear{20??}
\pagerange{\page
  ref{firstpage}--\pageref{lastpage}}
\volume{??}
\doi{??}

\author[R. Perera and J. Cheney]{
Roly Perera$^1$
\thanks{Supported by UK EPSRC grant
EP/K034413/1 and US AFOSR grant FA8655-13-1-3006.} James Cheney$^2$
\thanks{Supported by a Royal
  Society University Research Fellowship.}
  \\ $^1$University of Glasgow; University of Edinburgh
\addressbreak
$^2$University of Edinburgh
}

\begin{document}
\label{firstpage}
\maketitle

\begin{abstract}
We present a formalisation in Agda of the theory of concurrent
transitions, residuation, and causal equivalence of traces for the
\piCalculus. Our formalisation employs de Bruijn indices and
dependently-typed syntax, and aligns the ``proved transitions'' proposed
by Boudol and Castellani in the context of CCS with the proof terms
naturally present in Agda's representation of the labelled transition
relation. Our main contributions are proofs of the ``diamond lemma'' for
the residuals of concurrent transitions and a formal definition of
equivalence of traces up to permutation of transitions.

\quad In the \piCalculus transitions represent propagating binders
whenever their actions involve bound names. To accommodate these cases,
we require a more general diamond lemma where the target states of
equivalent traces are no longer identical, but are related by a
\emph{braiding} that rewires the bound and free names to reflect the
particular interleaving of events involving binders. Our approach may be
useful for modelling concurrency in other languages where transitions
carry metadata sensitive to particular interleavings, such as
dynamically allocated memory addresses.
\end{abstract}

\section{Introduction}

The \piCalculus~\cite{milner99,milner92} is an expressive model of concurrent
and mobile processes. It has been investigated extensively and many variants,
extensions and refinements proposed, including the asynchronous, polyadic, and
applied \piCalculus~\cite{sangiorgi01}. The
\piCalculus has also attracted considerable attention from the logical
frameworks and meta-languages community, and formalisations of its syntax and
semantics have been developed in most of the extant mechanised metatheory
systems, including HOL~\cite{melham94,aitmohamed95},
Coq~\cite{hirschkoff97,despeyroux00,honsell01},
Isabelle/HOL~\cite{rockl01a,gay01}, \begin{changebar}
Isabelle/FM~\cite{gabbay03}\end{changebar}, Nominal
Isabelle~\cite{bengtson09b}, Abella~\cite{baelde14}, CLF~\cite{cervesato02},
and Agda~\cite{orchard15}. Indeed, some early formalisations motivated or led
to important developments in mechanised metatheory, such as the
\begin{changebar}Theory of Contexts~\cite{bucalo06}\end{changebar}, or CLF's
support for monadic encapsulation of concurrent executions.

Prior formalisations have typically considered the syntax, semantics and
bisimulation theory of the \piCalculus. One interesting aspect of the
\piCalculus that has not been formally investigated, and remains to some
extent ill-understood informally, is its theory of \emph{causal
  equivalence}. Two transitions $t,t'$ that can be taken from a process
term $P$ are said to be \emph{concurrent}, written $t \concur t'$, if
they can be performed ``in either order'' --- that is, if after
performing $t$, there is a natural way to transform the other transition
$t'$ so that its effect is performed on the result of $t$, and vice
versa. The transformed version of the transition is said to be the
\emph{residual} of $t'$ after $t$, written $\residual{t'}{t}$. The key
property of this operation, called the ``diamond lemma'' \cite{levy80},
is that the two residuals $\residual{t}{t'}$ and $\residual{t'}{t}$
result in the same process. Finally, permutation of concurrent
transitions induces a \emph{causal equivalence} relation on pairs of
traces. This relation is the standard notion of permutation-equivalence
from the theory of traces over concurrent
alphabets~\cite{mazurkiewicz87}.\begin{changebar}\end{changebar}

In classical treatments of concurrency and residuation, starting with
\citet{levy80}, a transition is usually considered to be a triple
$(e,t,e')$ where $e$ and $e'$ are the source and target terms of the
transition and $t$ is some information about the step performed.
\citet{boudol89} introduced the \emph{proved transitions} approach for
CCS in which the labels of transitions are enriched with an
approximation of the derivation tree which proves that a particular
triple is in the transition relation. \citet{boreale98} and
\citet{degano99} developed theories of causal equivalence for the
\piCalculus, building indirectly on the proved transition approach;
\citet{danos04a} and \citet{cristescu13} developed notions of causality
in the context of reversible CCS and \piCalculus respectively.  

None of the above treatments has been mechanised, although the theory of
residuals for the \lambdaCalculus was formalised in Coq by
\citet{huet94} and in Abella by \citet{accattoli12}. In this paper, we
report on a formalisation of concurrency, residuation and causal equivalence
for the \piCalculus carried out in the dependently-typed programming language
Agda~\cite{norell09}. Our approach is inspired by the proved transitions
method of \citeauthor{boudol89}. However, by taking a ``Church-style'' view of
the labelled transition semantics and treating transitions as proof terms,
rather than triples $(e,t,e')$, we avoid the need for an auxiliary notion of
``proved transition''. Agda's dependent typing allows us to define the
concurrency relation on (compatibly-typed) transition proofs, and residuation
as a total function taking two transitions along with a proof that the
transitions are concurrent. Our formalisation employs de Bruijn
indices~\cite{debruijn72}, an approach with well-known strengths and
weaknesses compared, for example, to higher-order or nominal abstract syntax
techniques employed in existing formalisations;
\begin{changebar}some of these other techniques are discussed in
\secref{related-work}.\end{changebar}

Our definition of concurrency is not the only plausible one for the
\piCalculus. Indeed, there appears to be little consensus regarding the
characteristics of a canonical definition. For example,
\citet{cristescu13} write ``[in] the absence of an indisputable
definition of permutation equivalence for [labelled transition system]
semantics of the \piCalculus it is hard to assert the correctness of one
definition over another.'' We do, however, show that our definition of
concurrency is sound by proving the diamond property; to the best of our
knowledge, ours is the first mechanised version of this result for any
process calculus.

However, one key observation that emerges in our development is that
requiring residuals of concurrent transitions to reach exactly the same
state is too restrictive. When the action of a transition involves a
bound name, the transition represents a propagating binder. In such
cases equivalent traces no longer have identical target states, but
rather states which are equal up to a \emph{braiding} that rewires the
bound and free names to reflect the different order of events in the two
traces. Although typically unobservable to a program, such
interleaving-sensitive information may be important for other purposes,
such as memory locations in a debugger, or transaction ids in a
financial application. In these situations being able to robustly
translate between the target states of different interleavings may be
important. Our development may therefore be a useful case study for
formalising concurrency in other settings where transition labels carry
interleaving-sensitive metadata.\begin{changebar}

This is a substantially revised\end{changebar} version of a paper presented at
the \emph{Logical Frameworks and Meta-Languages: Theory and Practice} workshop
\cite{perera15}. This version extends the earlier work with graphical
proof-sketches for various lemmas, a more detailed comparison of related
formalisation efforts, extensive examples and discussion regarding the
generalised diamond property, \begin{changebar}a more precise definition of
cofinality, and a formalisation of composite braids. A companion paper
\cite{perera16d} uses the formalisation of concurrent transitions presented
here as the basis for ``causally consistent'' dynamic slicing of \piCalculus
programs.\end{changebar}

The paper is organised as follows. \secref{calculus} describes our
variant of the (synchronous) \piCalculus, including syntax, renamings,
and transitions. \secref{concurrent-transitions} defines concurrency and
residuation for transitions, and discusses the diamond lemma and the
notion of ``cofinal'' transitions. \secref{causal-equivalence} presents
our definition of causal equivalence. \secref{related-work} discusses
related work in more detail and \secref{conclusion} concludes and
discusses prospects for future work. \appref{module-structure}
summarises the Agda module structure; the source code can be found at
\url{https://github.com/rolyp/proof-relevant-pi}, release \ttt{0.3}.

\section{Synchronous \piCalculus}
\label{sec:calculus}

We present our formalisation in the setting of a first-order, synchronous,
monadic \piCalculus with recursion and internal choice, using a labelled
transition semantics.

Names are ranged over by $x$, $y$ and $z$. An input action is written
$\piInput{x}$. Output actions are written $\piOutput{x}{y}$ if $y$ is in scope
and $\piBoundOutput{x}$ if the action represents the output of a name whose
scope is extruding, in which case we say the action is a
\emph{bound} output. Bound outputs do not appear in \cbstart source programs
\cbend but arise during execution.

\begin{center}
\begin{minipage}[t]{0.45\linewidth}
\begin{syntaxfig}
\mbox{Name}
&
x, y, z & ::= & 0 \mid 1 \mid \cdots
\\[1mm]
\mbox{Action}
&
a
&
::=
&
\piInput{x}
&
\quad\text{input}
\\
&&&
\piOutput{x}{y}
&
\quad\text{output}
\\
&&&
\piBoundOutput{x}
&
\quad\text{bound output}
\\
&&&
\piTau
&
\quad\text{silent}
\end{syntaxfig}
\end{minipage}
\end{center}

\noindent \begin{changebar}Although it has become common practice to limit
attention to sums of guarded processes, here we study the calculus as
originally formulated by Milner \etal, which allows sums of arbitrary
processes. (Our basic approach should transfer to guarded choice, and other
common variants.)\end{changebar}

\vspace{-10pt}
\begin{center}
\begin{minipage}[t]{0.5\linewidth}
\begin{syntaxfig}
\mbox{Process}
&
P, Q, R, S
&
::=
&
\piZero
&
\quad\text{inactive}
\\
&&&
\piAction{\piInput{x}}{P}
&
\quad\text{input}
\\
&&&
\piAction{\piOutput{x}{y}}{P}
&
\quad\text{output}
\\
&&&
\piChoice{P}{Q}
&
\quad\begin{changebar}\text{non-guarded choice}\end{changebar}
\\
&&&
\piPar{P}{Q}
&
\quad\text{parallel}
\\
&&&
\piRestrict{P}
&
\quad\text{restriction}
\\
&&&
\piReplicate{P}
&
\quad\text{replication}
\end{syntaxfig}
\end{minipage}
\end{center}

Although the formal development uses de Bruijn indices, and we give
definitions and state properties in terms of this notation, we will sometimes
illustrate their meaning in terms of conventional
\piCalculus notation.  For example, the conventional $\pi$-calculus
term $\piRestrictN{x}{x(z).\compl{y}\langle{z}\rangle.\piZero \mid
\compl{x}\langle c \rangle.\piZero}$ would be represented using de Bruijn
indices as $\nu (\piInput{0}.\piOutput{\suc{n}}{0}.\piZero\mid
\piOutput{0}{\suc{m}}.\piZero)$, provided that $y$ and $c$ are
associated with indices $n$ and $m$. Here, the first $0$ represents the bound
variable $x$, the second $0$ the bound variable $z$, and the third refers to
$x$ again. Note that the symbol $\piZero$ denotes the inactive process, not a
de Bruijn index.

The syntax of actions and processes is defined more formally in
\figref{syntax:process-action} overleaf. Let $\Gamma$ and $\Delta$ range over
\emph{contexts}, which in an untyped setting are simply natural numbers. A
membership witness $x \in \Gamma$ is a proof that $x < \Gamma$. A context
$\Gamma$ \emph{closes} $P$ if $x \in \Gamma$ for every free variable $x$ of
$P$. We denote by $\Proc{\Gamma}$ the set of processes closed by $\Gamma$, as
defined below. We write $\Gamma \vdash P$ to mean $P \in \Proc{\Gamma}$.
Similarly, actions are well-formed only in closing contexts; we write $a:
\Action{\Gamma}$ to mean that $\Gamma$ is closing for $a$.

To specify the labelled transition semantics, it is convenient to
distinguish \emph{bound} actions $b$ from non-bound actions $c$. A bound
action $b: \Action{\Gamma}$ is of the form $\piInput{x}$ or
$\piBoundOutput{x}$, and shifts a process from $\Gamma$ to a target
context $\suc{\Gamma}$, freeing the index $0$. A non-bound action $c:
\Action{\Gamma}$ is of the form $\piOutput{x}{y}$ or $\piTau$, and has a
target context which is also $\Gamma$. Meta-variable $a$ ranges over all
actions, bound and non-bound. $\magnitude{a}$ denotes the amount by which
the action increments the context; thus $\magnitude{b} = 1$ and
$\magnitude{c} = 0$.

\begin{figure}[h]
\input{fig/syntax/process}
\input{fig/syntax/action}
\crossrule
\caption{Syntax of processes and actions}
\label{fig:syntax:process-action}
\end{figure}

\subsection{Renamings}

A de Bruijn indices formulation of \piCalculus makes extensive use of
renamings. A \emph{renaming} $\rho : \Gamma \to \Delta$ is any function
(injective or otherwise) from names in $\Gamma$ to names in $\Delta$. The
labelled transition semantics makes use of the lifting of the successor
function $\suc{\param}$ on natural numbers to renamings, which we call
$\pushR$ to avoid confusion with the $\suc{\param}$ operation on contexts;
$\pop{}{y}$, which undoes the effect of $\pushR$, replacing $0$ by $y$; and
$\swapR$, which transposes the roles of $0$ and $1$ but otherwise acts as the
identity. This de Bruijn treatment of \piCalculus is similar to that of
Hirschkoff's asynchronous $\mu s$ calculus
\cite{hirschkoff99}; \begin{changebar}in particular Hirschkoff's $\langle x\rangle$, $\phi$ and
$\psi$ operators correspond roughly to $\pop{}{x}$, $\pushR$ and
$\swapR$.\end{changebar} We give a late rather than early semantics; other
differences are discussed in
\secref{related-work} below.

\vspace{-10pt}
\begin{figure}[H]
\begin{minipage}[t]{0.32\linewidth}
{\small
\begin{align*}
\intertext{\shadebox{$\push{\Gamma}: \Gamma \longrightarrow \plus{\Gamma}{1}$}}
\push{}\;x
&=
x + 1
\end{align*}
}
\end{minipage}%
\begin{minipage}[t]{0.32\linewidth}
{\small
\begin{align*}
\intertext{\shadebox{$\pop{\Gamma}{y}: \plus{\Gamma}{1} \longrightarrow \Gamma$}}
\pop{}{y}\;0
&=
y
\\
\pop{}{y}\;(x + 1)
&=
x
\end{align*}
}
\end{minipage}%
\begin{minipage}[t]{0.32\linewidth}
{\small
\begin{align*}
\intertext{\shadebox{$\swap{\Gamma}: \plus{\Gamma}{2} \longrightarrow \plus{\Gamma}{2}$}}
\swapR\;0
&=
1
\\
\swapR\;1
&=
0
\\
\swapR\;(x + 2)
&=
x + 2
\end{align*}
}
\end{minipage}%
\vspace{-2mm}%
\\
\crossrule
\caption{$\pushR$, $\popR$ and $\swapR$ renamings}
\end{figure}

\noindent The $\Gamma$ subscripts that appear on $\push{\Gamma}$,
$\pop{\Gamma}{y}$ and $\swap{\Gamma}$ are shown in grey to indicate that
they may be omitted when their value is obvious or irrelevant; this is a
convention we use throughout the paper.

\subsubsection{Lifting renamings to processes and actions}
\label{sec:calculus:renamings:processes-and-actions}

The functorial extension $\renRaw{\rho}: \Proc{\Gamma} \longrightarrow
\Proc{\Delta}$ of a renaming $\rho: \Gamma \longrightarrow \Delta$ to
processes is defined in the usual way. Renaming under a binder utilises
the action of $\suc{\param}$ on renamings, which is also functorial.
Syntactically, $\renRaw{\rho}$ binds tighter than any process
constructor, and $\suc{\param}$ has higher precedence than composition,
so that (for example) $\pop{}{0} \after \suc{\push{}}$ means $\pop{}{0}
\after (\suc{\push{}})$, not $\suc{(\pop{}{0} \after \push{})}$.

\vspace{-5pt}
\begin{figure}[H]
\begin{minipage}[t]{0.49\linewidth}
\input{fig/renaming/process}
\end{minipage}%
\begin{minipage}[t]{0.49\linewidth}
\input{fig/renaming/action}
\input{fig/renaming/suc}
\end{minipage}%
\vspace{-2mm}\\
\crossrule
\caption{Renaming for processes and actions}
\end{figure}

\subsubsection{Properties of renamings}
\label{sec:calculus:properties-of-renamings}

Several equational properties of renamings are used throughout the
development; here we present the ones mentioned elsewhere in the paper.
For each lemma, we give the corresponding commutative diagram underneath
on the left, along with a string diagram that offers a graphical
intuition for why the lemma holds.

\begin{lemma}
\label{lem:pop-after-push}
$\pop{}{x} \after \push{} = \id$
\end{lemma}

\noindent Freeing the index $0$ and then immediately substituting $x$
for it is a no-op.

\begin{center}
\begin{minipage}{0.32\linewidth}
\scalebox{0.8}{
\begin{tikzpicture}[node distance=1.5cm, auto, baseline={([yshift=-.5ex]current bounding box.center)}]
  \node (GammaA) {$\Gamma$};
  \node (GammaB) [node distance=3cm, right of=GammaA] {$\suc{\Gamma}$};
  \node (GammaC) [below of=GammaB] {$\Gamma$};
  \draw[->] (GammaA) to node {$\push{}$} (GammaB);
  \draw[-,double distance=1pt] (GammaA) to node [xshift=0.8em, swap] {} (GammaC);
  \draw[->, shift left=0.2em] (GammaB) to node {$\pop{}{x}$} (GammaC);
\end{tikzpicture}
}
\end{minipage}
\begin{minipage}{0.67\linewidth}
\begin{center}
\scalebox{0.8}{
\begin{tikzpicture}[node distance=0.5cm, auto]
  \node [anchor=west] at (0,0) (LabelA) {$\Gamma$};
  \node [anchor=west, below of=LabelA] (0A) {0};
  \node [anchor=west, below of=0A] (1A) {1};
  \node [anchor=west, below of=1A] (2A) {$\vdots$};

  \node [anchor=east, right of=LabelA, node distance=2.5cm] (LabelB) {$\suc{\Gamma}$};
  \node [anchor=east, below of=LabelB] (0B) {0};
  \node [anchor=east, below of=0B] (1B) {1};
  \node [anchor=east, below of=1B] (2B) {2};
  \node [anchor=east, below of=2B] (3B) {$\vdots$};

  \node [anchor=east, right of=LabelB, node distance=2.5cm] (LabelC) {$\Gamma$};
  \node [anchor=east, below of=LabelC] (0C) {0};
  \node [anchor=east, below of=0C] (1C) {1};
  \node [anchor=east, below of=1C] (2C) {2};
  \node [anchor=east, below of=2C] (3C) {$\vdots$};
  \node [anchor=east, below of=3C] (4C) {$x$};

  \draw (LabelA) edge[out=0,in=180,->] node {$\push{}$} (LabelB);
  \draw (LabelB) edge[out=0,in=180,->] node {$\pop{}{x}$} (LabelC);
  \draw (0A) edge[out=0,in=180,|->] (1B);
  \draw (1A) edge[out=0,in=180,|->] (2B);
  \draw (0B) edge[out=0,in=180,|->] (4C);
  \draw (1B) edge[out=0,in=180,|->] (0C);
  \draw (2B) edge[out=0,in=180,|->] (1C);

  \node [right of=1C, node distance=1cm] {\Large $=$};

  \node [anchor=west, right of=LabelC, node distance=2cm] (LabelD) {$\Gamma$};
  \node [anchor=west, below of=LabelD] (0D) {0};
  \node [anchor=west, below of=0D] (1D) {1};
  \node [anchor=west, below of=1D] (2D) {$\vdots$};

  \node [anchor=east, right of=LabelD, node distance=2.5cm] (LabelE) {$\Gamma$};
  \node [anchor=east, below of=LabelE] (0E) {0};
  \node [anchor=east, below of=0E] (1E) {1};
  \node [anchor=east, below of=1E] (2E) {$\vdots$};

  \draw (LabelD) edge[out=0,in=180,->] node {$\id$} (LabelE);
  \draw (0D) edge[out=0,in=180,|->] (0E);
  \draw (1D) edge[out=0,in=180,|->] (1E);
\end{tikzpicture}
}
\end{center}
\end{minipage}
\end{center}

\begin{lemma}
\label{lem:pop-zero-after-suc-push}
$\pop{}{0} \after \suc{\push{}} = \id$
\begin{center}
\begin{minipage}{0.32\linewidth}
\scalebox{0.8}{
\begin{tikzpicture}[node distance=1.5cm, auto, baseline={([yshift=-.5ex]current bounding box.center)}]
  \node (GammaA) {$\suc{\Gamma}$};
  \node (GammaB) [node distance=3cm, right of=GammaA] {$\plus{\Gamma}{2}$};
  \node (GammaC) [below of=GammaB] {$\suc{\Gamma}$};
  \draw[->] (GammaA) to node {$\suc{\push{\Gamma}}$} (GammaB);
  \draw[-,double distance=1pt] (GammaA) to node [xshift=0.8em, swap] {} (GammaC);
  \draw[->, shift left=0.2em] (GammaB) to node {$\pop{\suc{\Gamma}}{0}$} (GammaC);
\end{tikzpicture}
}
\end{minipage}
\begin{minipage}{0.67\linewidth}
\begin{center}
\scalebox{0.8}{
\begin{tikzpicture}[node distance=0.5cm, auto]
  \node [anchor=west] at (0,0) (LabelA) {$\suc{\Gamma}$};
  \node [anchor=west, below of=LabelA] (0A) {0};
  \node [anchor=west, below of=0A] (1A) {1};
  \node [anchor=west, below of=1A] (2A) {2};
  \node [anchor=west, below of=2A] (3A) {$\vdots$};

  \node [anchor=east, right of=LabelA, node distance=2.5cm] (LabelB) {$\plus{\Gamma}{2}$};
  \node [anchor=east, below of=LabelB] (0B) {0};
  \node [anchor=east, below of=0B] (1B) {1};
  \node [anchor=east, below of=1B] (2B) {2};
  \node [anchor=east, below of=2B] (3B) {3};
  \node [anchor=east, below of=3B] (4B) {$\vdots$};

  \node [anchor=east, right of=LabelB, node distance=2.5cm] (LabelC) {$\suc{\Gamma}$};
  \node [anchor=east, below of=LabelC] (0C) {0};
  \node [anchor=east, below of=0C] (1C) {1};
  \node [anchor=east, below of=1C] (2C) {2};
  \node [anchor=east, below of=2C] (3C) {$\vdots$};

  \draw (LabelA) edge[out=0,in=180,->] node {$\suc{\push{}}$} (LabelB);
  \draw (LabelB) edge[out=0,in=180,->] node {$\pop{}{0}$} (LabelC);
  \draw (0A) edge[out=0,in=180,|->] (0B);
  \draw (1A) edge[out=0,in=180,|->] (2B);
  \draw (2A) edge[out=0,in=180,|->] (3B);
  \draw (0B) edge[out=0,in=180,|->] (0C);
  \draw (1B) edge[out=0,in=180,|->] (0C);
  \draw (2B) edge[out=0,in=180,|->] (1C);
  \draw (3B) edge[out=0,in=180,|->] (2C);

  \node [right of=1C, node distance=1cm] {\Large $=$};

  \node [anchor=west, right of=LabelC, node distance=2cm] (LabelD) {$\suc{\Gamma}$};
  \node [anchor=west, below of=LabelD] (0D) {0};
  \node [anchor=west, below of=0D] (1D) {1};
  \node [anchor=west, below of=1D] (2D) {2};
  \node [anchor=west, below of=2D] (3D) {$\vdots$};

  \node [anchor=east, right of=LabelD, node distance=2.5cm] (LabelE) {$\suc{\Gamma}$};
  \node [anchor=east, below of=LabelE] (0E) {0};
  \node [anchor=east, below of=0E] (1E) {1};
  \node [anchor=east, below of=1E] (2E) {2};
  \node [anchor=east, below of=2E] (3E) {$\vdots$};

  \draw (LabelD) edge[out=0,in=180,->] node {$\id$} (LabelE);
  \draw (0D) edge[out=0,in=180,|->] (0E);
  \draw (1D) edge[out=0,in=180,|->] (1E);
  \draw (2D) edge[out=0,in=180,|->] (2E);
\end{tikzpicture}
}
\end{center}
\end{minipage}
\end{center}
\end{lemma}

\begin{lemma}
\label{lem:swap-after-suc-swap-after-swap}
$\suc{\swap{}} \after \swap{} \after \suc{\swap{}} = \swap{} \after
\suc{\swap{}} \after \swap{}$

\begin{center}
\begin{minipage}{0.50\linewidth}
\scalebox{0.8}{
\begin{tikzpicture}[node distance=1.5cm, auto, baseline={([yshift=-.5ex]current bounding box.center)}]
  \node (GammaA) {$\plus{\Gamma}{3}$};
  \node (GammaB) [right of=GammaA, above of=GammaA] {$\plus{\Gamma}{3}$};
  \node (GammaC) [node distance=3cm, right of=GammaB] {$\plus{\Gamma}{3}$};
  \node (GammaD) [right of=GammaA, below of=GammaA] {$\plus{\Gamma}{3}$};
  \node (GammaE) [node distance=3cm, right of=GammaD] {$\plus{\Gamma}{3}$};
  \node (GammaF) [right of=GammaC, below of=GammaC] {$\plus{\Gamma}{3}$};
  \draw[->] (GammaA) to node {$\sub{\swapR}{\suc{\Gamma}}$} (GammaB);
  \draw[->] (GammaB) to node {$\suc{\sub{\swapR}{\Gamma}}$} (GammaC);
  \draw[->] (GammaA) to node [swap] {$\suc{\sub{\swapR}{\Gamma}}$} (GammaD);
  \draw[->] (GammaD) to node [swap] {$\sub{\swapR}{\suc{\Gamma}}$} (GammaE);
  \draw[->] (GammaC) to node {$\sub{\swapR}{\suc{\Gamma}}$} (GammaF);
  \draw[->] (GammaE) to node [swap] {$\suc{\sub{\swapR}{\Gamma}}$} (GammaF);
\end{tikzpicture}
}
\end{minipage}
\begin{minipage}{0.49\linewidth}
\begin{center}
\scalebox{0.8}{
\begin{tikzpicture}[node distance=0.5cm, auto]
  \node [anchor=west] at (0,0) (LabelA) {$\plus{\Gamma}{3}$};
  \node [anchor=west, below of=LabelA] (0A) {0};
  \node [anchor=west, below of=0A] (1A) {1};
  \node [anchor=west, below of=1A] (2A) {2};
  \node [anchor=west, below of=2A] (3A) {$\vdots$};

  \node [anchor=east, right of=LabelA, node distance=2.5cm] (LabelB) {$\plus{\Gamma}{3}$};
  \node [anchor=east, below of=LabelB] (0B) {0};
  \node [anchor=east, below of=0B] (1B) {1};
  \node [anchor=east, below of=1B] (2B) {2};
  \node [anchor=east, below of=2B] (3B) {$\vdots$};

  \node [anchor=east, right of=LabelB, node distance=2.5cm] (LabelC) {$\plus{\Gamma}{3}$};
  \node [anchor=east, below of=LabelC] (0C) {0};
  \node [anchor=east, below of=0C] (1C) {1};
  \node [anchor=east, below of=1C] (2C) {2};
  \node [anchor=east, below of=2C] (3C) {$\vdots$};

  \node [anchor=west, right of=LabelC, node distance=2.5cm] (LabelD) {$\plus{\Gamma}{3}$};
  \node [anchor=west, below of=LabelD] (0D) {0};
  \node [anchor=west, below of=0D] (1D) {1};
  \node [anchor=west, below of=1D] (2D) {2};
  \node [anchor=west, below of=2D] (3D) {$\vdots$};

  \draw (LabelA) edge[out=0,in=180,->] node {$\suc{\swap{\Gamma}}$} (LabelB);
  \draw (LabelB) edge[out=0,in=180,->] node {$\swap{\suc{\Gamma}}$} (LabelC);
  \draw (LabelC) edge[out=0,in=180,->] node {$\suc{\swap{\Gamma}}$} (LabelD);
  \draw (0A) edge[,out=0,in=180,|->] (1B);
  \draw (1A) edge[out=0,in=180,|->] (0B);
  \draw (2A) edge[,out=0,in=180,|->] (2B);
  \draw (0B) edge[out=0,in=180,|->] (0C);
  \draw (1B) edge[,out=0,in=180,|->] (2C);
  \draw (2B) edge[,out=0,in=180,|->] (1C);
  \draw (0C) edge[out=0,in=180,|->] (1D);
  \draw (1C) edge[,out=0,in=180,|->] (0D);
  \draw (2C) edge[,out=0,in=180,|->] (2D);

  \node [right of=2B, below of=0B, yshift=-0.5cm, node distance=1.25cm] {\Large $=$};

  \node [anchor=west] at (0,-3.25) (LabelAA) {$\plus{\Gamma}{3}$};
  \node [anchor=west, below of=LabelAA] (0AA) {0};
  \node [anchor=west, below of=0AA] (1AA) {1};
  \node [anchor=west, below of=1AA] (2AA) {2};
  \node [anchor=west, below of=2AA] (3AA) {$\vdots$};

  \node [anchor=east, right of=LabelAA, node distance=2.5cm] (LabelBB) {$\plus{\Gamma}{3}$};
  \node [anchor=east, below of=LabelBB] (0BB) {0};
  \node [anchor=east, below of=0BB] (1BB) {1};
  \node [anchor=east, below of=1BB] (2BB) {2};
  \node [anchor=east, below of=2BB] (3BB) {$\vdots$};

  \node [anchor=east, right of=LabelBB, node distance=2.5cm] (LabelCC) {$\plus{\Gamma}{3}$};
  \node [anchor=east, below of=LabelCC] (0CC) {0};
  \node [anchor=east, below of=0CC] (1CC) {1};
  \node [anchor=east, below of=1CC] (2CC) {2};
  \node [anchor=east, below of=2CC] (3CC) {$\vdots$};

  \node [anchor=west, right of=LabelCC, node distance=2.5cm] (LabelDD) {$\plus{\Gamma}{3}$};
  \node [anchor=west, below of=LabelDD] (0DD) {0};
  \node [anchor=west, below of=0DD] (1DD) {1};
  \node [anchor=west, below of=1DD] (2DD) {2};
  \node [anchor=west, below of=2DD] (3DD) {$\vdots$};

  \draw (LabelAA) edge[out=0,in=180,->] node {$\swap{\suc{\Gamma}}$} (LabelBB);
  \draw (LabelBB) edge[out=0,in=180,->] node {$\suc{\swap{\Gamma}}$} (LabelCC);
  \draw (LabelCC) edge[out=0,in=180,->] node {$\swap{\suc{\Gamma}}$} (LabelDD);
  \draw (0AA) edge[,out=0,in=180,|->] (0BB);
  \draw (1AA) edge[out=0,in=180,|->] (2BB);
  \draw (2AA) edge[,out=0,in=180,|->] (1BB);
  \draw (0BB) edge[,out=0,in=180,|->] (1CC);
  \draw (1BB) edge[,out=0,in=180,|->] (0CC);
  \draw (2BB) edge[out=0,in=180,|->] (2CC);
  \draw (0CC) edge[,out=0,in=180,|->] (0DD);
  \draw (1CC) edge[,out=0,in=180,|->] (2DD);
  \draw (2CC) edge[out=0,in=180,|->] (1DD);
\end{tikzpicture}
}
\end{center}
\end{minipage}
\end{center}
\end{lemma}

\noindent The above are two ways to swap indices 0 and 2.

\begin{lemma}
\label{lem:pop-swap}
$\pop{}{0} \after \swap{} = \pop{}{0}$
\begin{center}
\begin{minipage}{0.43\linewidth}
\scalebox{0.8}{
\begin{tikzpicture}[node distance=3cm, auto, baseline={([yshift=-.5ex]current bounding box.center)}]
  \node (P) {$\plus{\Gamma}{2}$};
  \node (PPrime) [right of=P] {$\plus{\Gamma}{2}$};
  \node (Q) [right of=PPrime] {$\suc{\Gamma}$};
  \draw[->, shift left=0.2em] (P) to node {$\swapR$} (PPrime);
  \draw[->, shift right=0.2em] (P) to node [swap] {$\id$} (PPrime);
  \draw[->] (PPrime) to node {$\pop{\suc{\Gamma}}{0}$} (Q);
\end{tikzpicture}
}
\end{minipage}
\begin{minipage}{0.56\linewidth}
\begin{center}
\scalebox{0.8}{
\begin{tikzpicture}[node distance=0.5cm, auto]
  \node [anchor=west] at (0,0) (LabelA) {$\plus{\Gamma}{2}$};
  \node [anchor=west, below of=LabelA] (0A) {0};
  \node [anchor=west, below of=0A] (1A) {1};
  \node [anchor=west, below of=1A] (2A) {2};
  \node [anchor=west, below of=2A] (3A) {$\vdots$};

  \node [anchor=east, right of=LabelA, node distance=2.5cm] (LabelB) {$\plus{\Gamma}{2}$};
  \node [anchor=east, below of=LabelB] (0B) {0};
  \node [anchor=east, below of=0B] (1B) {1};
  \node [anchor=east, below of=1B] (2B) {2};
  \node [anchor=east, below of=2B] (3B) {$\vdots$};

  \node [anchor=east, right of=LabelB, node distance=2.5cm] (LabelC) {$\suc{\Gamma}$};
  \node [anchor=east, below of=LabelC] (0C) {0};
  \node [anchor=east, below of=0C] (1C) {1};
  \node [anchor=east, below of=1C] (2C) {$\vdots$};

  \draw (LabelA) edge[out=0,in=180,->] node {$\swap{}$} (LabelB);
  \draw (LabelB) edge[out=0,in=180,->] node {$\pop{}{0}$} (LabelC);
  \draw (0A) edge[out=0,in=180,|->] (1B);
  \draw (1A) edge[out=0,in=180,|->] (0B);
  \draw (2A) edge[out=0,in=180,|->] (2B);
  \draw (0B) edge[out=0,in=180,|->] (0C);
  \draw (1B) edge[out=0,in=180,|->] (0C);
  \draw (2B) edge[out=0,in=180,|->] (1C);

  \node [right of=1C, node distance=1cm] {\Large $=$};

  \node [anchor=west, right of=LabelC, node distance=2cm] (LabelD) {$\plus{\Gamma}{2}$};
  \node [anchor=west, below of=LabelD] (0D) {0};
  \node [anchor=west, below of=0D] (1D) {1};
  \node [anchor=west, below of=1D] (2D) {2};
  \node [anchor=west, below of=1D] (3D) {$\vdots$};

  \node [anchor=east, right of=LabelD, node distance=2.5cm] (LabelE) {$\suc{\Gamma}$};
  \node [anchor=east, below of=LabelE] (0E) {0};
  \node [anchor=east, below of=0E] (1E) {1};
  \node [anchor=east, below of=1E] (2E) {$\vdots$};

  \draw (LabelD) edge[out=0,in=180,->] node {$\pop{}{0}$} (LabelE);
  \draw (0D) edge[out=0,in=180,|->] (0E);
  \draw (1D) edge[out=0,in=180,|->] (0E);
  \draw (2D) edge[out=0,in=180,|->] (1E);
\end{tikzpicture}
}
\end{center}
\end{minipage}
\end{center}
\end{lemma}

\begin{lemma}
\label{lem:swap-push}
$\swap{} \after \suc{\push{}} = \push{}$, $\swap{} \after \push{} = \suc{\push{}}$
\begin{center}
\begin{minipage}{0.31\linewidth}
\scalebox{0.8}{
\begin{tikzpicture}[node distance=1.5cm, auto, baseline={([yshift=-.5ex]current bounding box.center)}]
  \node (P) {$\suc{\Gamma}$};
  \node (PPrime) [node distance=3cm, right of=P] {$\plus{\Gamma}{2}$};
  \node (Q) [below of=PPrime] {$\plus{\Gamma}{2}$};
  \draw[->] (P) to node {$\suc{\push{\Gamma}}$} (PPrime);
  \draw[->] (P) to node [xshift=0.8em, swap] {$\push{\suc{\Gamma}}$} (Q);
  \draw[->, shift left=0.2em] (PPrime) to node {$\swapR$} (Q);
  \draw[->, shift left=0.2em] (Q) to node {$\swapR$} (PPrime);
\end{tikzpicture}
}
\end{minipage}
\begin{minipage}{0.66\linewidth}
\begin{center}
\scalebox{0.8}{
\begin{tikzpicture}[node distance=0.5cm, auto]
  \node [anchor=west] at (0,0) (LabelA) {$\suc{\Gamma}$};
  \node [anchor=west, below of=LabelA] (0A) {0};
  \node [anchor=west, below of=0A] (1A) {1};
  \node [anchor=west, below of=1A] (2A) {$\vdots$};

  \node [anchor=east, right of=LabelA, node distance=2.5cm] (LabelB) {$\plus{\Gamma}{2}$};
  \node [anchor=east, below of=LabelB] (0B) {0};
  \node [anchor=east, below of=0B] (1B) {1};
  \node [anchor=east, below of=1B] (2B) {2};
  \node [anchor=east, below of=2B] (3B) {$\vdots$};

  \node [anchor=east, right of=LabelB, node distance=2.5cm] (LabelC) {$\plus{\Gamma}{2}$};
  \node [anchor=east, below of=LabelC] (0C) {0};
  \node [anchor=east, below of=0C] (1C) {1};
  \node [anchor=east, below of=1C] (2C) {2};
  \node [anchor=east, below of=2C] (3C) {$\vdots$};

  \draw (LabelA) edge[out=0,in=180,->] node {$\push{\suc{\Gamma}}$} (LabelB);
  \draw (LabelB) edge[out=0,in=180,->] node {$\swap{\Gamma}$} (LabelC);
  \draw (0A) edge[out=0,in=180,|->] (1B);
  \draw (1A) edge[out=0,in=180,|->] (2B);
  \draw (0B) edge[out=0,in=180,|->] (1C);
  \draw (1B) edge[out=0,in=180,|->] (0C);
  \draw (2B) edge[out=0,in=180,|->] (2C);

  \node [right of=1C, node distance=1cm] {\Large $=$};

  \node [anchor=west, right of=LabelC, node distance=2cm] (LabelD) {$\suc{\Gamma}$};
  \node [anchor=west, below of=LabelD] (0D) {0};
  \node [anchor=west, below of=0D] (1D) {1};
  \node [anchor=west, below of=1D] (2D) {$\vdots$};

  \node [anchor=east, right of=LabelD, node distance=2.5cm] (LabelE) {$\plus{\Gamma}{2}$};
  \node [anchor=east, below of=LabelE] (0E) {0};
  \node [anchor=east, below of=0E] (1E) {1};
  \node [anchor=east, below of=1E] (2E) {2};
  \node [anchor=east, below of=2E] (3E) {$\vdots$};

  \draw (LabelD) edge[out=0,in=180,->] node {$\suc{\push{\Gamma}}$} (LabelE);
  \draw (0D) edge[out=0,in=180,|->] (0E);
  \draw (1D) edge[out=0,in=180,|->] (2E);
\end{tikzpicture}
}
\end{center}
\vspace{-5pt}
\begin{center}
\scalebox{0.8}{
\begin{tikzpicture}[node distance=0.5cm, auto]
  \node [anchor=west] at (0,0) (LabelA) {$\suc{\Gamma}$};
  \node [anchor=west, below of=LabelA] (0A) {0};
  \node [anchor=west, below of=0A] (1A) {1};
  \node [anchor=west, below of=1A] (2A) {$\vdots$};

  \node [anchor=east, right of=LabelA, node distance=2.5cm] (LabelB) {$\plus{\Gamma}{2}$};
  \node [anchor=east, below of=LabelB] (0B) {0};
  \node [anchor=east, below of=0B] (1B) {1};
  \node [anchor=east, below of=1B] (2B) {2};
  \node [anchor=east, below of=2B] (3B) {$\vdots$};

  \node [anchor=east, right of=LabelB, node distance=2.5cm] (LabelC) {$\plus{\Gamma}{2}$};
  \node [anchor=east, below of=LabelC] (0C) {0};
  \node [anchor=east, below of=0C] (1C) {1};
  \node [anchor=east, below of=1C] (2C) {2};
  \node [anchor=east, below of=2C] (3C) {$\vdots$};

  \draw (LabelA) edge[out=0,in=180,->] node {$\suc{\push{\Gamma}}$} (LabelB);
  \draw (LabelB) edge[out=0,in=180,->] node {$\swap{\Gamma}$} (LabelC);
  \draw (0A) edge[out=0,in=180,|->] (0B);
  \draw (1A) edge[out=0,in=180,|->] (2B);
  \draw (0B) edge[out=0,in=180,|->] (1C);
  \draw (1B) edge[out=0,in=180,|->] (0C);
  \draw (2B) edge[out=0,in=180,|->] (2C);

  \node [right of=1C, node distance=1cm] {\Large $=$};

  \node [anchor=west, right of=LabelC, node distance=2cm] (LabelD) {$\suc{\Gamma}$};
  \node [anchor=west, below of=LabelD] (0D) {0};
  \node [anchor=west, below of=0D] (1D) {1};
  \node [anchor=west, below of=1D] (2D) {$\vdots$};

  \node [anchor=east, right of=LabelD, node distance=2.5cm] (LabelE) {$\plus{\Gamma}{2}$};
  \node [anchor=east, below of=LabelE] (0E) {0};
  \node [anchor=east, below of=0E] (1E) {1};
  \node [anchor=east, below of=1E] (2E) {2};
  \node [anchor=east, below of=2E] (3E) {$\vdots$};

  \draw (LabelD) edge[out=0,in=180,->] node {$\push{\suc{\Gamma}}$} (LabelE);
  \draw (0D) edge[out=0,in=180,|->] (1E);
  \draw (1D) edge[out=0,in=180,|->] (2E);
\end{tikzpicture}
}
\end{center}
\end{minipage}
\end{center}
\end{lemma}

\begin{lemma}
$\push{}\after \rho= \suc{\rho} \after \push{}$
\label{lem:push-comm}
\end{lemma}

\begin{lemma}
$\rho \after \pop{}{x} = \pop{}{\rho x} \after \suc{\rho}$
\label{lem:pop-comm}
\end{lemma}

\begin{lemma}
\label{lem:swap-suc-suc}
$\swap{} \after \plus{\rho}{2} = \plus{\rho}{2} \after \swap{}$
\end{lemma}

\noindent These last three lemmas assert various naturality properties
of $\push{}$, $\pop{}{x}$ and $\swap{}$.

\begin{center}
\begin{minipage}[t]{0.5\linewidth}
\begin{nscenter}
\scalebox{0.8}{
\begin{tikzpicture}[node distance=1.5cm, auto]
  \node (P) {$\Gamma$};
  \node (Q) [below of=P] {$\Delta$};
  \node (PPrime) [node distance=2.6cm, right of=P] {$\suc{\Gamma}$};
  \node (QPrime) [below of=PPrime] {$\suc{\Delta}$};
  \node (PDoublePrime) [node distance=2.9cm, right of=PPrime] {$\Gamma$};
  \node (QDoublePrime) [below of=PDoublePrime] {$\Delta$};
  \draw[->] (P) to node {$\push{\Gamma}$} (PPrime);
  \draw[->] (Q) to node [swap] {$\push{\Delta}$} (QPrime);
  \draw[->] (QPrime) to node [swap] {$\pop{\Delta}{\rho x}$} (QDoublePrime);
  \draw[->] (P) to node [swap] {$\rho$} (Q);
  \draw[->] (PPrime) to node {$\suc{\rho}$} (QPrime);
  \draw[->] (PPrime) to node {$\pop{\Gamma}{x}$} (PDoublePrime);
  \draw[->] (PDoublePrime) to node {$\rho$} (QDoublePrime);
\end{tikzpicture}
}
\end{nscenter}
\end{minipage}
\begin{minipage}[t]{0.3\linewidth}
\begin{nscenter}
\scalebox{0.8}{
\begin{tikzpicture}[node distance=1.5cm, auto]
  \node (P) {$\plus{\Gamma}{2}$};
  \node (Q) [below of=P] {$\plus{\Delta}{2}$};
  \node (PPrime) [node distance=2.8cm, right of=P] {$\plus{\Gamma}{2}$};
  \node (QPrime) [below of=PPrime] {$\plus{\Delta}{2}$};
  \draw[->] (P) to node {$\swap{\Gamma}$} (PPrime);
  \draw[->] (Q) to node [swap] {$\swap{\Delta}$} (QPrime);
  \draw[->] (P) to node [swap] {$\plus{\rho}{2}$} (Q);
  \draw[->] (PPrime) to node {$\plus{\rho}{2}$} (QPrime);
\end{tikzpicture}
}
\end{nscenter}
\end{minipage}
\end{center}

\vspace{-5mm}
\subsection{Labelled transition semantics}
\label{sec:calculus:transitions}

An important feature of our presentation is that each transition rule
has an explicit constructor name. This allow derivations to be written
in a compact, expression-like form, similar to the \emph{proven
  transitions} used by \citet{boudol89} to define notions of concurrency
and residuation for CCS. However, rather than giving an additional
inductive definition describing the structure of a ``proof'' that $P
\transition{a} R$, we simply treat the inductive definition of
$\transition{}$ as a data type. This is a natural approach in a
dependently-typed setting.

The rule names are summarised below, and have been chosen to reflect,
where possible, the structure of the process triggering the rule. The
corresponding relation $P \transition{a} R$ is defined in
\figref{transition}, for any process $\Gamma \vdash P$, any $a:
\Action{\Gamma}$ and any $\plus{\Gamma}{\magnitude{a}} \vdash R$.

\begin{syntaxfig}
\mbox{Transition}
&
t, u
&
::=
&
\piAction{\piInput{x}}{P}
&
\text{input on $x$}
\\
&&&
\piAction{\piOutput{x}{y}}{P}
&
\text{output $y$ on $x$}
\\
&&&
\piChoiceL{t}{Q} \quad \piChoiceR{P}{u}
&
\text{choose left or right branch}
\\
&&&
\piParL{a}{t}{Q} \quad \piParR{a}{P}{u}
&
\text{propagate $a$ through parallel composition on the left or right}
\\
&&&
\piParLTau{t}{u}{y} \quad \piParRTau{t}{u}{y}
&
\text{synchronise (receiving $y$ on the left or right)}
\\
&&&
\piRestrictOutput{t}
&
\text{initiate extrusion of $\nu$}
\\
&&&
\piParLNu{t}{u} \quad \piParRNu{t}{u}
&
\text{$\nu$-synchronise (receiving $0$ on the left or right)}
\\
&&&
\piRestrictA{a}{t}
&
\text{propagate $a$ through binder}
\\
&&&
\piReplicate{t}
&
\text{replicate}
\end{syntaxfig}

\begin{figure}[h]
\noindent \shadebox{$P \transition{a} R$}
\begin{smathpar}
\inferrule*[left={\ruleName{$\piAction{\piInput{x}}{P}$}}]
{
}
{
   \piAction{\piInput{x}}{P} \transitionWithoutSmash{\piInput{x}} P
}
\and
\inferrule*[left={\ruleName{$\piAction{\piOutput{x}{y}}{P}$}}]
{
}
{
   \piAction{\piOutput{x}{y}}{P} \transitionWithoutSmash{\piOutput{x}{y}} P
}
\and
\inferrule*[left={\ruleName{$\piChoiceL{\param}{Q}$}}]
{
   P \transitionWithoutSmash{a} R
}
{
   \piChoice{P}{Q} \transitionWithoutSmash{a} R
}
\and
\inferrule*[left={\ruleName{$\piParL{c}{\param}{Q}$}}]
{
   P \transitionWithoutSmash{c} R
}
{
   \piPar{P}{Q} \transitionWithoutSmash{c} \piPar{R}{Q}
}
\and
\inferrule*[left={\ruleName{$\piParL{b}{\param}{Q}$}}]
{
   P \transitionWithoutSmash{b} R
}
{
   \piPar{P}{Q} \transitionWithoutSmash{b} \piPar{R}{\ren{\push{}}{Q}}
}
\and
\inferrule*[left={\ruleName{$\piParLTau{\param}{\param}{y}$}}]
{
   P \transitionWithoutSmash{\piInput{x}} R
   \\
   Q \transitionWithoutSmash{\piOutput{x}{y}} S
}
{
   \piPar{P}{Q} \transitionWithoutSmash{\piTau} \piPar{\ren{(\pop{}{y})}{R}}{S}
}
\and
\inferrule*[left={\ruleName{$\piRestrictOutput{\param}$}}]
{
   P \transitionWithoutSmash{\piOutput{(x + 1)}{0}} R
}
{
   \piRestrict{P} \transitionWithoutSmash{\piBoundOutput{x}} R
}
\and
\inferrule*[left={\ruleName{$\piParLNu{\param}{\param}$}}]
{
   P \transitionWithoutSmash{\piInput{x}} R
   \\
   Q \transitionWithoutSmash{\piBoundOutput{x}} S
}
{
   \piPar{P}{Q} \transitionWithoutSmash{\piTau} \piRestrict{(\piPar{R}{S})}
}
\and
\inferrule*[left={\ruleName{$\piRestrictA{c}{\param}$}}]
{
   P \transitionWithoutSmash{\ren{\push{}}{c}} R
}
{
   \piRestrict{P} \transitionWithoutSmash{c} \piRestrict{R}
}
\and
\inferrule*[left={\ruleName{$\piRestrictA{b}{\param}$}}]
{
   P \transitionWithoutSmash{\ren{\push{}}{b}} R
}
{
   \piRestrict{P} \transitionWithoutSmash{b} \piRestrict{(\ren{\swapR}{R})}
}
\and
\inferrule*[left={\ruleName{$\piReplicate{\param}$}}]
{
   \piPar{P}{\piReplicate{P}} \transitionWithoutSmash{a} R
}
{
   \piReplicate{P} \transitionWithoutSmash{a} R
}
\end{smathpar}
\crossrule
\caption{Labelled transition rules ($\piChoice{P}{\param}$,
  $\piParR{b}{P}{\param}$, $\piParR{c}{P}{\param}$,
  $\piParRNu{\param}{\param}$ and $\piParRTau{\param}{\param}{y}$
  variants omitted)}
\label{fig:transition}
\end{figure}

The constructor name for each rule is shown to the left of the rule.
There is an argument position, indicated by $\param$, for each premise
of the rule. Note that there are two forms of the transition
constructors $\piParL{a}{\param}{\param}$ and $\piRestrictA{a}{\param}$
distinguished by whether they are indexed by a bound action $b$ or by a
non-bound action $c$. Omitted from \figref{transition} are additional
(but symmetric) rules of the form $\piChoice{P}{\param}$,
$\piParR{b}{P}{\param}$ and $\piParR{b}{P}{\param}$ where the
sub-transition occurs on the opposite side of the operator, and also
$\piParRTau{\param}{\param}{y}$ (synchronise) and
$\piParRNu{\param}{\param}$ ($\nu$-synchronise) rules in which the
positions of sender and receiver are transposed. These are all
straightforward variants of the rules shown, and are omitted from the
figure to avoid clutter. Meta-variables $t$ and $u$ range over
transition derivations; if $t: P \transition{a} R$ then $\source{t}$
denotes $P$ and $\target{t}$ denotes $R$.

Although a de Bruijn formulation of \piCalculus requires a certain amount of
housekeeping, one pleasing consequence is that the usual side-conditions
associated with the \piCalculus transition rules are either subsumed by
syntactic constraints on actions, or ``operationalised'' using the renamings
above. In particular:

\begin{enumerate}
\item The use of $\pushR$ in the \ruleName{$\piParL{b}{\param}{Q}$} rule
  corresponds to the usual side-condition asserting that the binder
  being propagated by $P$ is not free in $Q$. In the de Bruijn setting
  every binder ``locally'' has the name 0, and so this requirement can
  be operationalised by rewiring $Q$ so that the name $0$ is reserved.
  The $\pushR$ will be matched by a later $\popR$ which substitutes for
  $0$, in the event that the action has a successful synchronisation.
\item The \ruleName{$\piRestrictOutput{\param}$} rule requires an
  extrusion to be initiated by an output of the form
  $\piOutput{\suc{x}}{0}$, capturing the usual side-condition that the
  name being extruded \emph{on} is distinct from the name being
  extruded.
\item The rules of the form \ruleName{$\piRestrictA{a}$} require that
  the action being propagated has the form $\ren{\pushR}{a}$, ensuring
  that it contains no uses of index $0$. This corresponds to the usual
  requirement that an action can only propagate through a binder that it
  does not mention.
\end{enumerate}

\noindent The use of $\swapR$ in the \ruleName{$\piRestrictA{b}$} case
follows \citet{hirschkoff99} and has no counterpart outside of the de
Bruijn setting. As a propagating binder passes through another binder,
their local indices are 0 and 1. Propagation transposes the binders, and
so to preserve naming we rewire $R$ with a ``braid'' that swaps $0$ and
$1$. Since binders are also reordered by \emph{permutations} that relate
causally equivalent executions, the $\swapR$ renaming will also play an
important role when we consider concurrent transitions
(\secref{concurrent-transitions}).

The following schematic derivation shows how the compact notation works.
Suppose $t: P \transition{\piOutput{\plus{z}{2}}{0}} R$ takes place
immediately under a $\nu$-binder, causing the scope of the binder to be
extruded. Then suppose the resulting bound output propagates through
another binder, giving the partial derivation on the left:

\begin{center}
\begin{minipage}[b]{0.3\linewidth}
\scalebox{\smathparscale}{
\begin{smathpar}
\inferrule*[left={\ruleName{$\piRestrictA{\piBoundOutput{z}}{\param}$}}]
{
  \inferrule*[left={\ruleName{$\piRestrictOutput{\param}$}}]
  {
    \inferrule*[left={\ruleName{$t$}}]
    {
      \vdots
    }
    {
      P \transitionWithoutSmash{\piOutput{\plus{z}{2}}{0}} R
    }
  }
  {
    \piRestrict{P} \transitionWithoutSmash{\piBoundOutput{\suc{z}}} R
  }
}
{
  \piRestrict{\piRestrict{P}} \transitionWithoutSmash{\piBoundOutput{z}} \piRestrict{R}
}
\end{smathpar}
}
\end{minipage}
\begin{minipage}[b]{0.3\linewidth}
\scalebox{\smathparscale}{
\begin{smathpar}
\inferrule*[left={\ruleName{$\piRestrictA{\piBoundOutput{z}}{\param}$}}]
{
  \inferrule*[left={\ruleName{$\piRestrictOutput{t}$}}]
  {
    \vdots
  }
  {
    \piRestrict{P} \transitionWithoutSmash{\piBoundOutput{\suc{z}}} R
  }
}
{
  \piRestrict{\piRestrict{P}} \transitionWithoutSmash{\piBoundOutput{z}} \piRestrict{R}
}
\end{smathpar}
}
\end{minipage}
\begin{minipage}[b]{0.3\linewidth}
\scalebox{\smathparscale}{
\begin{smathpar}
\inferrule*[left={\ruleName{$\piRestrictA{\piBoundOutput{z}}{\piRestrictOutput{t}}$}}]
{
  \vdots
}
{
  \piRestrict{\piRestrict{P}} \transitionWithoutSmash{\piBoundOutput{z}} \piRestrict{R}
}
\end{smathpar}
}
\end{minipage}
\end{center}

\noindent with $t$ standing in for the rest of the derivation. The
constructors annotating the left-hand side of the derivation tree
(shown in blue in the electronic version of this article) can be
thought of as a partially unrolled ``transition term'' representing the
proof. The $\param$ placeholders associated with each constructor are
conceptually filled by the transition terms annotating the premises of
that step. We can ``roll up'' the derivation by a single step, by moving
the premises into their corresponding placeholders, as shown in the
middle figure.

By repeating this process, we can write the whole derivation compactly
as $\piRestrictA{\piBoundOutput{z}}{\piRestrictOutput{t}}$, as shown on
the right. Thus the compact form is simply a flattened transition
derivation: similar to a simply-typed \lambdaCalculus term written as a
conventional expression, in a (Church-style) setting where a term is,
strictly speaking, a typing derivation.

\subsubsection{Residuals of transitions and renamings}

A transition $t$ with action $a$ survives any suitably-typed renaming
$\rho$. Moreover $\rho$ has an image in $t$, which is simply
$\plus{\rho}{\magnitude{a}}$.

\begin{lemma}
\label{lem:concurrent-transitions:renaming:transition}
Suppose $t: P \transition{a} Q$ and $\rho: \Gamma \to \Delta$, where
$\Gamma \vdash P$. Then there exists a transition $\ren{\rho}{t}:
\ren{\rho}{P} \transition{\ren{\rho}{a}}
\ren{(\plus{\rho}{\magnitude{a}})}{Q}$.

\begin{nscenter}
\vspace{2mm}
\scalebox{0.8}{
\begin{tikzpicture}[node distance=1.5cm, auto]
  \node (P) {$P$};
  \node (Q) [below of=P] {$\ren{\rho}{P}$};
  \node (PPrime) [node distance=3.4cm, right of=P] {$Q$};
  \node (QPrime) [below of=PPrime] {$\ren{\rho}{Q}$};
  \draw[->] (P) to node {$\cxtRaw{t}{c}$} (PPrime);
  \draw[dotted,->] (Q) to node [swap] {$\cxtRaw{(\ren{\rho}{t})}{\ren{\rho}{c}}$} (QPrime);
  \draw[->] (P) to node [swap] {$\ren{\rho}{}$} (Q);
  \draw[->] (PPrime) to node {$\ren{\rho}{}$} (QPrime);
\end{tikzpicture}%
\quad\quad%
\begin{tikzpicture}[node distance=1.5cm, auto]
  \node (P) {$P$};
  \node (Q) [below of=P] {$\ren{\rho}{P}$};
  \node (PPrime) [node distance=3.4cm, right of=P] {$Q$};
  \node (QPrime) [below of=PPrime] {$\ren{(\plus{\rho}{1})}{Q}$};
  \draw[->] (P) to node {$\cxtRaw{t}{b}$} (PPrime);
  \draw[dotted,->] (Q) to node [swap] {$\cxtRaw{(\ren{\rho}{t})}{\ren{\rho}{b}}$} (QPrime);
  \draw[->] (P) to node [swap] {$\ren{\rho}{}$} (Q);
  \draw[->] (PPrime) to node {$\ren{(\plus{\rho}{1})}{}$} (QPrime);
\end{tikzpicture}
}

\end{nscenter}
\end{lemma}

\noindent \emph{Proof.} By the following defining equations. The various
renaming lemmas needed to enable the induction hypothesis in each case
are omitted.

\vspace{8pt}
\begin{minipage}[t]{0.49\linewidth}
\noindent \shadebox{$\ren{\rho}{\cxtRaw{t}{c}}$}
{\small
\begin{align*}
\ren{\rho}{(\piAction{\piOutput{x}{y}}{P})}
&=
\piAction{\piOutput{\rho x}{\rho y}}{\ren{\rho}{P}}
\\
\ren{\rho}{(\piChoice{t}{Q})}
&=
\piChoice{\ren{\rho}{t}}{\ren{\rho}{Q}}
\\
\ren{\rho}{(\piChoice{P}{u})}
&=
\piChoice{\ren{\rho}{P}}{\ren{\rho}{u}}
\\
\ren{\rho}{(\piParR{c}{P}{u})}
&=
\piParR{\ren{\rho}{c}}{\ren{\rho}{P}}{\ren{\rho}{u}}
\\
\ren{\rho}{(\piParL{c}{t}{Q})}
&=
\piParL{\ren{\rho}{c}}{\ren{\rho}{t}}{\ren{\rho}{Q}}
\\
\ren{\rho}{(\piParLTau{t}{u}{y})}
&=
\piParLTau{\ren{\rho}{t}}{\ren{\rho}{u}}{\ren{\rho}{y}}
\\
\ren{\rho}{(\piParLNu{t}{u})}
&=
\piParLNu{\ren{\rho}{t}}{\ren{\rho}{u}}
\\
\ren{\rho}{(\piRestrictA{c}{t})}
&=
\piRestrictA{\ren{\rho}{c}}{\ren{(\suc{\rho})}{t}}
\\
\ren{\rho}{(\piReplicate{t})}
&=
\piReplicate{\ren{\rho}{t}}
\end{align*}
}
\end{minipage}
\begin{minipage}[t]{0.49\linewidth}
\noindent \shadebox{$\ren{\rho}{\cxtRaw{t}{b}}$}
{\small
\begin{align*}
\ren{\rho}{(\piAction{\piInput{x}}{P})}
&=
\piAction{\piInput{\rho x}}{\ren{(\suc{\rho})}{P}}
\\
\ren{\rho}{(\piChoice{t}{Q})}
&=
\piChoice{\ren{\rho}{t}}{\ren{\rho}{Q}}
\\
\ren{\rho}{(\piChoice{P}{u})}
&=
\piChoice{\ren{\rho}{P}}{\ren{\rho}{u}}
\\
\ren{\rho}{(\piParR{b}{P}{u})}
&=
\piParR{\ren{\rho}{b}}{\ren{\rho}{P}}{\ren{\rho}{u}}
\\
\ren{\rho}{(\piParL{b}{t}{Q})}
&=
\piParL{\ren{\rho}{b}}{\ren{\rho}{t}}{\ren{\rho}{Q}}
\\
\ren{\rho}{(\piRestrictOutput{t})}
&=
\piRestrictOutput{\ren{(\suc{\rho})}{t}}
\\
\ren{\rho}{(\piRestrictA{b}{t})}
&=
\piRestrictA{\ren{\rho}{b}}{\ren{(\suc{\rho})}{t}}
\\
\ren{\rho}{(\piReplicate{t})}
&=
\piReplicate{\ren{\rho}{t}}
\end{align*}
}
\end{minipage}

We would not expect $\ren{\rho}{t}$ to be \begin{changebar}a derivable
transition, and thus \lemref{concurrent-transitions:renaming:transition} to
hold,\end{changebar} for arbitrary $\rho$ in all extensions of the
\piCalculus. In particular, the mismatch operator $[x\neq y]P$ that steps to
$P$ if $x$ and $y$ are distinct names is only stable under injective
renamings.
  
\subsubsection{Structural congruences}

Our LTS semantics is standard and therefore closed under the usual
\piCalculus congruences. Structural congruences can be formalised as a
bisimulation, using an analogue of the notion of residuation with
respect to a transition used elsewhere in this paper. This remains out
of scope of the present development.

\section{Concurrency, residuals and cofinality}
\label{sec:concurrent-transitions}

Transitions $P \transition{a} R$ and $Q \transition{a'} S$ are
\emph{coinitial} when $P = Q$. In this section we formalise a symmetric,
irreflexive \emph{concurrency} relation $\concur$ over coinitial
transitions. Concurrent transitions $t \concur t'$ are independent, or
causally unordered. In an interleaving semantics, $t$ and $t'$ can
execute in either order without significant interference; in a true
concurrency setting, $t$ and $t'$ form a single, two-dimensional
``parallel move'' \cite*{curry58}. Concurrency was explored notably by
\citet{levy80} for the \lambdaCalculus, and later by \citet{stark89} for
arbitrary transition systems. The inspiration for the treatment
presented here is \citeauthor*{boudol89}'s concurrency relation for CCS
\citeyearpar{boudol89}.

\begin{figure}[h]
\begin{nscenter}
\scalebox{0.8}{
\begin{tikzpicture}[node distance=1.5cm, auto]
  \node (P) [node distance=2cm] {
    $P$
  };
  \node (R) [right of=P, above of=P] {
    $R$
  };
  \node (RPrime) [below of=P, right of=P] {
    $R'$
  };
  \node (PPrime) [right of=R, below of=R] {
    $Q$
  };
  \draw[->] (P) to node {$t$} (R);
  \draw[->] (P) to node [swap] {$t'$} (RPrime);
  \draw[dotted,->] (R) to node {$\residual{t'}{t}$} (PPrime);
  \draw[dotted,->] (RPrime) to node [swap] {$\residual{t}{t'}$} (PPrime);
\end{tikzpicture}
}
\end{nscenter}
\caption{Conventional diamond property for $t \concur t'$}
\label{fig:residual:square}
\end{figure}

The essence of $t \concur t'$ is illustrated in
\figref{residual:square}. If either execution step is taken, the other
remains valid, and moreover once both are taken, one ends up in
(essentially) the same state, regardless of which step is taken first.
However, concurrent transitions are not completely independent: the
location and indeed the nature of the redex acted on by one transition
may change as a consequence of the earlier transition being taken. This
idea is captured by the \emph{residual} $\residual{t'}{t}$ (``$t'$ after
$t$''), which specifies how $t'$ must be transformed to operate on
$\target{t}$ (sometimes called \emph{pseudocommutation}
\cite{angiuli14}).

The requirement that $\residual{t'}{t}$ and $\residual{t}{t'}$ are
\emph{cofinal} -- have the same target state -- is straightforward when
the transitions preserve the free variables of a term. This is trivially
the case in CCS since there are no binders, and is also true of the
\lambdaCalculus, where reductions are usually defined on closed terms.
In the late-style \piCalculus that we consider here, there are transition
rules that ``open'' a process with respect to a name, with the action on the
transition representing the upwards propagation of the binder through the
process term. In this setting the notion of cofinality is non-trivial;
\begin{changebar}while de Bruijn indices make this subtlety more explicit, we
note that the reordering of binders complicates things even in the named
setting.\end{changebar} We discuss this, with examples, in this section.
Permutation of concurrent transitions induces a congruence on traces called
\emph{causal equivalence}, which we turn to in
\secref{causal-equivalence}.

\subsection{Concurrent transitions}

In our setting, a transition $t: P \transition{a} R$ is a \emph{proof}
that locates a redex in $P$, witnessing the fact that $(P,a,R) \in
{\transition{}}$. The concurrency relation $\concur$ relates two such
proofs; it is defined as the symmetric closure of the relation defined
inductively by the rules in \figref{concurrent}. The figure makes use of
the compact notation for transitions introduced in
\secref{calculus:transitions}. As before, trivial variants of the rules
are omitted for clarity. For the transition constructors of the form
\ruleName{$\piParL{a}{\param}{Q}$} and
\ruleName{$\piRestrictA{a}{\param}$} that come in bound and non-bound
variants, we abuse notation a little and write a single $\concur$ rule
quantified over $a$ to mean that there are two separate (but otherwise
identical) cases.

\begin{figure}[h]
\shadebox{$t \concur t'$}
\begin{smathpar}
\inferrule*
{
  \strut
}
{
  \piParR{a}{P}{u}
  \concur
  \piParL{a'}{t}{Q}
}
\and
\inferrule*
{
  t \concur t'
}
{
  \piParL{a}{t}{Q}
  \concur
  \piParLTau{t'}{u}{y}
}
\and
\inferrule*
{
  t \concur t'
}
{
  \piParL{a}{t}{Q}
  \concur
  \piParRTau{t'}{u}{y}
}
\and
\inferrule*
{
  u \concur u'
}
{
  \piParR{a}{P}{u}
  \concur
  \piParLTau{t}{u'}{y}
}
\and
\inferrule*
{
  u \concur u'
}
{
  \piParR{a}{P}{u}
  \concur
  \piParRTau{t}{u'}{y}
}
\and
\inferrule*
{
  t \concur t'
}
{
  \piParL{a}{t}{Q}
  \concur
  \piParLNu{t'}{u}
}
\and
\inferrule*
{
  t \concur t'
}
{
  \piParL{a}{t}{Q}
  \concur
  \piParRNu{t'}{u}
}
\and
\inferrule*
{
  u \concur u'
}
{
  \piParR{a}{P}{u}
  \concur
  \piParLNu{t}{u'}
}
\and
\inferrule*
{
  u \concur u'
}
{
  \piParR{a}{P}{u}
  \concur
  \piParRNu{t}{u'}
}
\and
\inferrule*
{
  t \concur t'
}
{
  \piChoice{t}{Q}
  \concur
  \piChoiceL{t'}{Q}
}
\and
\inferrule*
{
  u \concur u'
}
{
  \piChoice{P}{u}
  \concur
  \piChoiceR{P}{u'}
}
\and
\inferrule*
{
  t \concur t'
}
{
  \piParR{a}{P}{t}
  \concur
  \piParR{a'}{P}{t'}
}
\and
\inferrule*
{
  t \concur t'
}
{
  \piParL{a}{t}{Q}
  \concur
  \piParL{a'}{t'}{Q}
}
\and
\inferrule*
{
  t \concur t'
  \\
  u \concur u'
}
{
  \piParLTau{t}{u}{y}
  \concur
  \piParLTau{t'}{u'}{z}
}
\and
\inferrule*
{
  t \concur t'
  \\
  u \concur u'
}
{
  \piParRTau{t}{u}{y}
  \concur
  \piParRTau{t'}{u'}{z}
}
\and
\inferrule*
{
  t \concur t'
  \\
  u \concur u'
}
{
  \piParLTau{t}{u}{y}
  \concur
  \piParRTau{t'}{u'}{z}
}
\and
\inferrule*
{
  t \concur t'
  \\
  u \concur u'
}
{
  \piParLTau{t}{u}{y}
  \concur
  \piParLNu{t'}{u'}
}
\and
\inferrule*
{
  t \concur t'
  \\
  u \concur u'
}
{
  \piParRTau{t}{u}{y}
  \concur
  \piParLNu{t'}{u'}
}
\and
\inferrule*
{
  t \concur t'
  \\
  u \concur u'
}
{
  \piParLTau{t}{u}{y}
  \concur
  \piParRNu{t'}{u'}
}
\and
\inferrule*
{
  t \concur t'
  \\
  u \concur u'
}
{
  \piParRTau{t}{u}{y}
  \concur
  \piParRNu{t'}{u'}
}
\and
\inferrule*
{
  t \concur t'
  \\
  u \concur u'
}
{
  \piParLNu{t}{u}
  \concur
  \piParLNu{t'}{u'}
}
\and
\inferrule*
{
  t \concur t'
  \\
  u \concur u'
}
{
  \piParRNu{t}{u}
  \concur
  \piParRNu{t'}{u'}
}
\and
\inferrule*
{
  t \concur t'
  \\
  u \concur u'
}
{
  \piParLNu{t}{u}
  \concur
  \piParRNu{t'}{u'}
}
\and
\inferrule*
{
  t \concur t'
}
{
  \piRestrictOutput{t}
  \concur
  \piRestrictOutput{t'}
}
\and
\inferrule*
{
  t \concur t'
}
{
  \piRestrictOutput{t}
  \concur
  \piRestrictA{a}{t'}
}
\and
\inferrule*
{
  t \concur t'
}
{
  \piRestrictA{a}{t}
  \concur
  \piRestrictA{a'}{t'}
}
\and
\inferrule*
{
  t \concur t'
}
{
  \piReplicate{t}
  \concur
  \piReplicate{t'}
}
\end{smathpar}
\crossrule
\caption{Concurrent transitions}
\label{fig:concurrent}
\end{figure}

Intuitively, transitions are concurrent when they pick out
non-overlapping redexes. The only axiom, $\piParR{a}{P}{u} \concur
\piParL{a'}{t}{Q}$, says that two transitions $t$ and $u$ are concurrent
if they take place on opposite sides of a parallel composition. The
remaining rules propagate concurrent sub-transitions up through
restriction, choice, parallel composition, and replication. There are no
rules allowing us to conclude that a transition which takes the left
branch of a choice is concurrent with a transition which takes the right
branch of the same choice; choices are mutually exclusive. Likewise,
there are no rules allowing us to conclude that an input or output
transition is concurrent with any other transition. Since $t$ and $t'$
are coinitial, if one of them picks out a prefix then the other
necessarily picks out the same prefix, and so they are equal and thus
not concurrent.

The $\piParLTau{t}{u}{y} \concur \piParLTau{t'}{u'}{z}$ rule says that a
synchronisation is concurrent with another, as long as the two input
transitions $t$ and $t'$ are concurrent on the left branch of the
parallel composition, and the two output transitions $u$ and $u'$ are
concurrent on the right. The $\piParLTau{t}{u}{y} \concur
\piParRTau{t'}{u'}{z}$ variant is similar, but permits concurrent input
and output transitions on the left, with their respective
synchronisation partners concurrent on the right. The
$\piParLTau{t}{u}{y} \concur \piParRNu{t'}{u'}$ rule and variants are
analogous, but permit a plain synchronisation to be concurrent with a
$\nu$-synchronisation. The main result of this section is that the
concurrency relation captured by $\concur$ is sound up to a suitable
notion of cofinality.

\begin{example}[Concurrent transitions]
\label{ex:concurrent-extrusions:same-binder}
Consider the \piCalculus process (using conventional named syntax)
$\piRestrictN{y}{\compl{x}\langle{y}\rangle.P \mid \compl{z}\langle y
  \rangle.Q}$. This can take two transitions, the first one sending $y$
on $x$, resulting in $P \mid \compl{z}\langle y \rangle.Q$, and the
second one sending $y$ on channel $z$, resulting in
$\compl{x}\langle{y}\rangle.P \mid Q$. Notice that $y$ becomes free in
both processes.

In de Bruijn notation, this process is written $\piRestrict{(\piPar{\piAction{
\piOutput{\suc{x}}{0}}{P}}{\piAction{\piOutput{\suc{z}}{0}}{Q}})}$. It can
take two transitions, each resulting in an extrusion of the $\nu$-binder; call
these $t$ and $t'$. The transition $t$ initiates the extrusion
$\piBoundOutput{x}$ on the left branch of the parallel composition:

\vspace{5pt}
\begin{nscenter}
\scalebox{\smathparscale}{
\begin{smathpar}
\inferrule*[left={\ruleName{$\piRestrictOutput{\param}$}}]
{
  \inferrule*[left={\ruleName{$\piParL{\piOutput{\suc{x}}{0}}{\param}{\piAction{\piOutput{\suc{z}}{0}}{Q}}$}}]
  {
    \suc{\Gamma} \vdash \piAction{\highlight{\piOutput{\suc{x}}{0}}}{P}
    \transitionWithoutSmash{\piOutput{\suc{x}}{0}}
    \suc{\Gamma} \vdash P
  }
  {
    \suc{\Gamma} \vdash \piPar{\piAction{\highlight{\piOutput{\suc{x}}{0}}}{P}}{\piAction{\piOutput{\suc{z}}{0}}{Q}}
    \transitionWithoutSmash{\piOutput{\suc{x}}{0}}
    \suc{\Gamma} \vdash \piPar{P}{\piAction{\piOutput{\suc{z}}{0}}{Q}}
  }
}
{
  \Gamma \vdash \piRestrict{(\piPar{\piAction{\highlight{\piOutput{\suc{x}}{0}}}{P}}{\piAction{\piOutput{\suc{z}}{0}}{Q}})}
  \transitionWithoutSmash{\piBoundOutput{x}}
  \suc{\Gamma} \vdash \piPar{P}{\piAction{\piOutput{\suc{z}}{0}}{Q}}
}
\end{smathpar}
}
\end{nscenter}

\vspace{1pt}

\noindent The transition $t'$ initiates an extrusion $\piBoundOutput{z}$
of the same binder on the right branch of the parallel composition:

\vspace{5pt}
\begin{nscenter}
\scalebox{\smathparscale}{
\begin{smathpar}
\inferrule*[left={\ruleName{$\piRestrictOutput{\param}$}}]
{
  \inferrule*[left={\ruleName{$\piParR{\piOutput{\suc{z}}{0}}{\piAction{\piOutput{\suc{x}}{0}}{Q}}{\param}$}}]
  {
    \suc{\Gamma} \vdash \piAction{\highlight{\piOutput{\suc{z}}{0}}}{Q}
    \transitionWithoutSmash{\piOutput{\suc{z}}{0}}
    \suc{\Gamma} \vdash Q
  }
  {
    \suc{\Gamma} \vdash \piPar{\piAction{\piOutput{\suc{x}}{0}}{P}}{\piAction{\highlight{\piOutput{\suc{z}}{0}}}{Q}}
    \transitionWithoutSmash{\piOutput{\suc{z}}{0}}
    \suc{\Gamma} \vdash \piPar{\piAction{\piOutput{\suc{x}}{0}}{P}}{Q}
  }
}
{
  \Gamma \vdash \piRestrict{(\piPar{\piAction{\piOutput{\suc{x}}{0}}{P}}{\piAction{\highlight{\piOutput{\suc{z}}{0}}}{Q}})}
  \transitionWithoutSmash{\piBoundOutput{z}}
  \suc{\Gamma} \vdash \piPar{\piAction{\piOutput{\suc{x}}{0}}{P}}{Q}
}
\end{smathpar}
}
\end{nscenter}

\vspace{1pt}

\noindent Since the two transitions are coinitial and arise on opposite
sides of a parallel composition, we can conclude $t \concur t'$ using
the rules in \figref{concurrent}. Here is the proof, writing the
derivations $t$ and $t'$ above using the compact notation for
transitions:

\vspace{5pt}
\begin{center}
\scalebox{\smathparscale}{
\begin{smathpar}
  \inferrule*
  {
    \piParL{\piOutput{\suc{x}}{0}}
           {\piAction{\piOutput{\suc{x}}{0}}{P}}
           {\piAction{\piOutput{\suc{z}}{0}}{Q}}
    \;\concur\;
    \piParR{\piOutput{\suc{z}}{0}}
                              {\piAction{\piOutput{\suc{x}}{0}}{P}}
                              {\piAction{\piOutput{\suc{z}}{0}}{Q}}
  }
  {
    \piRestrictOutput{(\piParL{\piOutput{\suc{x}}{0}}
                              {\piAction{\piOutput{\suc{x}}{0}}{P}}
                              {\piAction{\piOutput{\suc{z}}{0}}{Q}})}
    \;\concur\;
    \piRestrictOutput{(\piParR{\piOutput{\suc{z}}{0}}
                              {\piAction{\piOutput{\suc{x}}{0}}{P}}
                              {\piAction{\piOutput{\suc{z}}{0}}{Q}})}
  }
\end{smathpar}
}
\end{center}

\exampleend
\end{example}

\subsection{Residuals of concurrent transitions}
\label{sec:concurrent-transitions:residuals}

If two transitions are concurrent then their respective \emph{residuals}
provide a canonical way of merging or reconciling them.

\begin{definition}[Residual $\residual{t}{t'}$]
For any $t \concur t'$, the \emph{residual} of $t$ after $t'$, written
$\residual{t}{t'}$, is defined  by the
equations in \figref{residual}.
\end{definition}

\begin{figure}[H]
\noindent \shadebox{$\residual{t}{t'}$}
\begin{minipage}[t]{0.41\linewidth}
{\small
\begin{align*}
\residual{(\piParR{a}{P}{u})}
         {(\piParL{c}{t}{Q})}
&=
\piParR{a}{\target{t}}{u}
\\
\residual{(\piParR{a}{P}{u})}
         {(\piParL{b}{t}{Q})}
&=
\piParR{a}{\target{t}}{\ren{\push{}}{u}}
\\
\residual{(\piParL{a}{t}{Q})}
         {(\piParR{c}{P}{u})}
&=
\piParL{a}{t}{\target{u}}
\\
\residual{(\piParL{a}{t}{Q})}
         {(\piParR{b}{P}{u})}
&=
\piParL{a}{\ren{\push{}}{t}}{\target{u}}
\\
\residual{(\piParL{a}{t}{Q})}
         {(\piParLTau{t'}{u}{y})}
&=
\piParL{a}{\ren{(\pop{}{y})}{(\residual{t}{t'})}}{\target{u}}
\\
\residual{(\piParR{a}{P}{u})}
         {(\piParLTau{t}{u'}{y})}
&=
\piParR{a}{\ren{(\pop{}{y})}{\target{t}}}{\residual{u}{u'}}
\\
\residual{(\piParLTau{t}{u}{y})}
         {(\piParL{b}{t'}{Q})}
&=
\piParLTau{\residual{t}{t'}}{\ren{\push{}}{u}}{y}
\\
\residual{(\piParLTau{t}{u}{y})}
         {(\piParL{c}{t'}{Q})}
&=
\piParLTau{\residual{t}{t'}}{u}{y}
\\
\residual{(\piParLTau{t}{u}{y})}
         {(\piParR{b}{P}{u'})}
&=
\piParLTau{\ren{\push{}}{t}}{\residual{u}{u'}}{y}
\\
\residual{(\piParLTau{t}{u}{y})}
         {(\piParR{c}{P}{u'})}
&=
\piParLTau{t}{\residual{u}{u'}}{y}
\\
\residual{(\piParL{\piInput{x}}{t}{Q})}
         {(\piParLNu{t'}{u})}
&=
\piRestrictA{\piInput{x}}{(\piParL{\piInput{x+1}}{\residual{t}{t'}}{\target{u}})}
\\
\residual{(\piParL{\piBoundOutput{x}}{t}{Q})}
         {(\piParLNu{t'}{u})}
&=
\piRestrictOutput{(\piParL{\piOutput{x+1}{0}}{\residual{t}{t'}}{\target{u}})}
\\
\residual{(\piParL{c}{t}{Q})}
         {(\piParLNu{t'}{u})}
&=
\piRestrictA{c}{(\piParL{\ren{\push{}}{c}}{\residual{t}{t'}}{\target{u}})}
\\
\residual{(\piParR{\piInput{x}}{P}{u})}
         {(\piParLNu{t}{u'})}
&=
\piRestrictA{\piInput{x}}{(\piParR{\piInput{x+1}}{\target{t}}{\residual{u}{u'}})}
\\
\residual{(\piParR{\piBoundOutput{x}}{P}{u})}
         {(\piParLNu{t}{u'})}
&=
\piRestrictOutput{(\piParR{\piOutput{x+1}{0}}{\target{t}}{\residual{u}{u'}})}
\\
\residual{(\piParR{c}{P}{u})}
         {(\piParLNu{t}{u'})}
&=
\piRestrictA{c}{(\piParR{\ren{\push{}}{c}}{\target{t}}{\residual{u}{u'}})}
\\
\residual{(\piParLNu{t}{u})}
         {(\piParL{b}{t'}{Q})}
&=
\piParLNu{\residual{t}{t'}}{\ren{\push{}}{u}}
\\
\residual{(\piParLNu{t}{u})}
         {(\piParL{c}{t'}{Q})}
&=
\piParLNu{\residual{t}{t'}}{u}
\\
\residual{(\piParLNu{t}{u})}
         {(\piParR{\piInput{x}}{P}{u'})}
&=
\piParLNu{\ren{\push{}}{t}}{\residual{u}{u'}}
\\
\residual{(\piParLNu{t}{u})}
         {(\piParR{\piBoundOutput{x}}{P}{u'})}
&=
\piParLTau{\ren{\push{}}{t}}{\residual{u}{u'}}{0}
\\
\residual{(\piParLNu{t}{u})}
         {(\piParR{c}{P}{u'})}
&=
\piParLNu{t}{\residual{u}{u'}}
\\
\residual{(\piChoice{t}{Q})}
         {(\piChoice{t'}{Q})}
&=
\residual{t}{t'}
\\
\residual{(\piParR{\piInput{x}}{P}{u})}
         {(\piParR{b}{P}{u'})}
&=
\piParR{\piInput{x}}{\ren{\push{}}{P}}{\residual{u}{u'}}
\end{align*}
}
\end{minipage}%
\begin{minipage}[t]{0.5\linewidth}
{\small
\begin{align*}
\residual{(\piParR{b}{P}{u})}
         {(\piParR{\piInput{x}}{P}{u'})}
&=
\piParR{b}{\ren{\push{}}{P}}{\residual{u}{u'}}
\\
\residual{(\piParR{\piBoundOutput{x}}{P}{u})}
         {(\piParR{\piBoundOutput{u}}{P}{u'})}
&=
\piParR{\piOutput{x+1}{0}}{\ren{\push{}}{P}}{\residual{u}{u'}}
\\
\residual{(\piParR{c}{P}{u})}
         {(\piParR{b}{P}{u'})}
&=
\piParR{c}{\ren{\push{}}{P}}{\residual{u}{u'}}
\\
\residual{(\piParR{a}{P}{u})}
         {(\piParR{c}{P}{u'})}
&=
\piParR{a}{P}{\residual{u}{u'}}
\\
\residual{(\piParL{\piInput{x}}{t}{Q})}
         {(\piParL{b}{t'}{Q})}
&=
\piParL{\piInput{x}}{\residual{t}{t'}}{\ren{\push{}}{Q}}
\\
\residual{(\piParL{b}{t}{Q})}
         {(\piParL{\piInput{x}}{t'}{Q})}
&=
\piParL{b}{\residual{t}{t'}}{\ren{\push{}}{Q}}
\\
\residual{(\piParL{\piBoundOutput{x}}{t}{Q})}
         {(\piParL{\piBoundOutput{u}}{t'}{Q})}
&=
\piParL{\piOutput{x+1}{0}}{\residual{t}{t'}}{\ren{\push{}}{Q}}
\\
\residual{(\piParL{c}{t}{Q})}
         {(\piParL{b}{t'}{Q})}
&=
\piParL{c}{\residual{t}{t'}}{\ren{\push{}}{Q}}
\\
\residual{(\piParL{a}{t}{Q})}
         {(\piParL{c}{t'}{Q})}
&=
\piParL{a}{\residual{t}{t'}}{Q}
\\
\residual{(\piParLTau{t}{u}{y})}
         {(\piParLTau{t'}{u'}{z})}
&=
\piParLTau{\ren{(\pop{}{z})}{(\residual{t}{t'})}}{\residual{u}{u'}}{y}
\\
\residual{(\piParLTau{t}{u}{y})}
         {(\piParLNu{t'}{u'})}
&=
\piRestrictA{\tau}{(\piParLTau{\residual{t}{t'}}{\residual{u}{u'}}{y})}
\\
\residual{(\piParLNu{t}{u})}
         {(\piParLTau{t'}{u'}{z})}
&=
\piParLNu{\ren{(\pop{}{z})}{(\residual{t}{t'})}}{\residual{u}{u'}}
\\
\residual{(\piParLNu{t}{u})}
         {(\piParLNu{t'}{u'})}
&=
\piRestrictA{\tau}{(\piParLNu{\residual{t}{t'}}{\residual{u}{u'}})}
\\
\residual{(\piRestrictOutput{t})}
         {(\piRestrictOutput{t'})}
&=
\residual{t}{t'}
\\
\residual{(\piRestrictOutput{t})}
         {(\piRestrictA{b}{t'})}
&=
\piRestrictOutput{\;\ren{\swapR{}}{(\residual{t}{t'})}}
\\
\residual{(\piRestrictOutput{t})}
         {(\piRestrictA{c}{t'})}
&=
\piRestrictOutput{\;\residual{t}{t'}}
\\
\residual{(\piRestrictA{b}{t})}
         {(\piRestrictOutput{t'})}
&=
\residual{t}{t'}
\\
\residual{(\piRestrictA{c}{t})}
         {(\piRestrictOutput{t'})}
&=
\residual{t}{t'}
\\
\residual{(\piRestrictA{b}{t})}
         {(\piRestrictA{b}{t'})}
&=
\piRestrict{\;\residual{t}{t'}}
\\
\residual{(\piRestrictA{c}{t})}
         {(\piRestrictA{b}{t'})}
&=
\piRestrictA{c}{\;\ren{\swapR{}}{(\residual{t}{t'})}}
\\
\residual{(\piRestrictA{b}{t})}
         {(\piRestrictA{c}{t'})}
&=
\piRestrictA{b}{\;\residual{t}{t'}}
\\
\residual{(\piRestrictA{c}{t})}
         {(\piRestrictA{c}{t'})}
&=
\piRestrictA{c}{\;\residual{t}{t'}}
\\
\residual{(\piReplicate{t})}
         {(\piReplicate{t'})}
&=
\residual{t}{t'}
\end{align*}
}
\end{minipage} \\
\crossrule
\caption{Residual of $t$ after $t'$, omitting
  $\piParRTau{\param}{\param}{y}$ and $\piParRNu{\param}{\param}$ cases}
\label{fig:residual}
\end{figure}

\noindent The above definition is a total and terminating function on
concurrent transitions; in Agda, this is verified by the typechecker.
Syntactically, the operator $\residual{\param}{\param}$ has higher precedence
than any transition constructor. The definition makes use of the renaming
lemmas in \secref{calculus:properties-of-renamings} and the fact that the
transition system is closed under renamings
(\lemref{concurrent-transitions:renaming:transition}).

\begin{changebar}
While the definition is rather technical, the idea is quite simple: the
residual says how to update one transition to take into account the fact that
the other has taken place, for example by adjusting the path to the redex, or
applying an appropriate renaming. Several examples are included in the
sections which follow. \exref{concurrent-extrusions:same-binder:residuals}
below gives the basic idea, and \secref{concurrent-transitions:examples},
which explains the notion of cofinality, shows how these ``residual redexes''
are obtained in more complicated cases.
\end{changebar}

\begin{example}[Residuals of concurrent transitions]
\label{ex:concurrent-extrusions:same-binder:residuals}
First recall the named process in
\exref{concurrent-extrusions:same-binder} above, that is,
$\piRestrictN{y}{\compl{x}\langle{y}\rangle.P \mid \compl{z}\langle y
  \rangle.Q}$. Both of the transitions it can perform are bound
transitions, extruding $y$, which is no longer bound in the resulting
process. After the first transition, the second can be performed as a
free send of $y$ along $z$ and vice versa, and in both cases we obtain
the process $P \mid Q$, again containing $y$ free.

These observations are reflected in the de Bruijn representation. Since
$t$ and $t'$ are concurrent there should exist residual transitions
denoted $\residual{t'}{t}$ and $\residual{t}{t'}$, which are cofinal,
allowing us to complete the square

\vspace{5pt}
\begin{nscenter}
\scalebox{0.8}{
\begin{tikzpicture}[node distance=1.5cm, auto]
  \node (P) {
    \hspace{-2em}$\Gamma \vdash \piRestrict{(\piPar{\piAction{\piOutput{\suc{x}}{0}}{P}}{\piAction{\piOutput{\suc{z}}{0}}{Q}})}$
  };
  \node (R) [right of=P, above of=P] {
    $\suc{\Gamma} \vdash \piPar{P}{\piAction{\piOutput{\suc{z}}{0}}{Q}}$
  };
  \node (RPrime) [below of=P, right of=P] {
    $\suc{\Gamma} \vdash \piPar{\piAction{\piOutput{\suc{x}}{0}}{P}}{Q}$
  };
  \node (PPrime) [right of=R, below of=R] {
    $\plus{\Gamma}{\Delta} \vdash R$
  };
  \draw[->] (P) to node [yshift=-1ex] {$t$} (R);
  \draw[->] (P) to node [yshift=1ex,swap] {$t'$} (RPrime);
  \draw[dotted,->] (R) to node [yshift=-1ex] {$\residual{t'}{t}$} (PPrime);
  \draw[dotted,->] (RPrime) to node [yshift=1ex,swap] {$\residual{t}{t'}$} (PPrime);
\end{tikzpicture}
}
\end{nscenter}

\noindent for some $\Delta \in \set{1, 2}$ and some process $R$. In the
upper state $\target{t}$, the only candidate for $\residual{t'}{t}$ is
the output prefix $\piOutput{\suc{z}}{0}$. However, the $\nu$-binder to
which index $0$ refers no longer appears in $\target{t}$. Rather, that
binder is propagating and index $0$ is free, reflected by $\target{t}$
being in context $\suc{\Gamma}$. When the output transition is taken,
$\piOutput{\suc{z}}{0}$ therefore simply propagates as a non-bound
action, rather than causing a further extrusion:

\vspace{5pt}
\begin{center}
\scalebox{\smathparscale}{
\begin{smathpar}
\inferrule*[left={\ruleName{$\piParR{\piOutput{\suc{z}}{0}}{P}{\param}$}}]
{
  \suc{\Gamma} \vdash \piAction{\highlight{\piOutput{\suc{z}}{0}}}{Q}
  \transitionWithoutSmash{\piOutput{\suc{z}}{0}}
  \suc{\Gamma} \vdash Q
}
{
  \suc{\Gamma} \vdash \piPar{P}{\piAction{\highlight{\piOutput{\suc{z}}{0}}}{Q}}
  \transitionWithoutSmash{\piOutput{\suc{z}}{0}}
  \suc{\Gamma} \vdash \piPar{P}{Q}
}
\end{smathpar}
}
\end{center}

From the lower state $\target{t'}$ the only candidate for
$\residual{t}{t'}$ is the output prefix $\piOutput{\suc{x}}{0}$, and
similar reasoning applies. Thus for $\residual{t}{t'}$ we have
\begin{center}
\scalebox{\smathparscale}{
\begin{smathpar}
\inferrule*[left={\ruleName{$\piParL{\piOutput{\suc{x}}{0}}{\param}{Q}$}}]
{
  \suc{\Gamma} \vdash \piAction{\highlight{\piOutput{\suc{x}}{0}}}{P}
  \transitionWithoutSmash{\piOutput{\suc{x}}{0}}
  \suc{\Gamma} \vdash P
}
{
  \suc{\Gamma} \vdash \piPar{\piAction{\highlight{\piOutput{\suc{x}}{0}}}{P}}{Q}
  \transitionWithoutSmash{\piOutput{\suc{x}}{0}}
  \suc{\Gamma} \vdash \piPar{P}{Q}
}
\end{smathpar}
}
\end{center}

\noindent and therefore $\Delta = 1$ and $R = \piPar{P}{Q}$. In summary
when concurrent $t$ and $t'$ extrude the same binder, their respective
residuals are plain outputs, not bound outputs, because a given binder
can only be extruded once.

To relate this example to the defining equations of
$\residual{\param}{\param}$ we use the compact presentations of $t$ and
$t'$ from the end of \exref{concurrent-extrusions:same-binder}. It is
then easy to see that the rules in \figref{residual} indeed compute (in
compact form) the derivation above for $\residual{t}{t'}$:

\vspace{-5pt}
{\[\small
\begin{array}{cl}
& \residual{\piRestrictOutput{(\piParL{\piOutput{\suc{x}}{0}}
                                    {\piAction{\piOutput{\suc{x}}{0}}{P}}
                                    {\piAction{\piOutput{\suc{z}}{0}}{Q}})}}
         {\piRestrictOutput{(\piParR{\piOutput{\suc{z}}{0}}
                                    {\piAction{\piOutput{\suc{x}}{0}}{P}}
                                    {\piAction{\piOutput{\suc{z}}{0}}{Q}})}}
\\
=& \residual{(\piParL{\piOutput{\suc{x}}{0}}
                  {\piAction{\piOutput{\suc{x}}{0}}{P}}
                  {\piAction{\piOutput{\suc{z}}{0}}{Q}})}
         {(\piParR{\piOutput{\suc{z}}{0}}
                  {\piAction{\piOutput{\suc{x}}{0}}{P}}
                  {\piAction{\piOutput{\suc{z}}{0}}{Q}})}
\\
= & \piParL{\piOutput{\suc{x}}{0}}
       {\piAction{\piOutput{\suc{x}}{0}}{P}}
       {Q}
\end{array}\]}

\noindent and similarly for $\residual{t'}{t}$. \exampleend
\end{example}

\begin{changebar}\exref{concurrent-extrusions:same-binder:residuals}
\end{changebar} illustrated the basic idea of residuation, focusing on the
specific case where the residuals of transitions with bound actions have
actions that are not bound, a subtlety of residuation particular to
\piCalculus first noted by
\citet{cristescu13}. To capture this and other aspects of residuation,
it is useful to define a datatype of \emph{concurrent actions} $a
\concur a'$ and an associated notion of residual action
$\residual{a}{a'}$ and use these to index concurrent transitions and
their residuals.

We define both of these using the diagrams in
\figref{concurrent-actions} below. The datatype of concurrent actions,
ranged over by $\twoDim{a}$, has five constructors, one for each
diagram; the arrows diverging on the left represent the concurrent
actions $a$ and $a'$, and the arrows converging on the right define the
corresponding residuals $\residual{a'}{a}$ and $\residual{a}{a'}$.
Beneath each diagram is the \emph{braiding} relation
$\cofin{\twoDim{a}}$ which constitutes the notion of cofinality induced
by that form of concurrent action.
\begin{figure}[H]
\begin{nscenter}
\scalebox{0.65}{
\begin{tabular}{ccccc}
(i) & {\hspace{-0.5em}(ii)} & {\hspace{-1em}(iii)} & {\hspace{-1.5em}(iv)} & (v)
\\[2mm]
\begin{tikzpicture}[node distance=1.5cm, auto]
  \node (W) [node distance=2cm] {$\Gamma$};
  \node (N) [right of=W, above of=W] {$\Gamma$};
  \node (S) [below of=W, right of=W] {$\Gamma$};
  \node (E) [right of=N, below of=N] {$\Gamma$};
  \draw[->] (W) to node {$c$} (N);
  \draw[->] (W) to node [swap] {$c'$} (S);
  \draw[dotted,->] (N) to node {$c'$} (E);
  \draw[dotted,->] (S) to node [swap] {$c$} (E);
\end{tikzpicture}
&
\begin{tikzpicture}[node distance=1.5cm, auto]
  \node (W) [node distance=2cm] {$\Gamma$};
  \node (N) [right of=W, above of=W] {$\plus{\Gamma}{1}$};
  \node (S) [below of=W, right of=W] {$\Gamma$};
  \node (E) [right of=N, below of=N] {$\plus{\Gamma}{1}$};
  \draw[->] (W) to node {$b$} (N);
  \draw[->] (W) to node [swap] {$c$} (S);
  \draw[dotted,->] (N) to node {$\ren{\pushR}{c}$} (E);
  \draw[dotted,->] (S) to node [swap] {$b$} (E);
\end{tikzpicture}
&
\begin{tikzpicture}[node distance=1.5cm, auto]
  \node (W) [node distance=2cm] {$\Gamma$};
  \node (N) [right of=W, above of=W] {$\plus{\Gamma}{1}$};
  \node (S) [below of=W, right of=W] {$\plus{\Gamma}{1}$};
  \node (E) [right of=N, below of=N] {$\plus{\Gamma}{2}$};
  \draw[->] (W) to node {$b$} (N);
  \draw[->] (W) to node [swap] {$b'$} (S);
  \draw[dotted,->] (N) to node {$\ren{\pushR}{b'}$} (E);
  \draw[dotted,->] (S) to node [swap] {$\ren{\pushR}{b}$} (E);
\end{tikzpicture}
&
\begin{tikzpicture}[node distance=1.5cm, auto]
  \node (W) [node distance=2cm] {$\Gamma$};
  \node (N) [right of=W, above of=W] {$\plus{\Gamma}{1}$};
  \node (S) [below of=W, right of=W] {$\plus{\Gamma}{1}$};
  \node (E) [right of=N, below of=N] {$\plus{\Gamma}{1}$};
  \draw[->] (W) to node {$\piBoundOutput{x}$} (N);
  \draw[->] (W) to node [swap] {$\piBoundOutput{y}$} (S);
  \draw[dotted,->] (N) to node {$\piOutput{\plus{y}{1}}{0}$} (E);
  \draw[dotted,->] (S) to node [swap] {$\piOutput{\plus{x}{1}}{0}$} (E);
\end{tikzpicture}
&
\begin{tikzpicture}[node distance=1.5cm, auto]
  \node (W) [node distance=2cm] {$\Gamma$};
  \node (N) [right of=W, above of=W] {$\Gamma$};
  \node (S) [below of=W, right of=W] {$\Gamma$};
  \node (E) [right of=N, below of=N] {$\Gamma$};
  \draw[->] (W) to node {$\piTau_\nu$} (N);
  \draw[->] (W) to node [swap] {$\piTau_\nu$} (S);
  \draw[dotted,->] (N) to node {$\piTau_\nu$} (E);
  \draw[dotted,->] (S) to node [swap] {$\piTau_\nu$} (E);
\end{tikzpicture}
\\[2mm]
{\Large$=$} & {\hspace{-0.5em}\Large$=$} & {\hspace{-1em}\Large$\freeBraid$} & {\hspace{-1.5em}\Large$=$} & {\Large$\boundBraid$}
\end{tabular}
}
\end{nscenter}
\caption{Concurrent actions $\twoDim{a}: a \concur a'$, residuals
  $\residual{a'}{a}$ and $\residual{a}{a'}$, and braiding relation
  $\cofin{\twoDim{a}}$}
\label{fig:concurrent-actions}
\end{figure}

Diagrams (i) and (ii) capture the general pattern when at most one of the
actions is bound. In (ii), image of an action in a bound action is the
original action shifted under a binder; in both cases cofinality is simply
equality. Diagram (iii) is the general pattern when both actions are bound: in
this case the target states $P$ and $P'$ are related by a ``free braid'' $P
\freeBraid P'$ in the form of the permutation $\swapR$ which renames $0$ to
$1$ and $1$ to $0$, reflecting the transposition of the two binders. \begin{changebar}Free
braids are illustrated in some detail in
\exrefTwo{concurrent-extrusions:different-binders}{propagating-free-braid}
below.\end{changebar}

Diagram (iv) and (v) are specific to name extrusion. Diagram (iv) is an
exception to (iii) where the two bound actions happen to be extrusions of the
same binder, as in \exref{concurrent-extrusions:same-binder}; the
$\piRestrictOutput{t} \concur \piRestrictOutput{t'}$ rule is the only
concurrency rule that generates concurrent actions of this form. Diagram (v)
is an exception to (i) where the two non-bound actions happen to be
$\nu$-synchronisations of distinct binders. In this case the residual actions
will also be $\nu$-synchronisations (as suggested by the informal $\piTau_\nu$
notation) and the target states are related by a ``bound braid'' $P
\boundBraid P'$, essentially a free braid which has been ``closed'' by a pair
of $\nu$-binders, representing the transposition of those binders. The four
variants of the $\piParLNu{t}{u} \concur \piParLNu{t'}{u'}$ rule generate
concurrent actions of this form whenever the extruding binders are distinct.
\begin{changebar}Bound braids are illustrated in
\exref{concurrent-nu-synchronisations} below.\end{changebar}
 
Free and bound braids are now defined more formally.

\begin{definition}[Free braid]
\label{def:free-braid}
For any processes $\Gamma + 2 \vdash P, R$ define the symmetric relation
$P \freeBraid R$ as follows. The context $\Gamma$ is left implicit.
\[P \freeBraid R \quad\Leftrightarrow\quad P = \ren{\swapR}{R}\]
\end{definition}

\noindent Symmetry of $\freeBraid$ is immediate from the involutivity of
$\swapR$. Note that $\freeBraid$ is not irreflexive, since
$\ren{\swapR}{P} = P$ iff indices $0$ and $1$ are both unused in $P$.

\begin{definition}[Bound braid]
\label{def:bound-braid}
For any processes $\Gamma \vdash P, R$ inductively define the symmetric
relation $P \boundBraid R$ using the rules in \figref{bound-braid}.
Again the context $\Gamma$ is left implicit.
\end{definition}

\begin{figure}[h]
\noindent \shadebox{$P \boundBraid R$}
\begin{smathpar}
\begin{changebar}
\inferrule*[left={\ruleName{$\congRestrictSwap{P}$}}]
{
   P \freeBraid R
}
{
   \piRestrict{\piRestrict{P}}
   \boundBraid
   \piRestrict{\piRestrict{R}}
}
\end{changebar}
\and
\inferrule*[left={\ruleName{$\piChoice{\param}{Q}$}}]
{
   P \boundBraid R
}
{
   \piChoice{P}{Q}
   \boundBraid
   \piChoice{R}{Q}
}
\and
\inferrule*[left={\ruleName{$\piChoice{P}{\param}$}}]
{
   Q \boundBraid S
}
{
   \piChoice{P}{Q}
   \boundBraid
   \piChoice{P}{S}
}
\and
\inferrule*[left={\ruleName{$\piPar{\param}{Q}$}}]
{
   P \boundBraid R
}
{
   \piPar{P}{Q}
   \boundBraid
   \piPar{R}{Q}
}
\and
\inferrule*[left={\ruleName{$\piPar{P}{\param}$}}]
{
   Q \boundBraid S
}
{
   \piPar{P}{Q}
   \boundBraid
   \piPar{P}{S}
}
\and
\inferrule*[left={\ruleName{$\piRestrict{\param}$}}]
{
   P \boundBraid R
}
{
   \piRestrict{P}
   \boundBraid
   \piRestrict{R}
}
\and
\inferrule*[left={\ruleName{$\piReplicate{\param}$}}]
{
   P \boundBraid R
}
{
   \piReplicate{P}
   \boundBraid
   \piReplicate{R}
}
\end{smathpar}
\crossrule
\caption{Bound braid $P \boundBraid R$}
\label{fig:bound-braid}
\end{figure}

\noindent \begin{changebar}Note that the $\congRestrictSwap{P}$ rule requires $P$ and
$R$ to be related by a \emph{free} braid ($P \freeBraid R$) which it then
closes with a pair of $\nu$-binders. By contrast the $\piRestrict{\param}$
rule simply propagates a bound braid $(P \boundBraid R)$ through a
$\nu$-binder.\end{changebar}

We adopt a compact term-like notation for $\boundBraid$ proofs similar
to the convention introduced earlier for transitions. As before, rule
names are shown to the left of each rule, in blue. The symmetry of
$\freeBraid$ follows easily from the symmetry of $\boundBraid$; moreover
$\boundBraid$ is also not irreflexive, because $\freeBraid$ is not
irreflexive. Meta-variables $\phi$ and $\psi$ range over bound braids;
$\source{\phi}$ and $\target{\phi}$ denote $P$ and $R$ for any $\phi: P
\boundBraid R$. Bound braids are ``unobservable'' in the sense that two
processes related by a bound braid are strongly bisimilar. Indeed
$\piRestrict{\piRestrict{(\ren{\swapR}{P})}} \Cong
\piRestrict{\piRestrict{P}}$ is simply the de Bruijn counterpart of the
familiar congruence $\piRestrictN{xy}{P} \Cong \piRestrictN{yx}{P}$.
\begin{changebar}However in our constructive setting -- at least in the
absence of non-trivial techniques or extensions to type theory -- the usual
refrain ``work up to structural congruence!'' is of little help; representing
such congruences would still require explicit witnesses at least as complex as
bound braids.\end{changebar}

Concurrent actions and action residuals give transition residuals a more
precise type (omitted for simplicity from the definitions of $t \concur
t'$ and $\residual{t}{t'}$), making them somewhat easier to formally
define. But more important here is how they determine the appropriate
notion of cofinality relating $\target{\residual{t'}{t}}$ and
$\target{\residual{t}{t'}}$, namely the braiding relation $\cofin{a,a'}$
specified beneath each diagram in \figref{concurrent-actions}. A
braiding relation is a singleton type, whose unique inhabitant precisely
captures precisely the ``rewiring'' effect of reordering transitions
that involve binders.

\begin{definition}[Braiding]
\label{def:cofin}
For any context $\Gamma$, any $a, a' \in \Action{\Gamma}$ and any
$\twoDim{a}: a \concur a'$, define the following symmetric relation
$\cofin{\twoDim{a}}$ over processes in $\Gamma'$, where $\Gamma'$ is the
target context of $\twoDim{a}$.

\vspace{-15pt}
{\small
\begin{align*}
\cofin{\twoDim{a}}
& \eqdef
\text{one of $\freeBraid$, $\boundBraid$ or $=$ as defined in \figref{concurrent-actions}}%
\end{align*}%
}%

\end{definition}

Our key soundness result is that the targets of the residuals of
concurrent transitions $t \concur t'$ with actions $\twoDim{a}: a
\concur a'$ are cofinal in the sense of being related by
$\cofin{\twoDim{a}}$. We need first that bound braids are closed under
renamings, which we capture as a notion of residuation
$\residual{\phi}{\rho}$. The other residual $\residual{\rho}{\phi}$ is
always $\rho$.

\begin{lemma}
\label{lem:concurrent-transitions:renaming-preserves-cong}
For any $\Gamma \vdash P$, suppose $\phi: P \boundBraid Q$ and $\rho:
\Gamma \to \Delta$. Then there exists a bound braid
$\residual{\phi}{\rho}: \ren{\rho}{P} \boundBraid \ren{\rho}{Q}$.

\begin{nscenter}
\scalebox{0.8}{
\begin{tikzpicture}[node distance=1.5cm, auto]
  \node (P) {$P$};
  \node (Q) [below of=P] {$\ren{\rho}{P}$};
  \node (PPrime) [node distance=3.4cm, right of=P] {$Q$};
  \node (QPrime) [below of=PPrime] {$\ren{\rho}{Q}$};
  \draw[->] (P) to node {$\phi$} (PPrime);
  \draw[dotted,->] (Q) to node [swap] {$\residual{\phi}{\rho}$} (QPrime);
  \draw[->] (P) to node [swap] {$\ren{\rho}{}$} (Q);
  \draw[->] (PPrime) to node {$\ren{\rho}{}$} (QPrime);
\end{tikzpicture}
}

\end{nscenter}
\end{lemma}

\begin{theorem}[Cofinality of residuals]
\label{thm:concurrent-transitions:cofinality}
\begin{nscenter}
\scalebox{0.8}{
\begin{tikzpicture}[node distance=1.5cm, auto]
  \node (P) [node distance=2cm] {
    $\Gamma \vdash P$
  };
  \node (R) [right of=P, above of=P] {
    $\Gamma' \vdash R$
  };
  \node (RPrime) [below of=P, right of=P] {
    $\Gamma'' \vdash R'$
  };
  \node (Q) [node distance=3.25cm, right of=R] {
    $\plus{\Gamma}{\Delta} \vdash Q$
  };
  \node (QPrime) [node distance=3.25cm, right of=RPrime] {
    $\plus{\Gamma}{\Delta} \vdash Q'$
  };
  \draw[->] (P) to node [yshift=-1ex] {$t$} (R);
  \draw[->] (P) to node [yshift=1ex,swap] {$t'$} (RPrime);
  \draw[dotted,->] (R) to node {$\residual{t'}{t}$} (Q);
  \draw[dotted,->] (RPrime) to node [swap] {$\residual{t}{t'}$} (QPrime);
  \draw[dotted, -] (Q) to node {$\braiding{t,t'}$} (QPrime);
\end{tikzpicture}
}
\end{nscenter}

Suppose $t \concur t'$ with actions $\twoDim{a}: a \concur a'$. Then
there exists a unique $\braiding{t,t'}: \target{\residual{t'}{t}}
\cofin{\twoDim{a}} \target{\residual{t}{t'}}$.
\end{theorem}

\noindent We omit the $t,t'$ subscripts when the particular concurrent
actions are immaterial.

There is no analogous result to show that the definition of concurrency
is complete: that it includes every pair of coinitial transitions for
which a cofinal notion of residuation might be defined. It is not
entirely clear what form such a theorem might take, nor are we aware of
any such theorem in the literature. Choice in particular is potentially
problematic. Although by our (and \citeauthor{boudol89}'s) definition of
$\concur$ coinitial choices are never concurrent, in the following we
have distinct coinitial choices with ``obvious'' residuals, which are
indeed cofinal:

\vspace{3pt}
\begin{nscenter}
\scalebox{0.8}{
\begin{tikzpicture}[node distance=1.5cm, auto]
  \node (P) [node distance=2cm] {
    $\piChoice{(\piAction{\piInput{x}}{P})}{\piAction{\piInput{x}}{P}}$
  };
  \node (R) [right of=P, above of=P] {
    $\piAction{\piInput{x}}{P}$
  };
  \node (RPrime) [below of=P, right of=P] {
    $\piAction{\piInput{x}}{P}$
  };
  \node (PPrime) [right of=R, below of=R] {
    $P$
  };
  \draw[->] (P) to node {$\piChoice{\param}{\piAction{\piInput{x}}{P}}$} (R);
  \draw[->] (P) to node [swap] {$\piChoice{(\piAction{\piInput{x}}{P})}{\param}$} (RPrime);
  \draw[->] (R) to node {$\piAction{\piInput{x}}{P}$} (PPrime);
  \draw[->] (RPrime) to node [swap] {$\piAction{\piInput{x}}{P}$} (PPrime);
\end{tikzpicture}
}
\end{nscenter}

\noindent \begin{changebar}One avenue for justifying this (and
other) choices about the transition concurrency relation for
\piCalculus might be to prove results about the equivalence of ``proved
transition'' semantics and event structure semantics, analogous to the results
of \citet{boudol91} for CCS. We leave exploring canonical notions of
concurrency to future work.\end{changebar}

\subsection{Examples of braiding}
\label{sec:concurrent-transitions:examples}

\begin{example}[Free braid]
\label{ex:concurrent-extrusions:different-binders}

Free braids arise when there are concurrent bound actions. For example
the $\pushR$ injections used in the propagation rules
\ruleName{$\piParL{\piInput{x}}{\param}{Q}$} and
\ruleName{$\piParR{\piInput{x}}{P}{\param}$} open the process term with
respect to index $0$; if two of these happen consecutively, the order in
which they happen determines the roles of indices $0$ and $1$ in the
final process.

Concurrent name extrusions are analogous. In the process term
$\piRestrict{\piRestrict{(\piPar{(\piAction{\piOutput{\plus{x}{2}}{0}}{P})}
    {\piAction{\piOutput{\plus{z}{2}}{1}}{Q}})}}$ there are two binders
that can be extruded; call the outer one $\nu_1$ and the inner one
$\nu_2$. The output on the left extrudes $\nu_2$, and the output on the
right extrudes $\nu_1$. Let $t$ be the transition that extrudes
$\nu_2$:

\vspace{5pt}
\begin{nscenter}
\scalebox{\smathparscale}{
\begin{smathpar}
\inferrule*[left={\ruleName{$\piRestrictA{\piBoundOutput{x}}{\param}$}}]
{
  \inferrule*[left={\ruleName{$\piRestrictOutput{\param}$}}]
  {
    \inferrule*[left={\ruleName{$\piParL{\piOutput{\plus{x}{2}}{0}}{\param}{\piAction{\piOutput{\plus{z}{2}}{1}}{Q}}$}}]
    {
      \plus{\Gamma}{2} \vdash \piAction{\highlight{\piOutput{\plus{x}{2}}{0}}}{P}
      \transitionWithoutSmash{\piOutput{\plus{x}{2}}{0}}
      \plus{\Gamma}{2} \vdash P
    }
    {
      \plus{\Gamma}{2} \vdash
      \piPar{(\piAction{\highlight{\piOutput{\plus{x}{2}}{0}}}{P})}{\piAction{\piOutput{\plus{z}{2}}{1}}{Q}}
      \transitionWithoutSmash{\piOutput{\plus{x}{2}}{0}}
      \plus{\Gamma}{2} \vdash \piPar{P}{\piAction{\piOutput{\plus{z}{2}}{1}}{Q}}
    }
  }
  {
    \suc{\Gamma} \vdash
    \piRestrict{(\piPar{(\piAction{\highlight{\piOutput{\plus{x}{2}}{0}}}{P})}{\piAction{\piOutput{\plus{z}{2}}{1}}{Q}})}
    \transitionWithoutSmash{\piBoundOutput{\suc{x}}}
    \plus{\Gamma}{2} \vdash \piPar{P}{\piAction{\piOutput{\plus{z}{2}}{1}}{Q}}
  }
}
{
  \Gamma \vdash
  \piRestrict{\piRestrict{(\piPar{(\piAction{\highlight{\piOutput{\plus{x}{2}}{0}}}{P})}
                                 {\piAction{\piOutput{\plus{z}{2}}{1}}{Q}})}}
  \transitionWithoutSmash{\piBoundOutput{x}}
  \suc{\Gamma} \vdash \piRestrict{(\piPar{\ren{\swapR}{P}}{\piAction{\piOutput{\plus{z}{2}}{0}}{\ren{\swapR}{Q}}})}
}
\end{smathpar}
}
\end{nscenter}

\vspace{1pt} 

Here $\nu_2$ is extruded as the bound output $\piBoundOutput{\suc{x}}$,
propagating through the outer binder $\nu_1$ as $\piBoundOutput{x}$. In
$\target{t}$, index $0$ refers to the extruding $\nu_2$; the binder
remaining in the process term is $\nu_1$. The key detail here is that
the rule \ruleName{$\piRestrictA{\piBoundOutput{x}}{\param}$} moves
$\nu_1$ past $\nu_2$, explaining the use of $\swapR$ in $\target{t}$:
whenever a propagating binder moves past a static binder, a $\swapR$
must be applied under the static binder to preserve the local meaning of
indices $0$ and $1$ (\secref{calculus:transitions} above). This is also
why $\piOutput{\plus{z}{2}}{1}$ in $\source{t}$ becomes
$\piOutput{\plus{z}{2}}{0}$ in $\target{t}$.

Now let $t'$ be the transition that extrudes $\nu_1$:

\vspace{5pt}
\begin{nscenter}
\scalebox{\smathparscale}{
\begin{smathpar}
\inferrule*[left={\ruleName{$\piRestrictOutput{\param}$}}]
{
  \inferrule*[left={\ruleName{$\piRestrictA{\piOutput{\suc{z}}{0}}{\param}$}}]
  {
    \inferrule*[left={\ruleName{$\piParR{\piOutput{\plus{z}{2}}{1}}{(\piAction{\piOutput{\plus{x}{2}}{0}}{P})}{\param}$}}]
    {
      \plus{\Gamma}{2} \vdash \piAction{\highlight{\piOutput{\plus{z}{2}}{1}}}{Q}
      \transitionWithoutSmash{\piOutput{\plus{z}{2}}{1}}
      \plus{\Gamma}{2} \vdash Q
    }
    {
      \plus{\Gamma}{2} \vdash
      \piPar{(\piAction{\piOutput{\plus{x}{2}}{0}}{P})}{\piAction{\highlight{\piOutput{\plus{z}{2}}{1}}}{Q}}
      \transitionWithoutSmash{\piOutput{\plus{z}{2}}{1}}
      \plus{\Gamma}{2} \vdash \piPar{(\piAction{\piOutput{\plus{x}{2}}{0}}{P})}{Q}
    }
  }
  {
    \suc{\Gamma} \vdash
    \piRestrict{(\piPar{(\piAction{\piOutput{\plus{x}{2}}{0}}{P})}{\piAction{\highlight{\piOutput{\plus{z}{2}}{1}}}{Q}})}
    \transitionWithoutSmash{\piOutput{\suc{z}}{0}}
    \suc{\Gamma} \vdash \piRestrict{(\piPar{(\piAction{\piOutput{\plus{x}{2}}{0}}{P})}{Q})}
  }
}
{
  \Gamma \vdash
  \piRestrict{\piRestrict{(\piPar{(\piAction{\piOutput{\plus{x}{2}}{0}}{P})}
                                 {\piAction{\highlight{\piOutput{\plus{z}{2}}{1}}}{Q}})}}
  \transitionWithoutSmash{\piBoundOutput{z}}
  \suc{\Gamma} \vdash \piRestrict{(\piPar{(\piAction{\piOutput{\plus{x}{2}}{0}}{P})}{Q})}
}
\end{smathpar}
}
\end{nscenter}

\vspace{1pt} 

\noindent In this case, the output $\piOutput{\plus{x}{2}}{1}$
propagates through the inner binder $\nu_2$ as $\piOutput{\suc{z}}{0}$,
and then becomes the extrusion $\piBoundOutput{z}$ of the outer binder
$\nu_1$. In $\target{t'}$, index $0$ thus refers to the extruding
$\nu_1$, and the binder that remains in the process term is $\nu_2$.
The key detail here is that there is no $\swapR$ in $\target{t'}$
because this time the relative positions of $\nu_1$ and $\nu_2$ are
unchanged: the extruding $\nu_1$ is still the outer of the two binders.

The proof that $t$ and $t'$ are concurrent is straightforward because
the outputs occur under opposite sides of a parallel composition. The
notion of cofinality, however, is complicated by the use of indices to
refer to $\nu_1$ and $\nu_2$. Naively, our expectation would be to
derive $\residual{t'}{t}$ and $\residual{t}{t'}$ that complete the
square

\vspace{5pt}
\begin{nscenter}
\scalebox{0.8}{
\begin{tikzpicture}[node distance=1.5cm, auto]
  \node (P) {
    \hspace{-3em}$\Gamma \vdash
  \piRestrict{\piRestrict{(\piPar{(\piAction{\piOutput{\plus{x}{2}}{0}}{P})}
                                 {\piAction{\piOutput{\plus{z}{2}}{1}}{Q}})}}$
  };
  \node (R) [right of=P, above of=P] {
    $\suc{\Gamma} \vdash \piRestrict{(\piPar{\ren{\swapR}{P}}{\piAction{\piOutput{\plus{z}{2}}{0}}{\ren{\swapR}{Q}}})}$
  };
  \node (RPrime) [below of=P, right of=P] {
    $\suc{\Gamma} \vdash \piRestrict{(\piPar{(\piAction{\piOutput{\plus{x}{2}}{0}}{P})}{Q})}$
  };
  \node (PPrime) [right of=R, below of=R] {
    $\plus{\Gamma}{\Delta} \vdash R$
  };
  \draw[->] (P) to node [yshift=-1ex] {$t$} (R);
  \draw[->] (P) to node [yshift=1ex,swap] {$t'$} (RPrime);
  \draw[dotted,->] (R) to node [yshift=-1ex] {$\residual{t'}{t}$} (PPrime);
  \draw[dotted,->] (RPrime) to node [yshift=1ex,swap] {$\residual{t}{t'}$} (PPrime);
\end{tikzpicture}
}
\end{nscenter}

\noindent for some $\Delta \in \set{1, 2}$ and some process $R$.
However, the only candidate for $\residual{t'}{t}$ is to select the
output redex on the right, which becomes an extrusion of the remaining
binder, in this case $\nu_1$:

\vspace{5pt}
\begin{nscenter}
\scalebox{\smathparscale}{
\begin{smathpar}
\inferrule*[left={\ruleName{$\piRestrictOutput{\param}$}}]
{
  \inferrule*[left={\ruleName{$\piParR{\piOutput{\plus{z}{2}}{0}}{\ren{\swapR}{P}}{\param}$}}]
  {
    \plus{\Gamma}{2} \vdash \piAction{\highlight{\piOutput{\plus{z}{2}}{0}}}{\ren{\swapR}{Q}}
    \transitionWithoutSmash{\piOutput{\plus{z}{2}}{0}}
    \plus{\Gamma}{2} \vdash \ren{\swapR}{Q}
  }
  {
    \plus{\Gamma}{2} \vdash \piPar{\ren{\swapR}{P}}{\piAction{\highlight{\piOutput{\plus{z}{2}}{0}}}{\ren{\swapR}{Q}}}
    \transitionWithoutSmash{\piOutput{\plus{z}{2}}{0}}
    \plus{\Gamma}{2} \vdash \piPar{\ren{\swapR}{P}}{\ren{\swapR}{Q}}
  }
}
{
  \suc{\Gamma} \vdash \piRestrict{(\piPar{\ren{\swapR}{P}}{\piAction{\highlight{\piOutput{\plus{z}{2}}{0}}}{\ren{\swapR}{Q}}})}
  \transitionWithoutSmash{\piBoundOutput{\plus{z}{2}}}
  \plus{\Gamma}{2} \vdash \piPar{\ren{\swapR}{P}}{\ren{\swapR}{Q}}
}
\end{smathpar}
}
\end{nscenter}

\vspace{1pt}

\noindent leaving indices $0, 1$ referring to $\nu_1, \nu_2$
respectively in $\target{\residual{t'}{t}}$. Equally, the only candidate
for the other residual $\residual{t}{t'}$ is to select the output redex
on the left, which also becomes an extrusion of the remaining binder, in
this case $\nu_2$:

\vspace{5pt}
\begin{nscenter}
\scalebox{\smathparscale}{
\begin{smathpar}
\inferrule*[left={\ruleName{$\piRestrictOutput{\param}$}}]
{
  \inferrule*[left={\ruleName{$\piParL{\piOutput{\plus{x}{2}}{0}}{\param}{Q}$}}]
  {
    \plus{\Gamma}{2} \vdash \piAction{\highlight{\piOutput{\plus{x}{2}}{0}}}{P}
    \transitionWithoutSmash{\piOutput{\plus{x}{2}}{0}}
    \plus{\Gamma}{2} \vdash P
  }
  {
    \plus{\Gamma}{2} \vdash \piPar{(\piAction{\highlight{\piOutput{\plus{x}{2}}{0}}}{P})}{Q}
    \transitionWithoutSmash{\piOutput{\plus{x}{2}}{0}}
    \plus{\Gamma}{2} \vdash \piPar{P}{Q}
  }
}
{
  \suc{\Gamma} \vdash \piRestrict{(\piPar{(\piAction{\highlight{\piOutput{\plus{x}{2}}{0}}}{P})}{Q})}
  \transitionWithoutSmash{\piBoundOutput{\plus{x}{2}}}
  \plus{\Gamma}{2} \vdash \piPar{P}{Q}
}
\end{smathpar}
}
\end{nscenter}

\vspace{1pt}

\noindent leaving indices $0, 1$ in $\target{\residual{t}{t'}}$
referring to $\nu_2, \nu_1$ rather than $\nu_1, \nu_2$. So instead of
the expected square, we have the pentagon

\vspace{5pt}
\begin{nscenter}
\scalebox{0.8}{
\begin{tikzpicture}[node distance=1.5cm, auto,inner sep=2mm]
  \node (P) {
    \hspace{-3em}$\Gamma \vdash
  \piRestrict{\piRestrict{(\piPar{(\piAction{\piOutput{\plus{x}{2}}{0}}{P})}
                                 {\piAction{\piOutput{\plus{z}{2}}{1}}{Q}})}}$
  };
  \node (R) [right of=P, above of=P] {
    \hspace{-2em}$\suc{\Gamma} \vdash \piRestrict{(\piPar{\ren{\swapR}{P}}{\piAction{\piOutput{\plus{z}{2}}{0}}{\ren{\swapR}{Q}}})}$
  };
  \node (RPrime) [below of=P, right of=P] {
    $\suc{\Gamma} \vdash \piRestrict{(\piPar{(\piAction{\piOutput{\plus{x}{2}}{0}}{P})}{Q})}$
  };
  \node (PPrime) [node distance=5.5cm, right of=R] {
    $\plus{\Gamma}{2} \vdash \piPar{\ren{\swapR}{P}}{\ren{\swapR}{Q}}$
  };
  \node (PDoublePrime) [node distance=5.5cm, right of=RPrime] {
    $\plus{\Gamma}{2} \vdash \piPar{P}{Q}$
  };
  \draw[->] (P) to node [yshift=-1ex] {$t$} (R);
  \draw[->] (P) to node [yshift=1ex,swap] {$t'$} (RPrime);
  \draw[dotted,->] (R) to node [yshift=-1ex] {$\residual{t'}{t}$} (PPrime);
  \draw[dotted,->] (RPrime) to node [yshift=1ex,swap] {$\residual{t}{t'}$} (PDoublePrime);
  \draw[-] (PPrime) to node {$\ren{\swapR}{}$} (PDoublePrime);
\end{tikzpicture}
}
\end{nscenter}

\noindent with a $\swapR$ path between $\target{\residual{t'}{t}}$ and
$\target{\residual{t}{t'}}$ reflecting the reordering of the propagating
binders $\nu_1$ and $\nu_2$. \exampleend
\end{example}

\begin{example}[Propagating free braid]
\label{ex:propagating-free-braid}

Free braids are preserved by enclosing transitions as long as the
residual actions of those transitions are bound. In particular, if a
free braid propagates through a $\nu$-binder it remains a free braid.
Suppose $t \concur t'$ where the residual actions are both bound, so
that $\target{\residual{t'}{t}} \freeBraid \target{\residual{t}{t'}}$:

\begin{center}
\scalebox{0.8}{
\begin{tikzpicture}[node distance=1.5cm, auto]
  \node (PQ) [node distance=2cm] {
    $\suc{\Gamma} \vdash P$
  };
  \node (PPushQ) [right of=PQ, above of=PQ] {
    $\plus{\Gamma}{2} \vdash R$
  };
  \node (PushPQ) [below of=PQ, right of=PQ] {
    $\plus{\Gamma}{2} \vdash R'$
  };
  \node (PushPSucPushQ) [node distance=4cm, right of=PPushQ] {
    $\plus{\Gamma}{3} \vdash P'$
  };
  \node (SwapPSwapQ) [node distance=4cm, right of=PushPQ] {
    $\plus{\Gamma}{3} \vdash \ren{\swapR}{P'}$
  };
  \draw[->] (PQ) to node [yshift=-1ex] {$\cxtRaw{t}{\piInput{\suc{x}}}$} (PPushQ);
  \draw[->] (PQ) to node [yshift=1ex,swap] {$\cxtRaw{t'}{\piInput{\suc{z}}}$} (PushPQ);
  \draw[dotted,->] (PPushQ) to node {$\cxtRaw{(\residual{t'}{t})}{\piInput{\plus{z}{2}}}$} (PushPSucPushQ);
  \draw[dotted,->] (PushPQ) to node [swap] {$\cxtRaw{(\residual{t}{t'})}{\piInput{\plus{x}{2}}}$} (SwapPSwapQ);
  \draw[-] (PushPSucPushQ) to node {$\ren{\swapR}{}$} (SwapPSwapQ);
\end{tikzpicture}
}
\end{center}

\noindent Since both $\piInput{\suc{x}}$ and $\piInput{\suc{z}}$ are of
the form $\ren{\pushR}{b}$, we can use the
\ruleName{$\piRestrictA{b}{\param}$} rule to form the composite
transitions $\piRestrictA{\piInput{x}}{t}$ and
$\piRestrictA{\piInput{z}}{t'}$ which propagate the input actions of $t$
and $t'$ actions through a $\nu$-binder:

\vspace{5pt}
\begin{center}
\scalebox{\smathparscale}{
\begin{smathpar}
\inferrule*[left={\ruleName{$\piRestrictA{\piInput{x}}{\param}$}}]
{
  \inferrule*[left={\ruleName{$t$}}]
  {
    \vdots
  }
  {
    \suc{\Gamma} \vdash P
    \transitionWithoutSmash{\piInput{\suc{x}}}
    \plus{\Gamma}{2} \vdash R
  }
}
{
  \Gamma \vdash \piRestrict{P}
  \transitionWithoutSmash{\piInput{x}}
  \suc{\Gamma} \vdash \piRestrict{(\ren{\swapR}{R})}
}
\and
\inferrule*[left={\ruleName{$\piRestrictA{\piInput{z}}{\param}$}}]
{
  \inferrule*[left={\ruleName{$t'$}}]
  {
    \vdots
  }
  {
    \suc{\Gamma} \vdash P
    \transitionWithoutSmash{\piInput{\suc{z}}}
    \plus{\Gamma}{2} \vdash R'
  }
}
{
  \Gamma \vdash \piRestrict{P}
  \transitionWithoutSmash{\piInput{z}}
  \suc{\Gamma} \vdash \piRestrict{(\ren{\swapR}{R'})}
}
\end{smathpar}
}
\end{center}

\vspace{5pt}

Since $t \concur t'$ we can conclude $\piRestrictA{\piInput{x}}{t}
\concur \piRestrictA{\piInput{z}}{t'}$ by the rules in \figref{residual}
and compute the following composite residual
$\residual{(\piRestrictA{\piInput{z}}{t'})}{\piRestrictA{\piInput{x}}{t}}$:

\vspace{5pt}
\begin{center}
\scalebox{\smathparscale}{
\begin{smathpar}
\inferrule*[left={\ruleName{$\piRestrictA{\piInput{\suc{z}}}{\param}$}}]
{
  \inferrule*[left={\ruleName{$\ren{\swapR}{\param}$}}]
  {
    \inferrule*[left={\ruleName{$\residual{t'}{t}$}}]
    {
      \vdots
    }
    {
      \plus{\Gamma}{2} \vdash R
      \transitionWithoutSmash{\piInput{\plus{z}{2}}}
      \plus{\Gamma}{3} \vdash P'
    }
  }
  {
    \plus{\Gamma}{2} \vdash \ren{\swapR}{R}
    \transitionWithoutSmash{\piInput{\plus{z}{2}}}
    \plus{\Gamma}{3} \vdash \ren{(\suc{\swapR})}{P'}
  }
}
{
  \suc{\Gamma} \vdash \piRestrict{(\ren{\swapR}{R})}
  \transitionWithoutSmash{\piInput{\suc{z}}}
  \plus{\Gamma}{2} \vdash \piRestrict{(\ren{\swapR}{\ren{(\suc{\swapR})}{P'}})}
}
\end{smathpar}
}
\end{center}

\vspace{6pt}

\noindent noting that $\ren{\swapR}{(\piInput{\plus{z}{2}})} =
\piInput{\plus{z}{2}}$ by \lemref{swap-suc-suc}. The other residual
$\residual{(\piRestrictA{\piInput{x}}{t})}{\piRestrictA{\piInput{z}}{t'}}$
is similar but has an extra $\swapR$ inherited from
$\target{\residual{t}{t'}}$:

\vspace{5pt}
\begin{center}
\scalebox{\smathparscale}{
\begin{smathpar}
\inferrule*[left={\ruleName{$\piRestrictA{\piBoundOutput{\suc{x}}}{\param}$}}]
{
  \inferrule*[left={\ruleName{$\ren{\swapR}{\param}$}}]
  {
    \inferrule*[left={\ruleName{$\residual{t}{t'}$}}]
    {
      \vdots
    }
    {
      \plus{\Gamma}{2} \vdash R'
      \transitionWithoutSmash{\piInput{\plus{x}{2}}}
      \plus{\Gamma}{3} \vdash \ren{\swapR}{P'}
    }
  }
  {
    \plus{\Gamma}{2} \vdash \ren{\swapR}{R'}
    \transitionWithoutSmash{\piInput{\plus{x}{2}}}
    \plus{\Gamma}{3} \vdash \ren{(\suc{\swapR})}{\ren{\swapR}{P'}}
  }
}
{
  \suc{\Gamma} \vdash \piRestrict{(\ren{\swapR}{R'})}
  \transitionWithoutSmash{\piInput{\suc{x}}}
  \plus{\Gamma}{2} \vdash \piRestrict{(\ren{\swapR}{\ren{(\suc{\swapR})}{\ren{\swapR}{P'}}})}
}
\end{smathpar}
}
\end{center}

\vspace{6pt}

Nevertheless, the target states of the composite residuals are still
equated by $\swapR$, consistent with the fact that the residual actions
still bound.

\begin{center}
\scalebox{0.8}{
\begin{tikzpicture}[node distance=1.5cm, auto]
  \node (PQ) [node distance=2cm] {
    $\Gamma \vdash \piRestrict{P}$
  };
  \node (PPushQ) [right of=PQ, above of=PQ] {
    $\plus{\Gamma}{1} \vdash \piRestrict{(\ren{\swapR}{R})}$
  };
  \node (PushPQ) [below of=PQ, right of=PQ] {
    $\plus{\Gamma}{1} \vdash \piRestrict{(\ren{\swapR}{R'})}$
  };
  \node (PushPSucPushQ) [node distance=6.5cm, right of=PPushQ] {
    $\plus{\Gamma}{2} \vdash \piRestrict{(\ren{\swapR}{\ren{(\suc{\swapR})}{P'}})}$
  };
  \node (5) [below of=PushPSucPushQ] {
    $\plus{\Gamma}{2} \vdash \piRestrict{(\ren{(\suc{\swapR})}{\ren{\swapR}{\ren{(\suc{\swapR})}{P'}}})}$
  };
  \node (SwapPSwapQ) [node distance=6.5cm, right of=PushPQ] {
    $\plus{\Gamma}{2} \vdash \piRestrict{(\ren{\swapR}{\ren{(\suc{\swapR})}{\ren{\swapR}{P'}}})}$
  };
  \draw[->] (PQ) to node [yshift=-1ex] {$\piRestrictA{\piInput{x}}{t}$} (PPushQ);
  \draw[->] (PQ) to node [yshift=1ex,swap] {$\piRestrictA{\piInput{z}}{t'}$} (PushPQ);
  \draw[dotted,->] (PPushQ) to node {$\piRestrictA{\piInput{\suc{z}}}{(\ren{\swapR}{\residual{t'}{t}})}$} (PushPSucPushQ);
  \draw[dotted,->] (PushPQ) to node [swap] {$\piRestrictA{\piInput{\suc{x}}}{(\ren{\swapR}{\residual{t}{t'}})}$} (SwapPSwapQ);
  \draw[-] (PushPSucPushQ) to node {$\ren{\swapR}{}$} (5);
  \draw[-,double distance=1pt] (5) to node {$\piRestrict{\alpha}$} (SwapPSwapQ);
\end{tikzpicture}
}
\end{center}

\noindent Here $\alpha$ is the hexagon equating two ways of transposing
indices $0$ and $2$ (\lemref{swap-after-suc-swap-after-swap}) which
$\piRestrict{\alpha}$ lifts via congruence to an equality between one
target and the $\swapR$ image of the other. Thus $\freeBraid$ remains
the appropriate notion of cofinality.
\exampleend
\end{example}

\begin{example}[Bound braid]
\label{ex:concurrent-nu-synchronisations}

A bound braid arises when concurrent $\nu$-synchronisations have
residuals that also $\nu$-synchronise, which requires the underlying
extrusions to be distinct binders. The concurrent transitions $t \concur
t'$ and $u \concur u'$ below can be composed into concurrent
$\nu$-synchronisations that have this property; $u$ has an input
$\piInput{x}$ matching the bound output $\piBoundOutput{x}$ of $t$, and
$u'$ has a bound output $\piBoundOutput{z}$ matching the input
$\piInput{z}$ of $t'$. The extrusions $\piBoundOutput{x}$ and
$\piBoundOutput{z}$ are clearly of distinct binders since they arise on
opposite sides of a parallel composition.

\begin{center}
\scalebox{0.8}{
\begin{tikzpicture}[node distance=1.5cm, auto]
  \node (Q) [node distance=2cm] {$\Gamma \vdash P$};
  \node (SPrime) [below of=Q, right of=Q] {$\suc{\Gamma} \vdash R'$};
  \node (S) [right of=Q, above of=Q] {$\suc{\Gamma} \vdash R$};
  \node (QPrime) [node distance=3.5cm, right of=S] {$\plus{\Gamma}{2} \vdash P'$};
  \node (SwapQPrime) [node distance=3.5cm, right of=SPrime] {$\plus{\Gamma}{2} \vdash \ren{\swapR}{P'}$};
  \draw[->] (Q) to node [yshift=-1ex] {$\cxtRaw{t}{\piBoundOutput{x}}$} (S);
  \draw[->] (Q) to node [yshift=1ex,swap] {$\cxtRaw{t'}{\piInput{z}}$} (SPrime);
  \draw[dotted,->] (S) to node {$\cxtRaw{(\residual{t'}{t})}{\piInput{\suc{z}}}$} (QPrime);
  \draw[dotted,->] (SPrime) to node [swap] {$\cxtRaw{(\residual{t}{t'})}{\piBoundOutput{\suc{x}}}$} (SwapQPrime);
  \draw[-] (QPrime) to node {$\ren{\swapR}{}$} (SwapQPrime);
\end{tikzpicture}%
\qquad%
\begin{tikzpicture}[node distance=1.5cm, auto]
  \node (P) [node distance=2cm] {$\Gamma \vdash Q$};
  \node (RPrime) [below of=P, right of=P] {$\suc{\Gamma} \vdash S'$};
  \node (R) [right of=P, above of=P] {$\suc{\Gamma} \vdash S$};
  \node (PPrime) [node distance=3.5cm, right of=R] {$\plus{\Gamma}{2} \vdash Q'$};
  \node (SwapPPrime) [node distance=3.5cm, right of=RPrime] {$\plus{\Gamma}{2} \vdash \ren{\swapR}{Q'}$};
  \draw[->] (P) to node [yshift=-1ex] {$\cxtRaw{u}{\piInput{x}}$} (R);
  \draw[->] (P) to node [yshift=1ex,swap] {$\cxtRaw{u'}{\piBoundOutput{z}}$} (RPrime);
  \draw[dotted,->] (R) to node {$\cxtRaw{(\residual{u'}{u})}{\piBoundOutput{\suc{z}}}$} (PPrime);
  \draw[dotted,->] (RPrime) to node [swap] {$\cxtRaw{(\residual{u}{u'})}{\piInput{\suc{x}}}$} (SwapPPrime);
  \draw[-] (PPrime) to node {$\ren{\swapR}{}$} (SwapPPrime);
\end{tikzpicture}
}
\end{center}

\noindent The composites are the $\nu$-synchronisations
$\piParRNu{t}{u}: \piPar{P}{Q} \transitionWithoutSmash{\tau}
\piRestrict{(\piPar{R}{S})}$ and $\piParLNu{t'}{u'}: \piPar{P}{Q}
\transitionWithoutSmash{\tau} \piRestrict{(\piPar{R'}{S'})}$. Moreover
since $t \concur t'$ and $u \concur u'$ we can conclude $\piParRNu{t}{u}
\concur \piParLNu{t'}{u'}$ using the rules in \figref{concurrent}. The
equations in \figref{residual} determine the residual
$\residual{(\piParLNu{t'}{u'})}{(\piParRNu{t}{u})} =
\piRestrictA{\piTau}{(\piParLNu{\residual{t'}{t}}{\residual{u'}{u}})}$,
which we write down in full for clarity:

\begin{center}
\scalebox{\smathparscale}{
\begin{smathpar}
\inferrule*[left={\ruleName{$\piRestrictA{\piTau}{\param}$}}]
{
  \inferrule*[left={\ruleName{$\piParLNu{\param}{\param}$}}]
  {
    \inferrule*[left={\ruleName{$\residual{t'}{t}$}}]
    {
      \vdots
    }
    {
      \suc{\Gamma} \vdash S
      \transitionWithoutSmash{\piInput{\suc{z}}}
      \plus{\Gamma}{2} \vdash Q'
    }
    \inferrule*[left={\ruleName{$\residual{u'}{u}$}}]
    {
      \vdots
    }
    {
      \suc{\Gamma} \vdash R
      \transitionWithoutSmash{\piBoundOutput{\suc{z}}}
      \plus{\Gamma}{2} \vdash P'
    }
  }
  {
    \suc{\Gamma} \vdash \piPar{R}{S}
    \transitionWithoutSmash{\tau}
    \suc{\Gamma} \vdash \piRestrict{(\piPar{P'}{Q'})}
  }
}
{
  \Gamma \vdash \piRestrict{(\piPar{R}{S})}
  \transitionWithoutSmash{\tau}
  \Gamma \vdash \piRestrict{\piRestrict{(\piPar{P'}{Q'})}}
}
\end{smathpar}
}
\end{center}

The other residual $\residual{(\piParLNu{t'}{u'})}{(\piParRNu{t}{u})} =
\piRestrictA{\piTau}{(\piParLNu{\residual{t'}{t}}{\residual{u'}{u}})}$
is similar, except it inherits two extra $\swapR$ renamings from
$\residual{t}{t'}$ and $\residual{u}{u'}$:

\begin{center}
\scalebox{\smathparscale}{
\begin{smathpar}
\inferrule*[left={\ruleName{$\piRestrictA{\piTau}{\param}$}}]
{
  \inferrule*[left={\ruleName{$\piParRNu{\param}{\param}$}}]
  {
    \inferrule*[left={\ruleName{$\residual{t}{t'}$}}]
    {
      \vdots
    }
    {
      \suc{\Gamma} \vdash S'
      \transitionWithoutSmash{\piBoundOutput{\suc{x}}}
      \plus{\Gamma}{2} \vdash \ren{\swapR}{Q'}
    }
    \inferrule*[left={\ruleName{$\residual{u}{u'}$}}]
    {
      \vdots
    }
    {
      \suc{\Gamma} \vdash R'
      \transitionWithoutSmash{\piInput{\suc{x}}}
      \plus{\Gamma}{2} \vdash \ren{\swapR}{P'}
    }
  }
  {
    \suc{\Gamma} \vdash \piPar{R'}{S'}
    \transitionWithoutSmash{\tau}
    \suc{\Gamma} \vdash \piRestrict{(\piPar{\ren{\swapR}{P'}}{\ren{\swapR}{Q'}})}
  }
}
{
  \Gamma \vdash \piRestrict{(\piPar{R'}{S'})}
  \transitionWithoutSmash{\tau}
  \Gamma \vdash \piRestrict{\piRestrict{(\piPar{\ren{\swapR}{P'}}{\ren{\swapR}{Q'}})}}
}
\end{smathpar}
}
\end{center}

\noindent Thus each residual $\nu$-synchronises, and then propagates
through the binder reinserted by the first synchronisation, leaving a
double-$\nu$ in the final process. The residuals are related by the
pentagon

\begin{center}
\scalebox{0.8}{
\begin{tikzpicture}[node distance=1.5cm, auto, inner sep=1.5mm]
  \node (PQ) [node distance=2cm] {$\Gamma \vdash \piPar{P}{Q}$};
  \node (NuRPrimeSPrime) [below of=PQ, right of=PQ] {$\Gamma \vdash \piRestrict{(\piPar{R'}{S'})}$};
  \node (RS) [right of=PQ, above of=PQ] {$\Gamma \vdash \piRestrict{(\piPar{R}{S})}$};
  \node (NuPPrimeQPrime) [node distance=6cm, right of=RS] {
    $\Gamma \vdash \piRestrict{\piRestrict{(\piPar{P'}{Q'})}}$
  };
  \node (NuSwapPPrimeQPrime) [node distance=6cm, right of=NuRPrimeSPrime] {
    $\Gamma \vdash \piRestrict{\piRestrict{(\piPar{\ren{\swapR}{P'}}{\ren{\swapR}{Q'}})}}$
  };
  \draw[->] (PQ) to node [xshift=1ex,yshift=-1.5ex] {$\piParRNu{t}{u}$} (RS);
  \draw[->] (PQ) to node [yshift=1ex,swap] {$\piParLNu{t'}{u'}$} (NuRPrimeSPrime);
  \draw[dotted,->] (RS) to node {
    $\piRestrictA{\piTau}{(\piParLNu{\residual{t'}{t}}{\residual{u'}{u}})}$
  } (NuPPrimeQPrime);
  \draw[dotted,->] (NuRPrimeSPrime) to node [swap] {
    $\piRestrictA{\piTau}{(\piParRNu{\residual{t}{t'}}{\residual{u}{u'}})}$
  } (NuSwapPPrimeQPrime);
  \draw[-] (NuPPrimeQPrime) to node {$\congRestrictSwap{\piPar{P'}{Q'}}$} (NuSwapPPrimeQPrime);
\end{tikzpicture}
}
\end{center}

\noindent where $\congRestrictSwap{\piPar{P'}{Q'}}$ is the bound braid
that locates $\piPar{P'}{Q'} \freeBraid
\piPar{\ren{\swapR}{P'}}{\ren{\swapR}{Q'}}$ under the two binders,
representing the reordering of the binders.
\exampleend
\end{example}

\begin{example}[Braid erasure by synchronisation]
A free braid is erased if it is enclosed by a concurrent transition where the
notion of cofinality is equality. For example, consider a variant of
\exref{concurrent-nu-synchronisations} where the extrusions
$\piBoundOutput{x}$ and $\piBoundOutput{z}$ occur on the same side of the
parallel composition, and represent extrusions of the same binder.

\begin{center}
\scalebox{0.8}{
\begin{tikzpicture}[node distance=1.5cm, auto]
  \node (P) [node distance=2cm] {$\Gamma \vdash P$};
  \node (RPrime) [below of=P, right of=P] {$\suc{\Gamma} \vdash R'$};
  \node (R) [right of=P, above of=P] {$\suc{\Gamma} \vdash R$};
  \node (PPrime) [node distance=3.5cm, right of=R] {$\plus{\Gamma}{2} \vdash P'$};
  \node (SwapPPrime) [node distance=3.5cm, right of=RPrime] {$\plus{\Gamma}{2} \vdash \ren{\swapR}{P'}$};
  \draw[->] (P) to node [yshift=-1ex] {$\cxtRaw{t}{\piInput{x}}$} (R);
  \draw[->] (P) to node [yshift=1ex,swap] {$\cxtRaw{t'}{\piInput{z}}$} (RPrime);
  \draw[dotted,->] (R) to node {$\cxtRaw{(\residual{t'}{t})}{\piInput{\suc{z}}}$} (PPrime);
  \draw[dotted,->] (RPrime) to node [swap] {$\cxtRaw{(\residual{t}{t'})}{\piInput{x+1}}$} (SwapPPrime);
  \draw[-] (PPrime) to node {$\ren{\swapR}{}$} (SwapPPrime);
\end{tikzpicture}%
\qquad%
\begin{tikzpicture}[node distance=1.5cm, auto]
  \node (Q) [node distance=2cm] {$\Gamma \vdash Q$};
  \node (SPrime) [below of=Q, right of=Q] {$\suc{\Gamma} \vdash S'$};
  \node (S) [right of=Q, above of=Q] {$\suc{\Gamma} \vdash S$};
  \node (QPrime) [node distance=3.5cm, right of=S] {$\suc{\Gamma} \vdash Q'$};
  \node (QDoublePrime) [node distance=3.5cm, right of=SPrime] {$\suc{\Gamma} \vdash Q'$};
  \draw[->] (Q) to node [yshift=-1ex] {$\cxtRaw{u}{\piBoundOutput{x}}$} (S);
  \draw[->] (Q) to node [yshift=1ex,swap] {$\cxtRaw{u'}{\piBoundOutput{z}}$} (SPrime);
  \draw[dotted,->] (S) to node {$\cxtRaw{(\residual{u'}{u})}{\piOutput{\suc{z}}{0}}$} (QPrime);
  \draw[dotted,->] (SPrime) to node [swap] {$\cxtRaw{(\residual{u}{u'})}{\piOutput{x+1}{0}}$} (QDoublePrime);
  \draw[-,double distance=1pt] (QPrime) to node {} (QDoublePrime);
\end{tikzpicture}
}
\end{center}

\noindent \begin{changebar}(Using named syntax, the term $Q$ might be of the
form $\piRestrictN{y}{\compl{x}\langle{y}\rangle.Q_1 \mid \compl{z}\langle y
  \rangle.Q_2}$, as per \exref{concurrent-extrusions:same-binder}
  above.)\end{changebar}

The residuals $\residual{u'}{u}$ and $\residual{u}{u'}$ are
plain outputs, rather than bound outputs. While the composites
$\piParLNu{t}{u} \concur \piParLNu{t'}{u'}$ are concurrent
$\nu$-synchronisations as before, the residuals of the composites are
plain synchronisations, again propagated through the $\nu$-binder
reinserted by the preceding step.

\vspace{5pt}
\begin{center}
\scalebox{0.8}{
\begin{tikzpicture}[node distance=1.5cm, auto]
  \node (PQ) [node distance=2cm] {$\Gamma \vdash \piPar{P}{Q}$};
  \node (NuRPrimeSPrime) [below of=PQ, right of=PQ] {$\Gamma \vdash \piRestrict{(\piPar{R'}{S'})}$};
  \node (RS) [right of=PQ, above of=PQ] {$\Gamma \vdash \piRestrict{(\piPar{R}{S})}$};
  \node (NuSwapPPrimeQPrime) [node distance=6cm, right of=RS] {
    $\Gamma \vdash \piRestrict{(\piPar{\ren{(\pop{}{0})}{P'}}{Q'})}$
  };
  \node (NuPPrimeQPrime) [node distance=6cm, right of=NuRPrimeSPrime] {
    $\Gamma \vdash \piRestrict{(\piPar{\ren{(\pop{}{0})}{\ren{\swapR}{P'}}}{Q'})}$
  };
  \draw[->] (PQ) to node [yshift=-1.5ex] {$\piParLNu{t}{u}$} (RS);
  \draw[->] (PQ) to node [yshift=1ex,swap] {$\piParLNu{t'}{u'}$} (NuRPrimeSPrime);
  \draw[dotted,->] (RS) to node {
    $\piRestrictA{\piTau}{(\piParLTau{\residual{t'}{t}}{\residual{u'}{u}}{0})}$
  } (NuSwapPPrimeQPrime);
  \draw[dotted,->] (NuRPrimeSPrime) to node [swap] {
    $\piRestrictA{\piTau}{(\piParLTau{\residual{t}{t'}}{\residual{u}{u'}}{0})}$
  } (NuPPrimeQPrime);
  \draw[-,double distance=1pt] (NuSwapPPrimeQPrime) to node {$\piRestrict{(\piPar{\ren{\alpha}{\EqRefl{P'}}}{\EqRefl{Q'}})}$} (NuPPrimeQPrime);
\end{tikzpicture}
}
\end{center}

Since the residual actions are plain $\tau$ actions, cofinality is
simply equality. And indeed the substitution $\pop{}{0}$ erases the free
braid relating $P'$ and $\ren{\swapR}{P'}$, by mapping indices $0$ and
$1$ both to $0$. Here $\alpha$ is the equality $(\pop{}{0}) \after
\swapR = \pop{}{0}$ (\lemref{pop-swap}) and
$\piRestrict{(\piPar{\ren{\alpha}{P'}}{\EqRefl{Q'}})}$ uses congruence to lift
$\alpha$ to an equivalence on target states, where $\EqRefl{P'}$ and
$\EqRefl{Q'}$ denote \cbstart the canonical reflexivity proofs of $P'$ and
$Q'$\cbend.
\exampleend
\end{example}

This completes our formal treatment of concurrent transitions in
\piCalculus, including the counterpart of the diamond lemma. In our
setting, transitions may open terms with respect to variables, leading
to a non-trivial notion of cofinality when such transitions are
reordered. Like \citeauthor*{boudol89}, we omit a formalisation of
\citeauthor{levy80}'s ``cube'' property, which extends the notion of
concurrency to dimensions greater than two, since it is not required for
the formalisation of causal equivalence.

\section{Causal equivalence}
\label{sec:causal-equivalence}

We now turn to formalising \emph{causal equivalence}, the congruence over
sequences of transitions, or \emph{traces}, induced by the concurrency
relation for transitions. This is a standard concept from the theory of
concurrent alphabets \cite{mazurkiewicz87}, but is non-trivial in our setting
because of braidings, which \cbstart (as we shall see below) \cbend both
propagate horizontally and compose vertically.

An ``atom'' of causal equivalence equates $t \cons \residual{t'}{t}$ and
$t' \cons \residual{t}{t'}$ for concurrent transitions $t \concur t'$,
where $t \cons u$ denotes the composition of $t$ and $u$. When the
associated pentagon is composed horizontally into a larger computation,
the continuation must be transported through the braiding
$\braiding{t,t'}$ which relates the target states of $\residual{t'}{t}$
and $\residual{t}{t'}$. This requires two dimensions of closure, as
illustrated in \figref{causal-equivalence:illustrate:continuation}. For
coinitial $u$ and $\braiding{t,t'}$, the transition $u$ must have an
image $\residual{u}{\braiding{t,t'}}$ in $\braiding{t,t'}$, and the
braiding $\braiding{t,t'}$ must propagate as
$\residual{\braiding{t,t'}}{u}$:

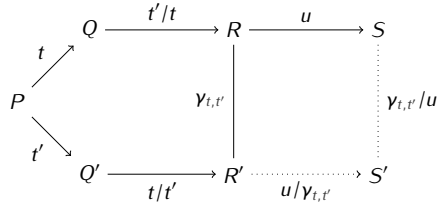
\begin{figure}[H]
\begin{center}
\scalebox{0.8}{
\begin{tikzpicture}[node distance=1.2cm, auto]
  \node (P) {$P$};
  \node (Q) [node distance=1.2cm, above of=P, right of=P] {$Q$};
  \node (Q') [node distance=1.2cm, below of=P, right of=P] {$Q'$};
  \node (R) [node distance=2.4cm, right of=Q] {$R$};
  \node (R') [node distance=2.4cm, right of=Q'] {$R'$};
  \node (S) [node distance=2.4cm, right of=R] {$S$};
  \node (S') [node distance=2.4cm, right of=R'] {$S'$};
  \draw[->] (P) to node {$t$} (Q);
  \draw[->] (P) to node [swap] {$t'$} (Q');
  \draw[->] (Q) to node {$\residual{t'}{t}$} (R);
  \draw[->] (Q') to node [swap] {$\residual{t}{t'}$} (R');
  \draw[->] (R) to node {$u$} (S);
  \draw[-] (R) to node [swap] {$\braiding{t,t'}$} (R');
  \draw[-,dotted] (S) to node {$\residual{\braiding{t,t'}}{u}$} (S');
  \draw[->,dotted] (R') to node [swap] {$\residual{u}{\braiding{t,t'}}$} (S');
\end{tikzpicture}}
\end{center}
\caption{Closure of transitions under braidings}
\label{fig:causal-equivalence:illustrate:continuation}
\end{figure}

\noindent \begin{changebar}The residual
$\residual{u}{\braiding{t,t'}}$ is a version of $u$ which takes into account
any braiding that arises from the concurrency of $t$ and $t'$, whereas
$\residual{\braiding{t,t'}}{u}$ represents the effect of the braiding on the
transition $u$.
\end{changebar}

For braidings to be preserved by transitions \begin{changebar}and vice-versa
requires two generalisations to the notion of braiding (\defref{cofin}). For
free braids, we need the renaming to be of the form $\plus{\swapR}{\Delta}$
rather than $\swapR$, so that braids can be preserved by subsequent bound
actions which further open up the process term. For bound braids, the effect
of doing more computation is that the unique pair of binders picked out by a
bound braid (\defref{bound-braid}) may end up being
\emph{dropped} (if it occurs on the discarded side of a choice) or
\emph{duplicated} (if it occurs under a replication). This requires a more
general notion of bound braid closed under reflexivity and parallel
composition.\end{changebar}

An additional requirement is that braidings compose vertically when
causal equivalences are composed via transitivity:

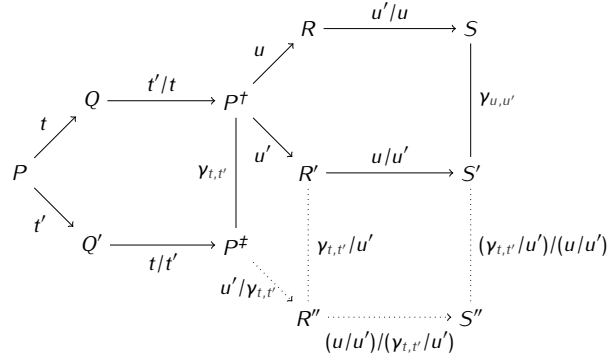
\begin{figure}[H]
\begin{center}
\scalebox{0.8}{
\begin{tikzpicture}[node distance=1.2cm, auto]
  \node (P) {$P$};
  \node (Q) [node distance=1.2cm, above of=P, right of=P] {$Q$};
  \node (Q') [node distance=1.2cm, below of=P, right of=P] {$Q'$};
  \node (PDag) [node distance=2.4cm, right of=Q] {$P^\dagger$};
  \node (PDDag) [node distance=2.4cm, right of=Q'] {$P^\ddagger$};
  \node (R) [node distance=1.2cm, above of=PDag, right of=PDag] {$R$};
  \node (R') [node distance=1.2cm, below of=PDag, right of=PDag] {$R'$};
  \node (S) [node distance=2.7cm, right of=R] {$S$};
  \node (S') [node distance=2.7cm, right of=R'] {$S'$};
  \node (R'') [node distance=1.2cm, below of=PDDag, right of=PDDag] {$R''$};
  \node (S'') [node distance=2.7cm, right of=R''] {$S''$};
  \draw[->] (P) to node {$t$} (Q);
  \draw[->] (P) to node [swap] {$t'$} (Q');
  \draw[->] (Q) to node {$\residual{t'}{t}$} (PDag);
  \draw[->] (Q') to node [swap] {$\residual{t}{t'}$} (PDDag);
  \draw[->] (PDag) to node {$u$} (R);
  \draw[->] (PDag) to node [xshift=1mm,swap] {$u'$} (R');
  \draw[->] (R) to node {$\residual{u'}{u}$} (S);
  \draw[->] (R') to node {$\residual{u}{u'}$} (S');
  \draw[->,dotted] (PDDag) to node [xshift=2mm,yshift=2mm,swap] {$\residual{u'}{\braiding{t,t'}}$} (R'');
  \draw[-] (PDag) to node [swap] {$\braiding{t,t'}$} (PDDag);
  \draw[-] (S) to node {$\braiding{u,u'}$} (S');
  \draw[-,dotted] (S') to node {$\residual{(\residual{\braiding{t,t'}}{u'})}{(\residual{u}{u'})}$} (S'');
  \draw[-,dotted] (R') to node {$\residual{\braiding{t,t'}}{u'}$} (R'');
  \draw[->,dotted] (R'') to node [yshift=-1mm,swap] {$\residual{(\residual{u}{u'})}{(\residual{\braiding{t,t'}}{u'})}$} (S'');
\end{tikzpicture}}
\end{center}
\caption{Sequential composition of concurrent transitions}
\label{fig:causal-equivalence:illustrate:vertical}
\end{figure}

\noindent This diagram represents the causal equivalence
\[
t \cons \residual{t'}{t} \cons u \cons \residual{u'}{u}
\permEq
t' \cons \residual{t}{t'} \cons \residual{u'}{\braiding{t,t'}} \cons \residual{(\residual{u}{u'})}{(\residual{\braiding{t,t'}}{u'})}
\]

\noindent with the targets $S$ and $S''$ related by the composite
braiding $\braiding{u,u'} \cons \residual{\braiding{t,t'}}{\residual{(\residual{\braiding{t,t'}}{u'})}{(\residual{u}{u'})}}$.%
\begin{changebar}It is worth reiterating that while the complexity of tracking
free braids is unique to the de Bruijn setting, the implications of bound
braids are not, since they arise from transposed binders.\end{changebar}

We proceed by defining traces $\vec{t}$
(\secref{causal-equivalence:traces}), and then showing that, suitably
generalised, braidings $\braiding{}$ ``commute'' with coinitial traces
$\vec{t}$, giving rise to residuals $\residual{\vec{t}}{\braiding{}}$
and $\residual{\braiding{}}{\vec{t}}$
(\secref{causal-equivalence:residual:braiding}). These are used to
define causal equivalences $\alpha: \vec{t} \permEq \vec{u}$ and
composite braidings $\braiding{\alpha}$ relating $\target{t}$ and
$\target{u}$ (\secref{causal-equivalence:causal-equivalence}).

\subsection{Traces}
\label{sec:causal-equivalence:traces}

Define $\vec{a}: \Actions{\Gamma}$ (bold $\vec{a}$) to be a finite
sequence of composable actions starting at $\Gamma$, where $a$ and $a'$
are composable iff $a \in \Action{\Gamma}$ and $a' \in
\Action{(\plus{\Gamma}{\target{a}})}$. $\magnitude{\vec{a}}$ denotes the
sum of $\magnitude{a}$ for every $a$ in $\vec{a}$. The empty sequence
(nil) at $\Gamma$ is written $\sub{\nil}{\Gamma}$; extension to the left
(cons) is written $a \cons \vec{a}$. A \emph{trace} $\vec{t}: P
\transition{\vec{a}} R$ (bold $\vec{t}$) is a finite sequence of
composable transitions, where $t$ and $u$ are composable iff $\source{u}
= \target{t}$. The nil trace at $P$ is written $\sub{\nil}{P}$; cons of
$t: P \transition{a} R$ onto $\vec{t}: R \transition{\vec{a}} S$ is
written $t \cons \vec{t}: P \transition{a \cons \vec{a}} S$.

The renamings $\ren{\rho}{a}$ and $\ren{\rho}{t}$ of an action and a
transition extend to action sequences and traces respectively.

\begin{lemma}[Lifting of renamings to action sequences and traces]
\item Suppose $\rho: \Gamma \to \Delta$ and $\vec{t}: P
  \transition{\vec{a}} R$, where $\Gamma \vdash P$,
  $\plus{\Gamma}{\Gamma'} \vdash R$ and $\vec{a}: \Actions{\Gamma}$.
\vspace{1mm}
\begin{nscenter}
\scalebox{0.8}{
\begin{tikzpicture}[node distance=1.5cm, auto]
  \node (Gamma) {$\Gamma \vdash P$};
  \node (Delta) [below of=Gamma] {$\Delta \vdash \ren{\rho}{P}$};
  \node (Gamma') [node distance=4cm, right of=Gamma] {$\plus{\Gamma}{\Gamma'} \vdash R$};
  \node (Delta') [below of=Gamma'] {$\plus{\Delta}{\Gamma'} \vdash \ren{(\plus{\rho}{\Gamma'})}{R}$};
  \draw[->] (Gamma) to node {$\cxtRaw{\vec{t}}{\vec{a}}$} (Gamma');
  \draw[dotted,->] (Delta) to node [swap] {$\cxtRaw{(\ren{\rho}{\vec{t}})}{\ren{\rho}{\vec{a}}}$} (Delta');
  \draw[->] (Gamma) to node [swap] {$\ren{\rho}{}$} (Delta);
  \draw[->] (Gamma') to node {$\ren{(\plus{\rho}{\Gamma'})}{}$} (Delta');
\end{tikzpicture}
}%
\end{nscenter}

\noindent Then there exist actions $\ren{\rho}{\vec{a}}:
\Actions{\Delta}$ and trace $\ren{\rho}{\vec{t}}: \ren{\rho}{P}
\transition{\ren{\rho}{\vec{a}}}
\ren{(\plus{\rho}{\Gamma'})}{R}$.
\end{lemma}

\begin{proof}
By the following defining equations.
\vspace{-14pt} 
\begin{nscenter}
\begin{minipage}[t]{0.29\linewidth}
\begin{align*}
\ren{\rho}{\sub{\nil}{\Gamma}}
&=
\sub{\nil}{\Delta}
\\[1mm]
\\
\ren{\rho}{\sub{\nil}{P}}
&=
\sub{\nil}{P}
\end{align*}
\end{minipage}%
\begin{minipage}[t]{0.29\linewidth}
\begin{align*}
\ren{\rho}{(b \cons \vec{a})}
&=
(\ren{\rho}{b}) \cons \ren{(\plus{\rho}{1})}{\vec{a}}
\\
\ren{\rho}{(c \cons \vec{a})}
&=
(\ren{\rho}{c}) \cons \ren{\rho}{\vec{a}}
\\[1mm]
\ren{\rho}{(\cxtRaw{t}{b} \cons \vec{t})}
&=
(\ren{\rho}{\cxtRaw{t}{b}}) \cons \ren{(\plus{\rho}{1})}{\vec{t}}
\\
\ren{\rho}{(\cxtRaw{t}{c} \cons \vec{t})}
&=
(\ren{\rho}{\cxtRaw{t}{c}}) \cons \ren{\rho}{\vec{t}}
\end{align*}
\end{minipage}
\end{nscenter}

\end{proof}

\subsection{Residuals of traces and braidings}
\label{sec:causal-equivalence:residual:braiding}

We now \begin{changebar} develop\end{changebar} a minimal generalisation of
our system of transitions and braidings sufficient to admit the following
notions of residuation:

\vspace{1mm}
\begin{nscenter}
\scalebox{0.8}{
\begin{tikzpicture}[node distance=1.5cm, auto]
  \node (P) {$\Gamma \vdash P$};
  \node (PPrime) [below of=P] {$\Gamma \vdash P'$};
  \node (R) [node distance=3cm, right of=P] {$\plus{\Gamma}{\Delta} \vdash R$};
  \node (RPrime) [below of=R] {$\plus{\Gamma}{\Delta} \vdash R'$};
  \draw[->] (P) to node {$t$} (R);
  \draw[dotted,->] (PPrime) to node [swap] {$\residual{t}{\gamma}$} (RPrime);
  \draw[-] (P) to node [swap] {$\gamma$} (PPrime);
  \draw[dotted,-] (R) to node {$\residual{\gamma}{t}$} (RPrime);
\end{tikzpicture}
}
\end{nscenter}

\noindent \begin{changebar}so that we can accommodate the scenario illustrated
earlier in
\figref{causal-equivalence:illustrate:continuation}. Here $\Delta \in
\set{0,1}$ and $\gamma$ is a braiding witnessing the
cofinality of the target states of an earlier concurrent transition. Recall
from
\defref{cofin} that $\gamma$ relates $P$ and $P'$ either by $\freeBraid$ (free
braid), $\boundBraid$ (bound braid) or $=$ (cofinality ``on the nose''); we
consider each case and explain how cofinality must be extended to support
$\residual{\gamma}{t}$. The final definitions of the two residuals are given
as the proof of
\lemref{residual:transition-braiding} below.\end{changebar}

\Paragraph{Case $P \freeBraid P'$.}
Then $P = \ren{\swapR}{P'}$ and $R = \ren{(\plus{\swapR}{\Delta})}{R'}$ by
\lemref{concurrent-transitions:renaming:transition}. If $\Delta = 1$ then the
free braid has shifted under a binder and thus $R \not\freeBraid R'$.
Therefore the first generalisation closes free braids under translations by an
arbitrary $\Delta$\begin{changebar}, allowing them to be preserved by
subsequent computation involving bound actions which open up the process
term.\end{changebar} We define the following relation, noting that $\freeBraid
= \FreeBraid{0}$.

\begin{definition}[Free braid, generalised]
\label{def:free-braid-shifted}
For any processes $\Gamma + 2 + \Delta \vdash P, R$ define the symmetric
relation $P \FreeBraid{\Delta} R$ as follows. The context $\Gamma$ is
left implicit.
\[P \FreeBraid{\Delta} R \quad\Leftrightarrow\quad P = \ren{(\plus{\swap{\Gamma}}{\Delta})}{R}\]
\end{definition}

\Paragraph{Case $P \boundBraid P'$.}
Whereas a free braid inserts a $\swapR$ renaming at the root of $P$, a bound
braid inserts a $\swapR$ under exactly one pair of adjacent binders in
$P$\begin{changebar}, and thus points to a specific location common to $P$ and
$P'$\end{changebar}. When a transition $t: P \transition{a} R$ is taken,
subterms of $P$ may be dropped or duplicated: in particular non-taken branches
of choices are discarded, and the bodies of replications are copied into both
sides of the resulting parallel compositions. It may therefore not be possible
to obtain $R'$ from $R$ by inserting exactly one bound $\swapR$,
\begin{changebar}since the braid might have been duplicated or thrown
away. The second generalisation thus closes bound braids under reflexivity (to
permit dropping) and parallel composition (to permit
duplication)\end{changebar}.
\figref{bound-braid-unrestricted} defines the new relation, also written
$\BoundBraid$.

\begin{figure}[h]
\noindent \shadebox{$P \BoundBraid R$}
\begin{smathpar}
\begin{changebar}
\inferrule*[left={\ruleName{$\congRestrictSwap{P}$}}]
{
   P \freeBraid R
}
{
   \piRestrict{\piRestrict{P}}
   \BoundBraid
   \piRestrict{\piRestrict{R}}
}
\end{changebar}
\and
\inferrule*[left={\ruleName{$\piZero$}}]
{
}
{
   \piZero
   \BoundBraid
   \piZero
}
\and
\inferrule*[left={\ruleName{$\piAction{\piInput{x}}{P}$}}]
{
}
{
   \piAction{\piInput{x}}{P}
   \BoundBraid
   \piAction{\piInput{x}}{P}
}
\and
\inferrule*[left={\ruleName{$\piAction{\piOutput{x}{y}}{P}$}}]
{
}
{
   \piAction{\piOutput{x}{y}}{P}
   \BoundBraid
   \piAction{\piOutput{x}{y}}{P}
}
\and
\inferrule*[left={\ruleName{$\piChoice{\param}{Q}$}}]
{
   P \BoundBraid R
}
{
   \piChoice{P}{Q}
   \BoundBraid
   \piChoice{R}{Q}
}
\and
\inferrule*[left={\ruleName{$\piChoice{P}{\param}$}}]
{
   Q \BoundBraid S
}
{
   \piChoice{P}{Q}
   \BoundBraid
   \piChoice{P}{S}
}
\and
\inferrule*[left={\ruleName{$\piPar{\param}{\param}$}}]
{
   P \BoundBraid R
   \\
   Q \BoundBraid S
}
{
   \piPar{P}{Q}
   \BoundBraid
   \piPar{R}{S}
}
\and
\inferrule*[left={\ruleName{$\piRestrict{\param}$}}]
{
   P \BoundBraid R
}
{
   \piRestrict{P}
   \BoundBraid
   \piRestrict{R}
}
\and
\inferrule*[left={\ruleName{$\piReplicate{\param}$}}]
{
   P \BoundBraid R
}
{
   \piReplicate{P}
   \BoundBraid
   \piReplicate{R}
}
\end{smathpar}
\crossrule
\caption{Bound braid $P \BoundBraid R$ that can be dropped or duplicated}
\label{fig:bound-braid-unrestricted}
\end{figure}

\Paragraph{Case $P = P'$.}
The situation is trivial\begin{changebar}, since $\residual{t}{\gamma}$ is
just $t$ and so\end{changebar} $\residual{\gamma}{t}$ is simply the
reflexivity proof that $R = R'$.

The three cases above determine a new braiding relation
$\cofin{\twoDim{a},\Delta}$ which is closed under transitions.

\begin{definition}[Braiding, generalised]
\label{def:cofin:generalised}
For any contexts $\Gamma, \Delta$, any $a, a' \in \Action{\Gamma}$ and
any $\twoDim{a}: a \concur a'$, define the following symmetric relation
$\cofin{\twoDim{a},\Delta}$ over processes in $\Gamma' + \Delta$, where
$\Gamma'$ is the target context of $\twoDim{a}$. There are only two
cases rather than three, since the $=$ case is now subsumed by the
reflexivity of bound braids.

\vspace{-10pt}
{\small
\begin{align*}
\cofin{\twoDim{a},\Delta}
& \eqdef
\begin{cases}
\FreeBraid{\Delta}
&
\text{if }{\cofin{\twoDim{a}}} = {\freeBraid}
\\
\BoundBraid
&
\text{otherwise}
\end{cases}
\end{align*}%
}%

\end{definition}

Since ${\freeBraid} = {\FreeBraid{0}}$, and there is an obvious
embedding, via reflexivity, of the old definition of $\BoundBraid$
(\figref{bound-braid}) into the new one, there is also an embedding of
$\cofin{\twoDim{a}}$ into $\cofin{\twoDim{a},0}$.

\begin{lemma}
${\cofin{\twoDim{a}}} \subseteq {\cofin{\twoDim{a},0}}$
\end{lemma}

The new braidings are \begin{changebar}sufficiently general to be closed under
transitions, so we can go ahead and define\end{changebar} the required
residuals $\residual{\gamma}{t}$ and $\residual{t}{\gamma}$. We start with the
case when $\gamma$ is a bound braid $\phi$. Note that subsuming the $=$ case
into the reflexivity of $\BoundBraid$ does not lose any precision, since for
any $t: P \transition{a} R$ we have $\residual{t}{\BoundBraidRefl{P}} = t$ and
thus $\residual{\BoundBraidRefl{P}}{t} = \BoundBraidRefl{R}$.

\begin{theorem}
\label{thm:causal-equivalence:residual:braiding-congruence}
Suppose $t: P \transition{a} R$ and $\phi: P \BoundBraid P'$. Then there
exists a process $R'$, transition $\residual{t}{\phi}: P' \transition{a}
R'$ and bound braid $\residual{\phi}{t}: R \BoundBraid R'$.

\vspace{2mm}
\begin{nscenter}
\scalebox{0.8}{
\begin{tikzpicture}[node distance=1.5cm, auto]
  \node (P) {$P$};
  \node (PPrime) [below of=P] {$P'$};
  \node (R) [node distance=3cm, right of=P] {$R$};
  \node (RPrime) [below of=R] {$R'$};
  \draw[->] (P) to node {$t$} (R);
  \draw[dotted,->] (PPrime) to node [swap] {$\residual{t}{\phi}$} (RPrime);
  \draw[-] (P) to node [swap] {$\phi$} (PPrime);
  \draw[dotted,-] (R) to node {$\residual{\phi}{t}$} (RPrime);
\end{tikzpicture}
}
\end{nscenter}

\end{theorem}

\noindent \emph{Proof.} By the defining equations in
\figref{residual:congruence}. Unlike residuals of the form
$\residual{t}{t'}$, the cofinality of $\residual{t}{\phi}$ and
$\residual{\phi}{t}$ is by construction. $\BoundBraidRefl{P}$ denotes
the reflexivity proof that $P \BoundBraid P$.

\begin{figure}[h]
\begin{nscenter}
\begin{minipage}[t]{0.49\linewidth}
\noindent \shadebox{$\residual{t}{\phi}$}
{\small
\begin{align*}
\smash{\residual{(\piRestrictOutput{\piRestrictA{\piOutput{\suc{x}}{0}}{t}})}{\congRestrictSwap{\source{t}}}}
&=
\smash{\piRestrictA{\piBoundOutput{x}}{\piRestrictOutput{(\ren{\swapR}{t})}}}
\\
\smash{\residual{(\piRestrictA{\piBoundOutput{x}}{\piRestrictOutput{t}})}{\congRestrictSwap{\source{t}}}}
&=
\smash{\piRestrictOutput{\piRestrictA{\piOutput{x+1}{0}}{(\ren{\swapR}{t})}}}
\\
\smash{\residual{(\piRestrictA{c}{\piRestrictA{c'}{t}})}{\congRestrictSwap{\source{t}}}}
&=
\smash{\piRestrictA{c}{\piRestrictA{c'}{(\ren{\swapR}{t}})}}
\\
\smash{\residual{(\piRestrictA{b}{\piRestrictA{b'}{t}})}{\congRestrictSwap{\source{t}}}}
&=
\smash{\piRestrictA{b}{\piRestrictA{b'}{(\ren{\swapR}{t}})}}
\\
\residual{(\piAction{\piInput{x}}{P})}{(\piAction{\piInput{x}}{\phi})}
&=
\piAction{\piInput{x}}{\target{\phi}}
\\
\residual{(\piAction{\piOutput{x}{y}}{P})}{(\piAction{\piOutput{x}{y}}{\phi})}
&=
\piAction{\piOutput{x}{y}}{\target{\phi}}
\\
\residual{(\piChoice{t}{Q})}{(\piChoice{\phi}{Q})}
&=
\piChoice{\residual{t}{\phi}}{Q}
\\
\residual{(\piChoice{t}{Q})}{(\piChoice{P}{\psi})}
&=
\piChoice{t}{\target{\psi}}
\\
\residual{(\piChoice{P}{u})}{(\piChoice{P}{\psi})}
&=
\piChoice{P}{\residual{u}{\psi}}
\\
\residual{(\piChoice{P}{u})}{(\piChoice{\phi}{Q})}
&=
\piChoice{\target{\phi}}{u}
\\
\residual{(\piParL{b}{t}{Q})}{(\piPar{\phi}{\psi})}
&=
\piParL{b}{\residual{t}{\phi}}{\target \psi}
\\
\residual{(\piParL{c}{t}{Q})}{(\piPar{\phi}{\psi})}
&=
\piParL{c}{\residual{t}{\phi}}{\target \psi}
\\
\residual{(\piParR{b}{P}{u})}{(\piPar{\phi}{\psi})}
&=
\piParR{b}{\target \phi}{\residual{u}{\psi}}
\\
\residual{(\piParR{c}{P}{u})}{(\piPar{\phi}{\psi})}
&=
\piParR{c}{\target \phi}{\residual{u}{\psi}}
\\
\residual{(\piParLTau{t}{u}{y})}{(\piPar{\phi}{\psi})}
&=
\piParLTau{\residual{t}{\phi}}{\residual{u}{\psi}}{y}
\\
\residual{(\piParRTau{t}{u}{y})}{(\piPar{\phi}{\psi})}
&=
\piParRTau{\residual{t}{\phi}}{\residual{u}{\psi}}{y}
\\
\residual{(\piParLNu{t}{u})}{(\piPar{\phi}{\psi})}
&=
\piParLNu{\residual{t}{\phi}}{\residual{u}{\psi}}
\\
\residual{(\piParRNu{t}{u})}{(\piPar{\phi}{\psi})}
&=
\piParRNu{\residual{t}{\phi}}{\residual{u}{\psi}}
\\
\residual{(\piRestrictOutput{t})}{(\piRestrict{\phi})}
&=
\piRestrictOutput{\;\residual{t}{\phi}}
\\
\residual{(\piRestrictA{b}{t})}{(\piRestrict{\phi})}
&=
\piRestrictA{b}{\residual{t}{\phi}}
\\
\residual{(\piRestrictA{c}{t})}{(\piRestrict{\phi})}
&=
\piRestrictA{c}{\residual{t}{\phi}}
\\
\residual{(\piReplicate{t})}{(\piReplicate{\phi})}
&=
\piReplicate{\residual{t}{(\piPar{\phi}{\piReplicate{\phi}})}}
\end{align*}}
\end{minipage}%
\begin{minipage}[t]{0.49\linewidth}
\noindent \shadebox{$\residual{\phi}{t}$}
{\small
\begin{align*}
\smash{\residual{\congRestrictSwap{\source{t}}}{(\piRestrictOutput{\piRestrictA{\piOutput{\suc{x}}{0}}{t}})}}
&=
\smash{\piRestrict{\;\BoundBraidRefl{\target{t}}}}
\\
\smash{\residual{\congRestrictSwap{\source{t}}}{(\piRestrictA{\piBoundOutput{x}}{\piRestrictOutput{t}})}}
&=
\smash{\piRestrict{\;\BoundBraidRefl{(\ren{\swapR}{\target{t}})}}}
\\
\smash{\residual{\congRestrictSwap{\source{t}}}{(\piRestrictA{c}{\piRestrictA{c'}{t}})}}
&=
\smash{\congRestrictSwap{\target{t}}}
\\
\smash{\residual{\congRestrictSwap{\source{t}}}{(\piRestrictA{b}{\piRestrictA{b'}{t}})}}
&=
\smash{\congRestrictSwap{\ren{\swapR}{\ren{(\suc{\swapR})}{\ren{\swapR}{\target{t}}}}}}
\\
\smash{\residual{(\piAction{\piInput{x}}{\phi})}{(\piAction{\piInput{x}}{P})}}
&=
\phi
\\
\residual{(\piAction{\piOutput{x}{y}}{\phi})}{(\piAction{\piOutput{x}{y}}{P})}
&=
\phi
\\
\residual{(\piChoice{\phi}{Q})}{(\piChoice{t}{Q})}
&=
\residual{\phi}{t}
\\
\residual{(\piChoice{P}{\psi})}{(\piChoice{t}{Q})}
&=
\smash{\BoundBraidRefl{\target{t}}}
\\
\residual{(\piChoice{P}{\psi})}{(\piChoice{P}{u})}
&=
\residual{\psi}{u}
\\
\residual{(\piChoice{\phi}{Q})}{(\piChoice{P}{u})}
&=
\smash{\BoundBraidRefl{\target{u}}}
\\
\residual{(\piPar{\phi}{\psi})}{(\piParL{b}{t}{Q})}
&=
\piPar{\residual{\phi}{t}}{\ren{\push{}}{\psi}}
\\
\residual{(\piPar{\phi}{\psi})}{(\piParL{c}{t}{Q})}
&=
\piPar{\residual{\phi}{t}}{\psi}
\\
\residual{(\piPar{\phi}{\psi})}{(\piParR{b}{P}{u})}
&=
\piPar{\ren{\push{}}{\phi}}{\residual{\psi}{u}}
\\
\residual{(\piPar{\phi}{\psi})}{(\piParR{c}{P}{u})}
&=
\piPar{\phi}{\residual{\psi}{u}}
\\
\residual{(\piPar{\phi}{\psi})}{(\piParLTau{t}{u}{y})}
&=
\piPar{\ren{(\pop{}{y})}{\residual{\phi}{t}}}{\residual{\psi}{u}}
\\
\residual{(\piPar{\phi}{\psi})}{(\piParRTau{t}{u}{y})}
&=
\piPar{\residual{\phi}{t}}{\ren{(\pop{}{y})}{\residual{\psi}{u}}}
\\
\residual{(\piPar{\phi}{\psi})}{(\piParLNu{t}{u})}
&=
\piRestrict{(\piPar{\residual{\phi}{t}}{\residual{\psi}{u}})}
\\
\residual{(\piPar{\phi}{\psi})}{(\piParRNu{t}{u})}
&=
\piRestrict{(\piPar{\residual{\phi}{t}}{\residual{\psi}{u}})}
\\
\residual{(\piRestrict{\phi})}{(\piRestrictOutput{t})}
&=
\residual{\phi}{t}
\\
\residual{(\piRestrict{\phi})}{(\piRestrictA{b}{t})}
&=
\piRestrict{\;\ren{\swapR{}}{\residual{\phi}{t}}}
\\
\residual{(\piRestrict{\phi})}{(\piRestrictA{c}{t})}
&=
\piRestrict{\;\residual{\phi}{t}}
\\
\residual{(\piReplicate{\phi})}{(\piReplicate{t})}
&=
\residual{(\piPar{\phi}{\piReplicate{\phi}})}{t}
\end{align*}}
\end{minipage}
\end{nscenter}
\crossrule
\caption{Residuals of transition $t$ and coinitial bound braid $\phi$}
\label{fig:residual:congruence}
\end{figure}

\figref{residual:congruence:nu-nu-swap-cases} illustrates
\thmref{causal-equivalence:residual:braiding-congruence} for the cases
where $\phi$ is of the form $\congRestrictSwap{P}$, omitting the various
renaming lemmas used as type-level coercions.

\begin{figure}[H]
\makebox[0.9\textwidth][c]{
\begin{minipage}[b]{.48\linewidth}
\scalebox{0.8}{
\begin{tikzpicture}[auto]
  \node (1) [node distance=2cm] {$\Gamma \vdash \piRestrict{\piRestrict{(\ren{\swapR}{P})}}$};
  \node (2) [node distance=2cm, below of=1] {$\Gamma \vdash \piRestrict{\piRestrict{P}}$};
  \node (3) [node distance=4.8cm, right of=1] {$\suc{\Gamma} \vdash \piRestrict{R}$};
  \node (4) [node distance=4.8cm, right of=2] {$\suc{\Gamma} \vdash \piRestrict{R}$};
  \draw[->] (1) to node {$\piRestrictOutput{\piRestrictA{\piOutput{\suc{x}}{0}}{t}}$} (3);
  \draw[-] (1) to node [swap] {$\congRestrictSwap{P}$} (2);
  \draw[-,double distance=1pt] (3) to node {} (4);
  \draw[dotted,->] (2) to node [swap] {$\piRestrictA{\piBoundOutput{x}}{\piRestrictOutput{(\ren{\swapR}{t})}}$} (4);
\end{tikzpicture}
}
\end{minipage}
\begin{minipage}[b]{.49\linewidth}
\scalebox{0.8}{
\begin{tikzpicture}[auto]
  \node (1) [node distance=2cm] {$\Gamma \vdash \piRestrict{\piRestrict{(\ren{\swapR}{P})}}$};
  \node (2) [node distance=2cm, below of=1] {$\Gamma \vdash \piRestrict{\piRestrict{P}}$};
  \node (3) [node distance=5.7cm, right of=1] {$\Gamma \vdash \piRestrict{\piRestrict{R}}$};
  \node (4) [node distance=5.7cm, right of=2] {$\Gamma \vdash \piRestrict{\piRestrict{(\ren{\swapR}{R})}}$};
  \draw[->] (1) to node {$\piRestrictA{c}{\piRestrictA{c'}{t}}$} (3);
  \draw[-] (1) to node [swap] {$\congRestrictSwap{P}$} (2);
  \draw[dotted,-] (3) to node {$\congRestrictSwap{R}$} (4);
  \draw[dotted,->] (2) to node [swap] {$\piRestrictA{c}{\piRestrictA{c'}{(\ren{\swapR}{t})}}$} (4);
\end{tikzpicture}
}
\end{minipage}
}
\\
\makebox[0.9\textwidth][c]{
\begin{minipage}[b]{.48\linewidth}
\scalebox{0.8}{
\begin{tikzpicture}[auto]
  \node (1) [node distance=2cm] {$\Gamma \vdash \piRestrict{\piRestrict{(\ren{\swapR}{P})}}$};
  \node (2) [node distance=2cm, below of=1] {$\Gamma \vdash \piRestrict{\piRestrict{P}}$};
  \node (3) [node distance=4.8cm, right of=1] {$\suc{\Gamma} \vdash \piRestrict{(\ren{\swapR}{R})}$};
  \node (4) [node distance=4.8cm, right of=2] {$\suc{\Gamma} \vdash \piRestrict{(\ren{\swapR}{R})}$};
  \draw[->] (1) to node {$\piRestrictA{\piBoundOutput{x}}{\piRestrictOutput{t}}$} (3);
  \draw[-] (1) to node [swap] {$\congRestrictSwap{P}$} (2);
  \draw[-,double distance=1pt] (3) to node {} (4);
  \draw[dotted,->] (2) to node [swap] {$\piRestrictOutput{\piRestrictA{\piOutput{x+1}{0}}{(\ren{\swapR}{t})}}$} (4);
\end{tikzpicture}
}
\end{minipage}
\begin{minipage}[b]{.49\linewidth}
\scalebox{0.8}{
\begin{tikzpicture}[auto]
  \node (1) [node distance=2cm] {$\Gamma \vdash \piRestrict{\piRestrict{(\ren{\swapR}{P})}}$};
  \node (2) [node distance=2cm, below of=1] {$\Gamma \vdash \piRestrict{\piRestrict{P}}$};
  \node (3) [node distance=5.7cm, right of=1] {$\suc{\Gamma} \vdash \piRestrict{\piRestrict{(\ren{(\suc{\swapR})}{\ren{\swapR}{R}})}}$};
  \node (4) [node distance=5.7cm, right of=2] {$\suc{\Gamma} \vdash \piRestrict{\piRestrict{(\ren{\swapR}{\ren{(\suc{\swapR})}{\ren{\swapR}{R}}})}}$};
  \draw[->] (1) to node {$\piRestrictA{b}{\piRestrictA{b'}{t}}$} (3);
  \draw[-] (1) to node [swap] {$\congRestrictSwap{P}$} (2);
  \draw[dotted,-] (3) to node {$\congRestrictSwap{{\ren{(\suc{\swapR})}{\ren{\swapR}{R}}}}$} (4);
  \draw[dotted,->] (2) to node [swap] {$\piRestrictA{b}{\piRestrictA{b'}{(\ren{\swapR}{t})}}$} (4);
\end{tikzpicture}
}
\end{minipage}
}
\makebox[\textwidth][c]{
  \crossrule
}
\caption{Cofinality of $\residual{\phi}{t}$ and $\residual{t}{\phi}$ in
  the $\congRestrictSwap{}$ cases}
\label{fig:residual:congruence:nu-nu-swap-cases}
\end{figure}

It is then straightforward to extend \begin{changebar}the bound braid cases
$\residual{t}{\phi}$ and $\residual{\phi}{t}$ to arbitrary braidings $\gamma$
and sequences of transitions $\vec{t}$\end{changebar}.

\begin{lemma}[Residuals of transition $t$ and $\gamma$]
\label{lem:residual:transition-braiding}
Suppose $t: P \transition{a} R$ and $\gamma : P
\cofin{\twoDim{a},\Delta} P'$. Then there exists process $R'$, action
$\residual{a}{\gamma}$, transition $\residual{t}{\gamma}: P'
\transition{\residual{a}{\gamma}} R'$ and braiding
$\residual{\gamma}{t}: R \cofin{\twoDim{a},\Delta'} R'$, where $\Delta'
= \plus{\Delta}{\magnitude{a}}$.
\vspace{1mm}
\begin{nscenter}
\scalebox{0.8}{
\begin{tikzpicture}[node distance=1.5cm, auto]
  \node (P) {$\plus{\Gamma}{\Delta} \vdash P$};
  \node (PPrime) [below of=P] {$\plus{\Gamma}{\Delta} \vdash P'$};
  \node (R) [node distance=3cm, right of=P] {$\plus{\Gamma}{\Delta'} \vdash R$};
  \node (RPrime) [below of=R] {$\plus{\Gamma}{\Delta'} \vdash R'$};
  \draw[->] (P) to node {$t$} (R);
  \draw[dotted,->] (PPrime) to node [swap] {$\residual{t}{\gamma}$} (RPrime);
  \draw[-] (P) to node [swap] {$\gamma$} (PPrime);
  \draw[dotted,-] (R) to node {$\residual{\gamma}{t}$} (RPrime);
\end{tikzpicture}
}
\end{nscenter}

\end{lemma}

\vspace{-2mm}
\noindent \emph{Proof.} By the following defining equations, which are
given for $\residual{t}{\gamma}$ and $\residual{\gamma}{t}$ simultaneously. As
\begin{changebar}before\end{changebar} $\EqRefl{P}$ denotes the reflexivity
proof that $P = P$.

\vspace{-5pt}
\begin{nscenter}
{\small
\begin{align*}
(\residual{t}{\gamma},\residual{\gamma}{t})
&=
\begin{cases}
  (\ren{(\plus{\swapR}{\Delta})}{t}, \EqRefl{(\ren{(\plus{\swapR}{\Delta'})}{R})})
  & \text{if } P \FreeBraid{\Delta} P'
  \\
  (\residual{t}{\phi}, \residual{\phi}{t})
  & \text{if } P \BoundBraid P'\text{ and }\gamma = \phi
\end{cases}
\end{align*}
}
\end{nscenter}

The diagram for \lemref{residual:trace-braiding} is the same as for
\lemref{residual:transition-braiding} but with $\vec{t}$ instead of $t$.

\begin{lemma}[Residuals of trace $\vec{t}$ and $\gamma$]
\label{lem:residual:trace-braiding}
\item
Suppose $\vec{t}: P \transition{\vec{a}} R$ and $\gamma: P
\cofin{\twoDim{a},\Delta} P'$. Then there exists process $R'$, action
sequence $\residual{\vec{a}}{\gamma}$, trace
$\residual{\vec{t}}{\gamma}: P' \transition{\residual{\vec{a}}{\gamma}}
R'$ and braiding $\residual{\gamma}{\vec{t}}: R
\cofin{\twoDim{a},\Delta'} R'$, where $\Delta' = \Delta +
\magnitude{\vec{a}}$.
\end{lemma}

\noindent \emph{Proof.} By the following defining equations.

\begin{nscenter}
\begin{minipage}[t]{0.35\linewidth}
\begin{nscenter}
\scalebox{0.8}{
\begin{tikzpicture}[node distance=1.5cm, auto]
  \node (P) {$P$};
  \node (PPrime) [below of=P] {$P'$};
  \node (R) [node distance=2cm, right of=P] {$P$};
  \node (RPrime) [below of=R] {$P'$};
  \draw[-,double distance=1pt] (P) to node {$\sub{\nil}{P}$} (R);
  \draw[-,double distance=1pt] (PPrime) to node [swap] {$\sub{\nil}{P'}$} (RPrime);
  \draw[-] (P) to node [swap] {$\gamma$} (PPrime);
  \draw[-] (R) to node {$\gamma$} (RPrime);
\end{tikzpicture}
}
\end{nscenter}
\end{minipage}%
\begin{minipage}[t]{0.35\linewidth}
\begin{nscenter}
\scalebox{0.8}{
\begin{tikzpicture}[node distance=1.5cm, auto]
  \node (P) {$P$};
  \node (PPrime) [below of=P] {$P'$};
  \node (R) [node distance=2.5cm, right of=P] {$R$};
  \node (RPrime) [below of=R] {$R'$};
  \node (S) [node distance=2.5cm, right of=R] {$S$};
  \node (SPrime) [below of=S] {$S'$};
  \draw[->] (P) to node {$t$} (R);
  \draw[->] (R) to node {$\vec{t}$} (S);
  \draw[dotted,->] (PPrime) to node [swap] {$\residual{t}{\gamma}$} (RPrime);
  \draw[dotted,->] (RPrime) to node [swap] {$\residual{\vec{t}}{(\residual{\gamma}{t})}$} (SPrime);
  \draw[-] (P) to node [swap] {$\gamma$} (PPrime);
  \draw[dotted,-] (R) to node {$\residual{\gamma}{t}$} (RPrime);
  \draw[dotted,-] (S) to node {$\residual{(\residual{\gamma}{t})}{\vec{t}}$} (SPrime);
\end{tikzpicture}
}
\end{nscenter}
\end{minipage}
\begin{minipage}[t]{0.35\linewidth}
{\small
\begin{align*}
\residual{\sub{\nil}{P}}{\gamma}
&=
\sub{\nil}{P'}
\\
\residual{\gamma}{\sub{\nil}{P}}
&=
\gamma
\end{align*}}
\end{minipage}%
\begin{minipage}[t]{0.35\linewidth}
{\small
\begin{align*}
\residual{(t \cons \vec{t})}{\gamma}
&=
\residual{t}{\gamma} \cons {\residual{\vec{t}}{(\residual{\gamma}{t})}}
\\
\residual{\gamma}{(t \cons \vec{t})}
&=
\residual{(\residual{\gamma}{t})}{\vec{t}}
\end{align*}}
\end{minipage}
\end{nscenter}

\subsection{Causal equivalence}
\label{sec:causal-equivalence:causal-equivalence}

A causal equivalence $\alpha: \vec{t} \permEq \vec{u}$ reorders a trace
$\vec{t}$ into an equal-length, coinitial trace $\vec{u}$ by permuting
concurrent transitions. Meta-variables $\alpha$, $\beta$ range over
causal equivalences. If $\alpha: \vec{t} \permEq \vec{u}$ then
$\target{t}$ and $\target{u}$ are related by a unique braiding
$\braiding{\alpha}$.

In what follows, rules which mention a trace of the form $t \cons
\vec{t}$ have an implicit side-condition asserting $\target{t} =
\source{\vec{t}}$, and rules which mention a braiding $\braiding{t,t'}$
have an implicit side-condition asserting $t \concur t'$.

\begin{figure}[H]
\begin{center}
\scalebox{0.8}{
\begin{tikzpicture}[node distance=1.5cm, auto]
  \node (P) {$P$};
  \node (R) [above of=P, right of=P] {$R$};
  \node (R') [below of=P, right of=P] {$R'$};
  \node (Q) [node distance=2.5cm, right of=R] {$Q$};
  \node (Q') [node distance=2.5cm, right of=R'] {$Q'$};
  \node (S) [node distance=2.5cm, right of=Q] {$S$};
  \node (S') [node distance=2.5cm, right of=Q'] {$S'$};
  \draw[->] (P) to node {$t$} (R);
  \draw[->] (P) to node [swap] {$t'$} (R');
  \draw[->] (R) to node {$\residual{t'}{t}$} (Q);
  \draw[->] (R') to node [swap] {$\residual{t}{t'}$} (Q');
  \draw[->] (Q) to node {$\vec{t}$} (S);
  \draw[->,dotted] (Q') to node [swap,xshift=1mm,yshift=1mm] {$\residual{\vec{t}}{\braiding{t,t'}}$} (S');
  \draw[-] (Q) to node [swap] {$\braiding{t,t'}$} (Q');
  \draw[-,dotted] (S) to node {$\residual{\braiding{t,t'}}{\vec{t}}$} (S');
\end{tikzpicture}}
\end{center}
\caption{Causal equivalence, transposition rule}
\label{fig:causal-equivalence:transpose-rule}
\end{figure}
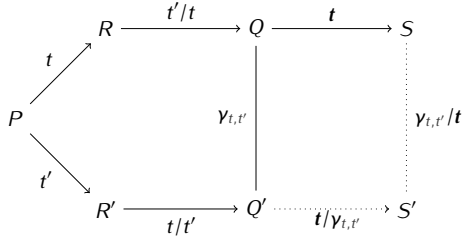

\begin{definition}
Inductively define the relation $\permEq$ using the rules in
\figref{causal-equivalence}, where syntactically $\permEq$ has lower
priority than $\cons$.

\begin{figure}[h]
\noindent \shadebox{$\vec{t} \permEq \vec{u}$}
\begin{smathpar}
\inferrule*[left={\ruleName{$\permEqNil{P}$}}]
{
  \strut
}
{
  \sub{\nil}{P} \permEq \sub{\nil}{P}
}
\and
\inferrule*[left={\ruleName{$\permEqCons{t}{\param}$}}]
{
  \vec{t} \permEq \vec{u}
}
{
  t \cons \vec{t} \permEq t \cons \vec{u}
}
\\
\inferrule*[
  left={\ruleName{$\permEqSwap{t}{t'}{\vec{t}}$}}
]
{
  \strut
}
{
  t \cons \residual{t'}{t} \cons \vec{t}
  \permEq
  t' \cons \residual{t}{t'} \cons \residual{\vec{t}}{\braiding{t,t'}}
}
\and
\inferrule*[left={\ruleName{$\permEqTrans{\param}{\param}$}}]
{
  \vec{t}' \permEq \vec{u}
  \\
  \vec{t} \permEq \vec{t}'
}
{
  \vec{t} \permEq \vec{u}
}
\end{smathpar}
\crossrule
\caption{Causal equivalence}
\label{fig:causal-equivalence}
\end{figure}

\end{definition}
The $\sub{\nil}{P}$ and $t \cons \param$ rules are the congruence cases.
The $\permEqTrans{\param}{\param}$ rule closes under transitivity, which
is a form of vertical composition and which also causes braidings to
compose vertically. The transposition rule $\permEqSwap{t}{t'}{\vec{t}}$
composes a concurrent pair $t \concur t'$ with a continuation $\vec{t}$
for $\residual{t'}{t}$, transporting $\vec{t}$ through the braiding
$\braiding{t,t'}$ witnessing the cofinality of $t$ and $t'$ to obtain
the continuation $\residual{\vec{t}}{\braiding{t,t'}}$ for
$\residual{t}{t'}$, as shown in
\figref{causal-equivalence:transpose-rule}.

\begin{theorem}
$\permEq$ is an equivalence relation.
\end{theorem}

\begin{proof}
Reflexivity is a trivial induction, using the $\permEqNil{P}$ and
$\permEqCons{t}{\alpha}$ rules. Transitivity is immediate from the
\cbstart $\permEqTrans{\param}{\param}$ \cbend rule. Symmetry is trivial in the
$\sub{\nil}{P}$, $t \cons \alpha$ and $\permEqTrans{\alpha}{\beta}$
cases. The $\permEqSwap{t}{t'}{\vec{t}}$ case requires the symmetry of
$\concur$ and that
$\residual{(\residual{\vec{t}}{\braiding{}})}{\braiding{}} = \vec{t}$.
\end{proof}

A causal equivalence $\alpha: \vec{t} \permEq \vec{u}$ determines a
composite braiding relation $\cofin{\alpha}$ which precisely sequences
the atomic braidings required to relate $\target{\vec{t}}$ to
$\target{\vec{u}}$.

\begin{definition}[Braiding for equivalent traces]
Inductively define the family \cbstart $\cofin{\alpha}$ of relations between processes\cbend, for any $a:
\vec{t} \permEq \vec{u}$, using the rules in
\figref{braiding:causal-equivalence}.
\end{definition}

\begin{figure}[H]
\noindent \shadebox{$P \cofin{\alpha} R$}
\begin{smathpar}
\inferrule*
{
  \strut
}
{
  P \cofin{\sub{\nil}{P}} P
}
\and
\inferrule*[right={$\residual{\braiding{t,t'}}{\vec{t}}: P \cofin{\twoDim{a},\Delta} R$}]
{
  \strut
}
{
  P \cofin{\permEqSwap{t}{t'}{\vec{t}}} R
}
\and
\inferrule*
{
  P \cofin{\alpha} R
}
{
  P \cofin{\permEqCons{t}{\alpha}} R
}
\and
\inferrule*
{
  P \cofin{\beta} R
  \\
  R \cofin{\alpha} S
}
{
  P \cofin{\permEqTrans{\alpha}{\beta}} S
}
\end{smathpar}
\crossrule
\caption{Braiding relation $\cofin{\alpha}$ relating $\target{t}$ and
  $\target{u}$ for any $\alpha: \vec{t} \permEq \vec{u}$}
\label{fig:braiding:causal-equivalence}
\end{figure}

\noindent As with $\cofin{\twoDim{a},\Delta}$, the relation
$\cofin{\alpha}$ is a singleton, inhabited by a unique path
$\braiding{\alpha}$ between $\target{\vec{t}}$ and $\target{\vec{u}}$.
(However $\alpha$ itself is not unique, since there are many ways of
proving $\vec{t} \permEq \vec{u}$.) The $P \cofin{\permEqNil{P}} P$ case
is an empty composite braiding. The $P
\cofin{\permEqSwap{t}{t'}{\vec{t}}} R$ case turns an atomic braiding
$\braiding{t,t'}$ into one step of a composite braiding, after
transporting it through the continuation $\vec{t}$. The $P
\cofin{\permEqCons{t}{\alpha}} R$ case simply recognises that $\target{t
  \cons \vec{t}} = \target{\vec{t}}$. Finally $P
\cofin{\permEqTrans{\alpha}{\beta}} R$ is the composition rule, closing
under transitivity.

\begin{theorem}
Suppose $\alpha: \vec{t} \permEq \vec{u}$. Then there exists a unique
$\braiding{\alpha}: \target{\vec{t}} \cofin{\alpha} \target{\vec{u}}$.
\end{theorem}

\begin{theorem}
$\cofin{\alpha}$ is a $\permEq$-indexed family of equivalence relations.
\end{theorem}

\section{Related work}
\label{sec:related-work}

The $\mu s$ calculus \cite{hirschkoff99} has a similar treatment of de
Bruijn indices. Its renaming operators $\langle x \rangle$, $\phi$ and
$\psi$ are effectively our $\pop{}{x}$, $\push{}$ and $\swapR{}$
renamings, but fused with the $\ren{\param}{}$ operator which applies a
renaming to a process. Hirschkoff's operators are also syntactic forms
in the $\mu s$ calculus, rather than meta-operations, and therefore the
operational semantics also includes rules for reducing occurrences of
the renaming operators that arise during a process reduction step.

As noted earlier in the paper, our approach to defining causal
equivalence of traces is influenced by a line of work stemming from
the study of optimal reduction in the \lambdaCalculus~\cite{levy80},
via the ``proved transition'' semantics of CCS~\cite{boudol89}.

\citet*{boreale98} and \citet*{degano99} investigate causality in the
context of the \piCalculus. Similar ideas (from which we also drew
inspiration) appear in work on reversible CCS, such as
RCCS~\cite{danos04a}, and reversible \piCalculi, such as
$\rho\pi$~\cite{lanese10} and R$\pi$~\cite{cristescu13}). Reversible
calculi equip process terms with additional structure to support undoing
actions; causal equivalence and permutation of transitions is necessary
here to allow undoing actions in a different (sequential) order than
they were performed. However, this additional structure changes the
metatheory: for example, in R$\pi$ two traces are coinitial and cofinal
if and only if they are equivalent, which does not hold in our setting.
To the best of our knowledge, there is no prior work that presents a
proved transition semantics for a ``vanilla'' \piCalculus, rather than
an augmented variant.

\begin{changebar}Another related concept for concurrency calculi,
\emph{confluence}, has been studied for CCS~\cite{milner80} and for the
\piCalculus~\cite{philippou97}.  A process is confluent if none of its
possible actions interfere with each other. Intuitively, this should be the
case if the process has only one possible trace modulo causal equivalence.
However, to the best of our knowledge, confluence has not been studied using
the proved transitions approach and the formal relationship between confluence
and causal equivalence is unclear. Our formalisation provides a platform for
future study of this matter.\end{changebar}

\cbstart\subsection{Mechanised treatments}\cbend

Formalisations of the \piCalculus have been undertaken in several
theorem provers used for mechanised metatheory, including Coq, HOL,
Isabelle/HOL, Nominal Isabelle, CLF, Abella, and Agda.

\Paragraph{HOL}
\citet*{melham94} reports on a formalisation of the \piCalculus in HOL,
using names axiomatised as an unspecified, infinite set, and following
\citet*{milner92} closely. Substitution is parameterised over a choice
function specifying how to choose a name fresh for a given set of names,
which is used to rename bound names to avoid capture.
\citet*{aitmohamed95} formalised the \piCalculus in HOL using concrete
syntax and verified proof rules for early bisimulation checking.

\Paragraph{Coq}
An early mechanisation of residuation theory was \citeauthor{huet94}'s
formalisation in Coq of residuals for \lambdaCalculus \cite{huet94},
which also uses de Bruijn indices. Huet's chief contribution is an
inductive definition of residual, a proof that residuals commute with
substitution, and a ``prism'' theorem that generalises
\citeauthor{levy80}'s cube lemma.

\citet*{hirschkoff97} formalised the \piCalculus in Coq using de Bruijn
indices, and verified properties such as congruence and structural
equivalence laws of bisimulation. \citet*{despeyroux00} formalised the
\piCalculus in Coq using weak higher-order abstract syntax, assuming a
decidable type of names, and using two separate transitions, for
ordinary, input and output transitions respectively; for input and
output transitions the right-hand side is a function of type
$\typename{name} \to \typename{proc}$. This formalisation included a
simple type system and proof of type soundness. \citet*{honsell01}
formalised the \piCalculus in Coq, also using weak higher-order abstract
syntax. The type of names \typename{name} is a type parameter assumed to
admit decidable equality and freshness ($\typename{notin}$) relations.
Transitions are encoded using two inductive definitions, for free and
bound actions, which differ in the type of the third argument
($\typename{proc}$ vs.~$\typename{name} \to \typename{proc}$). Numerous
results from \citet*{milner92} are verified, using the \emph{theory of
  contexts} (whose axioms are assumed in their formalisation, but have
been validated semantically by~\citet*{bucalo06}).

\citet*{affeldt08} developed a library based on a variant of the
\piCalculus (with channels typed using Coq types) for representing and
reasoning about concurrent processes. Processes are represented using
higher-order abstract syntax, and exotic terms are allowed; some lemmas
are not formally proved but introduced as axioms with semantic
justifications.

\Paragraph{Isabelle/HOL}
\citet*{rockl01a} and \citet*{rockl03} formalised the \piCalculus in
Isabelle/HOL and verified properties such as adequacy, following the
theory of contexts approach to higher-order abstract syntax introduced
by \citet{honsell01}, and using well-formedness predicates to rule out
exotic terms. \citet*{gay01} developed a framework for formalising
(linear) type systems for the \piCalculus in Isabelle/HOL, using de
Bruijn indices for binding syntax and a reduction-style semantics rather
than labelled transitions.

\Paragraph{Abella}
\citet*{tiu10} encode the syntax and semantics of the \piCalculus using
the $\lambda$-term abstract syntax variant of higher-order abstract
syntax; like a number of other approaches they split the transition
relation into two relations to handle scope extrusion. Their
formalisations employ the meta-logic $\textsf{FOL}^{\Delta\nabla}$ which
forms the basis of the Abella theorem prover, and similar specifications
have been used as the basis for verification of properties of the
\piCalculus in Abella~\cite{baelde14}.

\citet{accattoli12} adapts \citeauthor{huet94}'s Coq formalisation of
residuals from de Bruijn indices to Abella's higher-order abstract
syntax and nominal quantifier $\nabla$, yielding a significant
simplification of Huet's proof. \citeauthor{accattoli12} also proves the
cube lemma directly, rather than introducing an intermediate prism
theorem.
\cbstart 
It may be that reformalizing our approach using Abella would make it possible
to simplify our proof in a similar way.
\cbend

\Paragraph{Nominal Isabelle}
The Nominal Datatype Package extension to Isabelle/HOL~\cite{urban08}
supports the
Gabbay-Pitts style ``nominal'' approach to abstract syntax modulo
name-binding~\cite*{gabbay02}, and has been used in several
formalisations. Two early contributions using similar ideas predate its
development: \citet*{rockl01b} formalised the syntax of \piCalculus and
$\alpha$-equivalence in Isabelle/HOL. \citet*{gabbay03} described how to
use Gabbay-Pitts nominal abstract syntax to represent the \piCalculus,
without giving a mechanised formalisation or proofs of properties.

\citet*{bengtson09b} report on an extensive formalisation in Nominal
Isabelle, including inversion principles up to structural congruence,
properties of strong and weak bisimulation, and a proof that an
axiomatisation of strong late bisimilarity is sound and complete. They
use a single inductively-defined transition relation, whose third
argument is a sum type allowing either an ordinary process or a residual
process with a distinguished bound name.

\Paragraph{CLF}
\citet*{cervesato02} formalise synchronous and asynchronous versions
of \piCalculus in the Concurrent Logical Framework (CLF), and Watkins
\etal.~[\citeyear{watkins08}] develop a static type system and
operational semantics modeled on that of Gordon and
Jeffrey~[\citeyear{gordon03}] for checking correspondence properties of
protocols specified in the \piCalculus.  CLF employs higher-order
abstract syntax, linearity and a monadic encapsulation of certain
linear constructs that can identify objects such as traces up to
causal equivalence. Thus, CLF's \piCalculus encodings naturally induce
equivalences on traces satisfying commuting conversions among
synchonous operations. However, a non-trivial effort appears necessary
to compare CLF's notion of trace equivalence with others, because
traces are quotiented by a definitional equality by default and there
is no explicit notion of concurrency or residuation.

\Paragraph{Agda}
\cbstart \citet*{orchard15} \cbend present a translation from a functional
language with effects to a \piCalculus with session types and verify some
type-preservation properties of the translation in Agda.

\section{Conclusions and future work}
\label{sec:conclusion}

To the best of our knowledge, we are the first to report on a mechanised
formalisation of concurrency, residuation and causal equivalence for the
\piCalculus. We employed de Bruijn indices to represent binders and
names. Formalisations of \lambdaCalculi often employ this technique, but to
our knowledge only \citeauthor*{orchard15} \begin{changebar}also employ de
Bruijn indices in a mechanised formalisation of \piCalculus.\end{changebar}
Whilst de Bruijn indices incur a certain level of administrative overhead,
the use of dependent types helps tame their complexity: many invariants are automatically
checked by the type system rather than requiring additional explicit
reasoning.

Our work appears to be the first to align the notion of ``proved
transitions'' from \citeauthor*{boudol89}'s work on CCS with
``transition proofs'' in the \piCalculus. This hinges on the capability
to manipulate and perform induction or recursion over derivations, and
means we can leverage dependent typing so that residuation is defined
only for concurrent transitions, rather than on all pairs of
transitions. It is worth noting that while CLF's approach to encoding
\piCalculus automatically yields an equivalence on traces, it is unclear
(at least to us) whether this equivalence is similar to the one we
propose, or whether such traces can be manipulated explicitly as proof
objects if desired.

The most notable aspect of our development is the generalised diamond
lemma, which allows causally equivalent traces to have target states
which are not equal ``on the nose'' but only up to a precise braiding
which captures how binders were reordered. These braidings are more
explicit in a de Bruijn indices setting, since free as well as bound
names must be rewired when binders are transposed. Generalised
cofinality may be relevant to modelling concurrency in other languages
where concurrent transitions have effects which commute only up to some
equivalence relation, such as dynamic memory allocation.

\subsection{Future work}

One possible future direction would be to explore trace structures
explicitly quotiented by causal equivalence, such as dependence
graphs~\cite{mazurkiewicz87}, event structures~\cite{boudol89}, or rigid
families~\cite{cristescu2015}. We are also interested in extending our
approach to accommodate structural congruences, and in understanding
whether ideas from homotopy type theory~\cite{HoTTbook}, such as
quotients or higher inductive types, could be applied to ease reasoning
about \piCalculus traces modulo causal equivalence and structural
congruence.

\begin{changebar}
An interesting possibility would be to separately formalise the abstract
notion of a ``residuation system'' parameterised on a notion of cofinality.
One could then show that the \piCalculus (equipped with a particular notion of
name binding) admits such a residuation system, with cofinality suitably
instantiated. This would shed light on which aspects of concurrency and
causality are specific to the choice of name-binding formalism. Potentially
this modular approach would also make it easier to study variants of
\piCalculus where interaction arises from different communication patterns,
such as the join-calculus~\cite{fournet02} or polyadic
\piCalculus~\cite{carbone03}. Again, it might be possible to model concurrency
and causality in these settings independently of the rewiring issues
associated with permuting transitions that manipulate scope.
\end{changebar}

\section*{Acknowledgements}
We are grateful to our colleagues in the Programming Languages Interest
Group at Edinburgh for useful discussions, to V\'it \v{S}efl for
assistance with the Agda formalisation, and to the anonymous referees
for comments on the paper. Effort sponsored by the Air Force Office of
Scientific Research, Air Force Material Command, USAF, under grant
number FA8655-13-1-3006, and EPSRC, grant number EP/K034413/1. The U.S.
Government and University of Edinburgh are authorized to reproduce and
distribute reprints for their purposes notwithstanding any copyright
notation thereon.

\bibliographystyle{apalike} 

\pagebreak
\begin{appendix}

\section{Agda module structure}
\label{app:module-structure}

\figref{modules} summarises the module structure of the Agda
formalisation.

\begin{figure}[H]
\begin{nscenter}
{\small
\noindent \begin{tabular}{lp{10cm}}
\text{\emph{Utilities}}
\\
\tt{Ext}
& Extensions to Agda library, \url{https://github.com/rolyp/agda-stdlib-ext}
\\
\\
\text{\emph{Core modules}}
\\
\tt{Action}
& Actions $a$
\\
\tt{Action.Concur}
& Concurrent actions $a \concur a'$; residuals $\residual{a}{a'}$
\\
\tt{Action.Seq}
& Action sequences $\vec{a}$
\\
\tt{Braiding.Proc}
& Bound braids $\phi: P \boundBraid P'$
\\
\tt{Braiding.Transition}
& Residuals $\residual{t}{\phi}$ and $\residual{\phi}{t}$
\\
\tt{Name}
& Contexts $\Gamma$; names $x$
\\
\tt{Proc}
& Processes $P$
\\
\tt{ProofRelevantPi}
& Include everything; compile to build project
\\
\tt{ProofRelevantPiCommon}
& Common imports from standard library
\\
\tt{Ren}
& Renamings $\rho: \Gamma \to \Gamma'$
\\
\tt{Ren.Properties}
& Additional properties relating to renamings
\\
\tt{Transition}
& Transitions $t: P \transition{a} R$
\\
\tt{Transition.Concur}
& Concurrent transitions $t \concur t'$; residuals $\residual{t}{t'}$
\\
\tt{Transition.Concur.Cofinal}
& Cofinality witnesses $\gamma$
\\
\tt{Transition.Concur.Cofinal.Transition}
& Residuals $\residual{t}{\gamma}$ and $\residual{\gamma}{t}$
\\
\tt{Transition.Seq}
& Transition sequences
\\
\tt{Transition.Seq.Cofinal}
& Residuals $\residual{\vec{t}}{\gamma}$ and $\residual{\gamma}{\vec{t}}$; permutation equivalence $\alpha: \vec{t} \permEq \vec{u}$
\\
\tt{Transition.Seq.Cofinal.Cofinal}
& Proof that $\residual{\vec{t}}{\gamma}$ and $\residual{\gamma}{\vec{t}}$ are (heterogeneously) cofinal
\\
\\
\text{\emph{Common sub-modules}}
\\
\tt{.Ren}
& Renaming lifted to entity defined in parent module
\\ 
\\
\end{tabular}
}
\end{nscenter}
\crossrule
\caption{Module overview, release \ttt{0.3}}
\label{fig:modules}
\end{figure}

\end{appendix}

\end{document}